\documentclass[11pt]{amsart}
\usepackage{geometry}                % See geometry.pdf to learn the layout options. There are lots.
\usepackage{graphicx}
\usepackage{amssymb}
\usepackage{epstopdf}
\newtheorem{theorem}{Theorem}[section]
\newtheorem{lemma}{Lemma}[section]
\newtheorem{proposition}{Proposition}[section]
\newtheorem{definition}{Definition}[section]
\newtheorem{remark}{\emph{Remark}}[section]
\title{BKL conjecture in Bianchi VIII and IX with the ultrarelativistic fluid}
\author{Raisa Galimova}
\address{Department of Mathematics, ETH Zurich, Switzerland}
\email{raisa.galimova@math.ethz.ch}
\date{\today}                                           % Activate to display a given date or no date

\begin{document}
\maketitle
\begin{abstract}
We rigorously verify that in the spatially homogeneous spacetimes (Bianchi VIII and IX), the presence of matter does not affect the oscillatory behavior of the solutions to the Einstein field equations as first conjectured in \cite{BKL}. This paper is an extension of \cite{RT} and uses the same formalism. We use the ultrarelativistic equation of state.
\end{abstract}
\tableofcontents
\newpage
\section*{Overview}
We rigorously analyse the system of ordinary differential equations (ODE)
\\
\begin{eqnarray}
\label{eq:first}
-\frac{d}{d\tau}\alpha_{i}-\beta^{2}_{i}+\beta^{2}_{j}+\beta^{2}_{k}-2\beta_{j}\beta_{k}+\frac{2}{3}\gamma^{2}&=&0 \\
-\frac{d}{d\tau}\beta_{i}+\beta_{i}\alpha_{i}&=&0 \\
\frac{d}{d\tau}\gamma-\frac{1}{6}\gamma(\alpha_{1}+\alpha_{2}+\alpha_{3}) &=&0
\end{eqnarray}
\\
with $(i,j,k)\in C=\lbrace(1,2,3),(2,3,1),(3,1,2)\rbrace$, that are subject to the quadratic constraint
\\
\begin{equation}
\label{eq:last}
\alpha_{1}\alpha_{2}+\alpha_{2}\alpha_{3}+\alpha_{3}\alpha_{1}-\beta_{1}^{2}-\beta_{2}^{2}-\beta_{3}^{2}+2\beta_{1}\beta_{2}+2\beta_{2}\beta_{3}+2\beta_{3}\beta_{1}=4\gamma^{2} 
\end{equation}
\\
where $\alpha=\alpha(\tau)$, $\beta=\beta(\tau)$ and $\gamma=\gamma(\tau)$, and the construction is carried out on \linebreak $\tau\in[0,+\infty)$\footnote{We assume the singularity to be located at $\tau=+\infty$.}. The equations (\ref{eq:first})-(\ref{eq:last}) are equivalent to the spatially homogeneous Einstein field equations with the ultrarelativistic fluid, see Section 1 and the beginning of Section 3.
\\
\\
In 1970, Belinski, Khalatnikov and Lifshitz (BKL) conjectured that there is a generic class of spacelike singularities, near which generic spacetimes behave like spatially homogeneous spacetimes\footnote{For spatially homogeneous spacetimes the Einstein field equations reduce to a system of ODEs.} \cite{BKL}. The latter exhibit an oscillatory behaviour described by the Gauss map, which is known to be chaotic. BKL's idea is that there is a generic class of solutions to the Einstein field equations that are described by decomposing the semi-infinte time axis into an infinite number of finite subintervals and by using the Kasner evolution on each subinterval\footnote{Such behaviour is also called Mixmaster dynamics.}. These subintervals they called epochs. BKL conjectured that the presence of an ultra relativistic fluid does not affect this regime. The latter statement is rigorously verified in the present paper.
\\
\\
We reduce the analysis of (\ref{eq:first})-(\ref{eq:last}) to the analysis of a Poincar$\acute{\mathrm{e}}$ map on a five-dimensional Poincar$\acute{\mathrm{e}}$ section. Modulo scaling symmetry of the solutions to (\ref{eq:first})-(\ref{eq:last}), the set of parameters is $(\mathrm{\bf{h}},w,q,z)$. The parameters $\mathrm{\bf{h}}$, $w$ and $q$ already appear in the vacuum case, see \cite{RT}. The parameter $z$ describes the fluid. The important observation of \cite{BKL}, that the transition period between the Kasner epochs is small compared to the duration of the epochs, is encapsulated in the smallness of the parameter $\mathrm{\bf{h}}$. Using the scale invariance of the solutions to (\ref{eq:first})-(\ref{eq:last}), every orbit of the 4-dimensional discrete dynamical system is lifted to a \emph{unique} orbit of the 5-dimensional discrete dynamical system through the map $\Lambda$, see Proposition \ref{prop:3.3}. 
\\
\\
The oscillatory behaviour of the solutions to (\ref{eq:first})-(\ref{eq:last}) can be understood in terms of the dynamics of $\beta_{1},\beta_{2},\beta_{3}$. Within each epoch, the logarithms of $|\beta_{1}|,|\beta_{2}|,|\beta_{3}|$ are approximately linear functions with slopes $\alpha_{1},\alpha_{2},\alpha_{3}$, with one component of $\alpha$ being positive and the other two negative at any point. At any time, at least two of the $\beta$'s are so small that they can be neglected to obtain the leading order approximation.
\\
\\
In Section 3, we show the existence of the transfer maps (i.e. maps from Poincar$\acute{\mathrm{e}}$ section to Poincar$\acute{\mathrm{e}}$ section), that map the state $\Phi(\tau_{i})$ to an \emph{earlier} state $\Phi(\tau_{i-1})$, where $\Phi=\alpha\oplus\beta\oplus\gamma\in\mathbb{R}^{3}\oplus\mathbb{R}^{3}\oplus\mathbb{R}$ is a solution to (\ref{eq:first})-(\ref{eq:last}). We also show that the transfer maps $(\mathcal{P}_{L},\Pi)$ are close to the approximate transfer maps $(\mathcal{P}_{L},\mathcal{Q}_{L})$ in Definition \ref{def:3.16}, and give explicit error bounds. See Proposition \ref{prop:3.3}.
\\
\\
In Section 4, we analyse the dynamics of the approximate transfer maps and show that part of it is related to the dynamics of the Gauss map. The Gauss map is the left-shift on the continued fraction expansion of $w$. We use the domain of definition of the approximate transfer maps with full Lebesgue measure in $w$, where the continued fraction expansion grows at most polynomially. See Proposition \ref{prop:4.4}. 
\\
\\
In Section 6, we construct semi-global solutions by combining the results of Proposition \ref{prop:3.3} and the theorem in Section 5. See Theorems \ref{main_theorem_2}, \ref{main_theorem_3}.
\\
\\
\\
Starting from Section 3, the notation and enumeration of the paper follows those in \cite{RT}. Every statement has an analogue for the vacuum case in \cite{RT}.
\\
\\
\\
I thank Michael Reiterer and Eugene Trubowitz for their support and encouragement.
%PRIMARY SYSTEM OF EQUATIONS%
\newpage
\section{Primary system of equations}
We start with arbitrary frame fields $e_{0}$, $e_{i}$, $e_{j}$, $e_{k}$, where $(i,j,k)$ belong to the set of cyclic permutations of $(1,2,3)$. The goal is then to construct a spacetime, given that some $\alpha_{i}=\alpha_{i}(\tau)$ and $\beta_{i}=\beta_{i}(\tau)$ are $\emph{defined}$ through their commutators (no summation convention):

\begin{equation}
\label{eq:commutator_def_space}
[e_{j},e_{k}]=:e^{\zeta}\beta_{i}e_{i}
\end{equation}

and similarly for $\alpha$
\\
\begin{equation}
\label{eq:commutator_def_time}
[e_{0},e_{i}]=: -\frac{1}{2}e^{\zeta}\alpha_{i}\: e_{i},
\end{equation}
\\
where $\zeta$ is some (not yet explicitly defined) function. 
Ensuring that the Jacobi Identity holds brings
\\
\begin{equation}
0=-\frac{d}{d\tau}\beta_{i}+\beta_{i}\alpha_{i}
\end{equation}
\\
which is exactly the original equation (1.1b) in \cite{RT}. 
\\
Now, assuming that $e_{i}$ are an orthonormal frame, we have the relation
\\
\[g(\nabla_{e_{a}}e_{b},e_{c})=\frac{1}{2}(-g(e_{a},[e_{b},e_{c}])-g(e_{b},[e_{a},e_{c}])+g(e_{c},[e_{a},e_{b}]))\]
\\
Further,
\begin{equation}
R_{abcd}=g([\nabla_{e_{c}},\nabla_{e_{d}}]e_{b}-\nabla_{[e_{c},e_{d}]}e_{b},e_{a})
\end{equation}
\\
Therefore, for the \emph{vacuum} equations  we get
\\
\[-\frac{d}{d\tau}\alpha_{i}-\beta^{2}_{i}+\beta^{2}_{j}+\beta^{2}_{k}-2\beta_{j}\beta_{k} =0\]
\\
and the constraint
\\
\[\alpha_{1}\alpha_{2}+\alpha_{2}\alpha_{3}+\alpha_{3}\alpha_{1}-\beta_{1}^{2}-\beta_{2}^{2}-\beta_{3}^{2}+2\beta_{1}\beta_{2}+2\beta_{2}\beta_{3}+2\beta_{3}\beta_{1}=0.\]
\\
These are the exact equations that appear in \cite{RT}.
\\
\subsection{Putting in the relativistic fluid}
We have $R_{ab}=T_{ab}-\frac{1}{2}Tg_{ab}$, where $T_{ab}$ is the energy-momentum tensor given by
\\
\begin{equation}
\label{eq:energy_momentum_tensor}
T_{ab}=(p+\epsilon)u_{a}u_{b}+pg_{ab}
\end{equation}
\\
with pressure $p=p(\epsilon)$ and total energy density $\epsilon$ . 
\\
The equation (\ref{eq:energy_momentum_tensor}) is the energy-momentum tensor of a perfect fluid. We consider the special case of the fluid at rest with a 4-velocity $u^{a}=(1,0,0,0)=e_{0}$. Raising an index, we get $T^{\mu}_{\nu}=diag(-\epsilon,p,p,p)$. Therefore, for $T$ (the trace of the energy-momentum tensor) we have
\\
\begin{equation}
g^{ab}T_{ab}=T=T^{0}_{0}+T^{1}_{1}+T^{2}_{2}+T^{3}_{3}=3p-\epsilon
\end{equation}
\\
Now, using the obtained results for the Ricci tensor components, we get:
\\
\begin{equation}
\label{eq:ricci_time}
R_{00}=T_{00}-\frac{1}{2}Tg_{00}=\frac{1}{2}(3p+\epsilon)
\end{equation}
\\
and (summation over $i$ is not implied)
\\
\begin{equation}
\label{eq:ricci_space}
R_{ii}=T_{ii}-\frac{1}{2}Tg_{ii}=-\frac{1}{2}(p-\epsilon)
\end{equation}
\\
Also, $R_{0i}=0$ and $R_{ij}=0$ for $i\neq j$. Using the change of variables $P:=e^{-2\zeta}p$ and $E:=e^{-2\zeta}\epsilon$, the equation (\ref{eq:ricci_space}), the new version of the equation (1.1a) in \cite{RT} is
\\
\begin{equation}
\label{eq:1.1a}
-\frac{d}{d\tau}\alpha_{i}-\beta^{2}_{i}+\beta^{2}_{j}+\beta^{2}_{k}-2\beta_{j}\beta_{k}=P-E
\end{equation}
\\
And the equation (1.1c) in \cite{RT}, i.e. the constraint equation, is now
\\
\begin{equation}
-\frac{d}{d\tau}(\alpha_{1}+\alpha_{2}+\alpha_{3})+\alpha_{1}\alpha_{2}+\alpha_{2}\alpha_{3}+\alpha_{3}\alpha_{1}=3P+E
\end{equation}
\\
\\
From now on we assume that $p=\frac{1}{3}\epsilon$, i.e. ultra relativistic equation of state.\footnote{The ultrarelativistic equation of state $p=\frac{1}{3}\epsilon$ follows directly from the general equation for gas pressure $p=\frac{1}{3}\int_{0}^{\infty}vkn(k)dk$, where $n(k)dk$ is the number density of particles in the momentum interval $k$ to $k+dk$. Using $E=kc$ for ultrarelativistic particles, we get the equation.}.\\
\\
\\
\\
Now, we also have the conservation equation for the energy-momentum tensor, i.e.
\\
\begin{equation}
\nabla_{\alpha}T^{\alpha\beta}=0
\end{equation}
\\
Substituting the definition for $T^{\alpha\beta}$ we get
\begin{equation}
\nabla_{\alpha}((p+\epsilon)u^{\alpha}u^{\beta}+pg^{\alpha\beta})=(p+\epsilon)u^{\alpha}\nabla_{\alpha}u^{\beta}+g^{\alpha\beta}\nabla_{\alpha}p+u^{\beta}\nabla_{\alpha}[(p+\epsilon)u^{\alpha}]=0
\end{equation}
\\
Using the fact that $u_{\beta}u^{\beta}=-1$ and contracting the equation above with $u_{\beta}$, we get
\\
\begin{equation}
u^{\alpha}\partial_{\alpha}\epsilon+(p+\epsilon)\nabla_{\alpha}u^{\alpha}=0
\end{equation}
\\
Using the definitions of $P$ and $E$  and the homogeneity assumption (i.e. only makes sense to take time derivatives) we obtain for the conservation equation:
\\
\begin{equation}
\label{eq:working}
e^{\zeta}\frac{d}{d\tau}(e^{2\zeta}E)+e^{2\zeta}(P+E)\frac{1}{2}e^{\zeta}(\alpha_{i}+\alpha_{j}+\alpha_{k})=0,
\end{equation}
\\
where $\nabla_{e_{i}}e_{0}=\frac{1}{2}e^{\zeta}\alpha_{i}e_{i}$ was used to compute $\nabla_{\alpha}u^{\alpha}=\mathrm{tr}(x\mapsto\nabla_{x}u)=\frac{1}{2}e^{\zeta}(\alpha_{i}+\alpha_{j}+\alpha_{k})$. Now, using the definition of $\zeta$, we have $\frac{d}{d\tau}\zeta=-\frac{1}{2}(\alpha_{i}+\alpha_{j}+\alpha_{k})$. Substituting into (\ref{eq:working}), we get
\\
\begin{equation}
e^{\zeta}(-e^{2\zeta}(\alpha_{i}+\alpha_{j}+\alpha_{k})E+e^{2\zeta}\frac{d}{d\tau}E)+\frac{1}{2}e^{3\zeta}(P+E)(\alpha_{i}+\alpha_{j}+\alpha_{k})=0
\end{equation}
\\
which simplifies to
\\
\begin{equation}
\frac{d}{d\tau}E+\frac{1}{2}(\alpha_{1}+\alpha_{2}+\alpha_{3})(P-E)=0.
\end{equation}
\\
Therefore, the new field equations, i.e. the new (1.1a) and (1.1b) are:
\\
\begin{eqnarray*}
-\frac{d}{d\tau}\alpha_{i}-\beta^{2}_{i}+\beta^{2}_{j}+\beta^{2}_{k}-2\beta_{j}\beta_{k}&=&P-E \\ 
-\frac{d}{d\tau}\beta_{i}+\beta_{i}\alpha_{i}&=&0. 
\end{eqnarray*}
\\
The new constraint (i.e. new (1.1c)) and conservation equation:
\\
\begin{eqnarray*}
\alpha_{1}\alpha_{2}+\alpha_{2}\alpha_{3}+\alpha_{3}\alpha_{1}-\beta_{1}^{2}-\beta_{2}^{2}-\beta_{3}^{2}+2\beta_{1}\beta_{2}+2\beta_{2}\beta_{3}+2\beta_{3}\beta_{1} &=&4E \\
\frac{d}{d\tau}E+\frac{1}{2}(P-E)(\alpha_{i}+\alpha_{j}+\alpha_{k}) &=&0.
\end{eqnarray*}
\\
By our ultrarelativistic assumption, the last equation simplifies to \linebreak $\frac{d}{d\tau}E-\frac{1}{3}(\alpha_{1}+\alpha_{2}+\alpha_{3})E=0$.
\\
\\
These four equations are equivalent to the Einstein field equations for the relativistic fluid.

%SECTION 2%
\newpage
\section{Reformulation of the conservation equation}
We have $\frac{d}{d\tau}\beta_{i}=\beta_{i}\alpha_{i}$. Therefore, $\frac{d}{d\tau}\vert\beta_{i}\beta_{j}\beta_{k}\vert=(\alpha_{i}+\alpha_{j}+\alpha_{k})\vert\beta_{i}\beta_{j}\beta_{k}\vert$. Now, we can generalise it  and write
\\
\begin{equation}
\label{eq:section2_first}
\frac{d}{d\tau}(\vert\beta_{i}\beta_{j}\beta_{k}\vert^{\chi})=\chi(\alpha_{i}+\alpha_{j}+\alpha_{k})\vert\beta_{i}\beta_{j}\beta_{k}\vert^{\chi}
\end{equation}
\\
where $\chi$ is a constant. Next step is to show that (\ref{eq:section2_first}) is equivalent to the conservation equation 
\\
\begin{equation}
\label{eq:conservation}
\frac{d}{d\tau}E+\frac{1}{2}(P-E)(\alpha_{i}+\alpha_{j}+\alpha_{k})=0
\end{equation}
\\
We're working in Bianchi VIII and IX, therefore we can multiply both sides of (\ref{eq:conservation}) by $\vert\beta_{i}\beta_{j}\beta_{k}\vert\neq0$. We get
\\
\[\left(\frac{d}{d\tau}E\right)\vert\beta_{i}\beta_{j}\beta_{k}\vert+\frac{1}{2}(P-E)\underbrace{(\alpha_{i}+\alpha_{j}+\alpha_{k})\vert\beta_{i}\beta_{j}\beta_{k}\vert}_\mathrm{\frac{d}{d\tau}\vert\beta_{i}\beta_{j}\beta_{k}\vert}=0\]
\\
Using $P=\frac{E}{3}$ and (\ref{eq:section2_first}), we find that $\chi=-\frac{1}{3}$, and we can write
\\
\[\left(\frac{d}{d\tau}E\right)\vert\beta_{i}\beta_{j}\beta_{k}\vert^{-1/3}-\frac{1}{3}E(\alpha_{i}+\alpha_{j}+\alpha_{k})\vert\beta_{i}\beta_{j}\beta_{k}\vert^{-1/3}=0\]
\\
which by the Leibnitz rule is equivalent to
\\
\begin{equation}
\label{eq:section2_two}
\frac{d}{d\tau}(E\vert\beta_{i}\beta_{j}\beta_{k}\vert^{-1/3})=0.
\end{equation}
%\\
%For a general equation of state $P=\kappa E$, equation (\ref{eq:section2_two}) becomes
%\\
%\begin{equation}
%\label{eq:conservation_final}
%\frac{d}{d\tau}(E\vert\beta_{i}\beta_{j}\beta_{k}\vert^\frac{\kappa-1}{2})=0.
%\end{equation}
%\\
%The main conclusion following from (\ref{eq:conservation_final}): $E\vert\beta_{i}\beta_{j}\beta_{k}\vert^\frac{\kappa-1}{2}$ is a conserved quantity.

%SECTION3%
\newpage
\section{Construction of transfer maps}
\noindent Since E is a positive quantity, set $E=:\gamma^{2}$. 
\\
Introduce $n=(n_{1},n_{2},n_{3},m)$. Then we have:
%Definition 3.1%
\begin{definition} 
\label{def:3.1}
$\forall\Phi=\alpha\oplus\beta\oplus\gamma\in C^{\infty}((\tau_{0},\tau_{1}),\mathbb{R}^{6}\oplus\mathbb{R})$, $\forall\mathrm{\bf{h}}>0$, $\forall n\in\mathbb{R}^{4}$, associate a field
\[\frak a\left[\Phi,\mathrm{\bf{h}},n\right]\oplus\frak b\left[\Phi,\mathrm{\bf{h}},n\right]\oplus c\left[\Phi,\mathrm{\bf{h}},n\right]\oplus\frak d\left[\Phi,\mathrm{\bf{h}},n\right]:(\tau_{0},\tau_{1})\rightarrow\mathbb{R}^{3}\oplus\mathbb{R}^{3}\oplus\mathbb{R}\oplus\mathbb{R}\]
by
\\
\begin{eqnarray}
\label{eq:frak_a}
\mathfrak{a}_{i}\left[\Phi,\mathrm{\bf{h}},n\right]&=&-\mathrm{\bf{h}}\frac{\mathrm{d}}{\mathrm{d}\tau}\alpha_{i}-(n_{i}\beta_{i})^{2}+(n_{j}\beta_{j}-n_{k}\beta_{k})^{2}+\frac{2}{3}(m\gamma)^{2}\\
\label{eq:frak_b}
\mathfrak{b}_{i}\left[\Phi,\mathrm{\bf{h}},n\right]&=&-\mathrm{\bf{h}}\frac{\mathrm{d}}{\mathrm{d}\tau}\beta_{i}+\beta_{i}\alpha_{i}\\
\label{eq:frak_c}
c\left[\Phi,\mathrm{\bf{h}},n\right]&=&-4(m\gamma)^{2}+\sum_{(i,j,k)}(\alpha_{j}\alpha_{k}-(n_{i}\beta_{i})^{2}+2n_{j}n_{k}\beta_{j}\beta_{k}) \\
\label{eq:frak_d}
\mathfrak{d}\left[\Phi,\mathrm{\bf{h}},n\right]&=&\mathrm{\bf{h}}\frac{\mathrm{d}}{\mathrm{d}\tau}\gamma-\frac{1}{6}\gamma (\alpha_{1}+\alpha_{2}+\alpha_{3})
\end{eqnarray}
where $(i,j,k)\in C$, set of cyclic permutations of (1,2,3). Define 
\begin{equation*}
\frak a_{i}\left[\Phi,\mathrm{\bf{h}},n_{1},n_{2}\right]:=\frak a_{i}\left[\Phi,\mathrm{\bf{h}},n_{1}\right]-\frak a_{i}\left[\Phi,\mathrm{\bf{h}},n_{2} \right]
\end{equation*}
\end{definition}
%Definition 3.2%
\begin{definition}
\label{def:3.2}
Introduce the vectors $B_{1}=(1,0,0,0)$, $B_{2}=(0,1,0,0)$, $B_{3}=(0,0,1,0)$ and $Z=(1,1,1,1)$ that will play the role of $n$. 
\end{definition}

\begin{proposition} 
\label{prop:3.1}
($\mathrm{\bf{Global Symmetries}}$): 
%Since $\alpha$ scales as $A$, $\beta$ scales as $A$, and $\gamma$ scales as $\sqrt{E}$ which also scales as $A$. $\chi(\tau)=p\tau+q$, with $p>0$, and $A>0$ a constant.
Set $\chi(\tau)=p\tau+q$ with $p>0$, then
\\
\begin{equation}
\label{eq:prop:3.1}
(\mathfrak{a},\mathfrak{b},c,\mathfrak{d})\left[A(\Phi\circ\chi),\frac{1}{p}A\mathrm{\bf{h}},n\right]=A^{2}\left((\mathfrak{a},\mathfrak{b},c,\mathfrak{d})\left[\Phi,\mathrm{\bf{h}},n\right]\circ\chi\right),
\end{equation}
\end{proposition}
\begin{remark}
The equations $(\frak a, \frak b, c, \frak d)[\Phi,\mathrm{\bf{h}},Z]=0$ are equivalent to (\ref{eq:first})-(\ref{eq:last}) for any $\mathrm{\bf{h}}>0$.
\end{remark}
\begin{proposition} 
\label{prop:3.2}
Recall Definition \ref{def:3.1}. For all $\Phi=\alpha\oplus\beta\oplus\gamma\in C^{\infty}((\tau_{0},\tau_{1}),\mathbb{R}^{6}\oplus\mathbb{R})$, all $\mathrm{\bf{h}}>0$, all $n\in\mathbb{R}^{4}$, we have
\\
\begin{equation}
\label{eq:prop3.2}
0=-\mathrm{\bf{h}}\frac{d}{d\tau}c+\sum_{(i,j,k)\in C}(-\alpha_{j}\frak a_{k}-\alpha_{k}\frak a_{j}+2(n_{i})^{2}\beta_{i}\frak b_{i}-2n_{j}n_{k}\beta_{j}\frak b_{k}-2n_{j}n_{k}\beta_{k}\frak b_{j})-8m^{2}\gamma\frak d
\end{equation}
\end{proposition}
\noindent \underline{Proof}: Straightforward.
\\
\\
%Definition 3.3%
\begin{definition}
\label{def:3.3}
$\forall \mathrm{\bf{h}}\in (0,\infty)$, $\forall \Phi=\alpha\oplus\beta\oplus\gamma$ with $\beta_{1},\beta_{2},\beta_{3}\neq0$ define
\\
\begin{eqnarray*}
A_{m}\left[ \Phi \right]&:=&\sqrt{\vert\alpha_{m}\vert^{2}+\vert\beta_{m}\vert^{2}} \\
\phi_{m}\left[\Phi\right]&:=&-\mathrm{arcsinh}\frac{\alpha_{m}}{\vert\beta_{m}\vert} \\
\xi_{m}\left[\Phi,\mathrm{\bf{h}}\right]&:=&\mathrm{\bf{h}}\mathrm{log}\vert\frac{1}{2}\beta_{m}\vert \\
\alpha_{m,n}\left[\Phi\right]&:=&\alpha_{m}+\alpha_{n},\:\:\:\:\:\:\:\:\:\xi_{m,n}:=\xi_{m}\left[\Phi,\mathrm{\bf{h}}\right]+\xi_{n}\left[\Phi,\mathrm{\bf{h}}\right] \\
\rho&:=&\gamma\vert\beta_{1}\beta_{2}\beta_{3}\vert^{-1/6}
\end{eqnarray*}
with $m=1,2,3$.
\end{definition}
\begin{remark} 
The same as Remark 3.2 in \cite{RT}
\end{remark}

\begin{lemma}
\label{lemma:3.1} 
The same as Lemma 3.1 in \cite{RT}
\end{lemma}
%Definition 3.4%
\begin{definition}
\label{def:3.4}
Set $S_{3}=$the set of all permutations of $(1,2,3)$.
\end{definition}
%Definition 3.5%
\begin{definition} 
\label{def:3.5}
For all $\sigma_{*}\in\lbrace-1,+1\rbrace^{3}$ let $\mathcal{D}(\sigma_{*})$ be the set of all $\Phi=\alpha\oplus\beta\oplus\gamma\in\mathbb{R}^{3}\oplus\mathbb{R}^{3}\oplus\mathbb{R}$ with $(\mathrm{sgn}\beta_{1},\mathrm{sgn}\beta_{2},\mathrm{sgn}\beta_{3})=\sigma_{*}$. For all $\tau_{0},\tau_{1}\in\mathbb{R}$ with $\tau_{0}<\tau_{1}$ let $\mathcal{E}(\sigma_{*};\tau_{0},\tau_{1})$ be the set of all continuous maps $\Phi:[\tau_{0},\tau_{1}]\rightarrow\mathcal{D}(\sigma_{*})$.
\end{definition}
%Definition 3.6%
\begin{definition}
\label{def:3.6}
$\pi=(a,b,c)\in S_{3}$, $\mathrm{\bf{h}}>0$, $\sigma_{*}\in\lbrace-1,+1\rbrace^{3}$. Define $d_{\mathcal{D}(\sigma_{*}),(\pi,\mathrm{\bf{h}})}:\mathcal{D}(\sigma_{*})\times\mathcal{D}(\sigma_{*})\rightarrow [0,\infty)$ by
\begin{eqnarray*}
d_{\mathcal{D}(\sigma_{*}),(\pi,\mathrm{\bf{h}})}(\Phi,\Psi)=\mathrm{max}\lbrace \vert A_{a}[\Phi]-A_{a}[\Psi]\vert&,& \vert\mathrm{\bf{h}}\frac{\phi_{a}[\Phi]}{A_{a}[\Phi]}-\mathrm{\bf{h}}\frac{\phi_{a}[\Psi]}{A_{a}[\Psi]}\vert, \\
\vert\alpha_{b,a}[\Phi]-\alpha_{b,a}[\Psi]\vert&,&\vert\xi_{b,a}[\Phi,\mathrm{\bf{h}}]-\xi_{b,a}[\Psi,\mathrm{\bf{h}}]\vert, \\
 \vert\alpha_{c,a}[\Phi]-\alpha_{c,a}[\Psi]\vert&,& \vert\rho^{2}[\Phi]-\rho^{2}[\Psi]\vert\rbrace
\end{eqnarray*}
\\
and
\\
\begin{equation*}
\not{d}_{\mathcal{D}(\sigma_{*})}(\Phi,\Psi)=\underset{i=1,2,3}{\mathrm{max}}\lbrace\vert\alpha_{i}[\Phi]-\alpha_{i}[\Psi]\vert, \vert\xi_{i}[\Phi,\mathrm{\bf{h}}]-\xi_{i}[\Psi,\mathrm{\bf{h}}]\vert, \vert \rho^{2}[\Phi]-\rho^{2}[\Psi]\vert\rbrace
\end{equation*}
\\
Both $d(\Phi,\Psi)$ and $\not{d}(\Phi,\Psi)$ satisfy the three defining properties of a metric ($d(\Phi,\Psi)\geq 0$, with equality iff $\Phi=\Psi$; $d(\Phi,\Psi)=d(\Psi,\Phi)$ and $d(\Phi,\Psi)\leq d(\Phi,\Psi')+d(\Psi',\Phi)$. Analogue for $\not{d}$). Therefore, $(\mathcal{D}(\sigma_{*}),d_{\mathcal{D}(\sigma_{*}),(\pi,\mathrm{\bf{h}})})$ and $(\mathcal{D}(\sigma_{*}),\not{d}_{\mathcal{D}(\sigma_{*}),\mathrm{\bf{h}}})$ are metric spaces.
\end{definition}
\begin{definition}
\label{def:3.7}
Stays the same as Definition 3.7 in \cite{RT}.
\end{definition}
\begin{lemma}
\label{lemma:3.2}
Stays the same as in \cite{RT}. The addition of $\rho^{2}$ terms to both $d$ and $\not{d}$ doesn't make a difference because $C,D\geq1$ so both old inequalities a) and b) also hold for the case with the ultra relativistic fluid. 
\end{lemma}
\begin{definition}
\label{def:3.8}
Let $\mathcal{X}=\mathcal{D}(\sigma_{*})$ or $\mathcal{X}=\mathcal{E}(\sigma_{*};\tau_{0},\tau_{1})$. For all $\delta\geq0$ and $\Phi\in\mathcal{X}$ and $\pi\in S_{3}$ and $\mathrm{\bf{h}}>0$, set $B_{\mathcal{X},(\pi,\mathrm{\bf{h}})}[\delta,\Phi]=\lbrace\Psi\in\mathcal{X} | d_{\mathcal{X},(\pi,\mathrm{\bf{h}})}(\Phi,\Psi)\leq\delta\rbrace$.
\end{definition}
%Definition 3.9 REFERENCE FIELD%
\begin{definition} 
\label{def:3.9}
($\textbf{Reference field}$) Recall Definition \ref{def:3.3}. For all $\pi\in S_{3}$, $(\mathrm{\bf{h}},w,q,z)\in(0,\infty)^{4}$, the reference field $\Phi_{0}=\Phi_{0}(\pi,(\mathrm{\bf{h}},w,q,z),\sigma_{*})$ is given by
\\
\begin{eqnarray}
\label{ref_field_A}
A_{a}\left[\Phi_{0}\right](\tau)&=&1\\
\label{ref_field_theta}
\theta_{a}\left[\Phi_{0},\mathrm{\bf{h}}\right](\tau)&=&0\\
\label{ref_field_alpha_b}
\alpha_{b,a}\left[\Phi_{0}\right](\tau)&=&-(1+w)^{-1}\\
\label{ref_field_alpha_c}
\alpha_{c,a}\left[\Phi_{0}\right](\tau)&=&-(1+w)\\
\label{ref_field_xi_b}
\xi_{b,a}\left[\Phi_{0},\mathrm{\bf{h}}\right](\tau)&=&-1-\mathrm{\bf{h}}\mathrm{log}2-(1+w)^{-1}\tau\\
\label{ref_field_xi_c}
\xi_{c,a}\left[\Phi_{0},\mathrm{\bf{h}}\right](\tau)&=&-(1+w)q-\mathrm{\bf{h}}\mathrm{log}2-(1+w)\tau\\
\label{ref_field_gamma}
\gamma\left[\Phi_{0},h\right](\tau)&=&\gamma_{0}\mathrm{exp}\left[-\frac{1}{\mathrm{\bf{h}}}\frac{1}{6}\frac{2+w(2+w)}{1+w}\tau\right]\left(\mathrm{Cosh}(\frac{1}{\mathrm{\bf{h}}}\tau)\right)^{1/6}
\end{eqnarray}
Define $\sqrt{z}:=\rho[\Phi_{0}]$
\end{definition}

\begin{lemma}
\label{lemma:3.3}
Let $\Phi_{0}$ be as in Definition \ref{def:3.9}. Then $(\frak a,\frak b, c, \frak d)[\Phi_{0},\mathrm{\bf{h}},B_{a}]=0$.
\end{lemma}

%Definition 3.10%
\begin{definition}
\label{def:3.10}
$\forall \mathrm{\bf{f}}=(\mathrm{\bf{h}},w,q,z) \in(0,\infty)^{4}$ set
\begin{eqnarray*}
\tau_{-}(f)&=&-\left(1-\frac{1}{2+w} \right)\mathrm{min}(1,q)<0 \\
\tau_{+}(f)&=&1+\frac{1}{w}>0
\end{eqnarray*}
\end{definition}
\emph{Motivation for Definition \ref{def:3.10}}.
The bounce of $\beta_{a}$ happens at $\tau=0$, then the next bounce (towards the right) will be by $\beta_{b}$, i.e. $\beta_{b}=1$ and $\xi_{b}=0$ (per definition). This is exactly $\tau_{+}$. We have: $\xi_{b}=-1-\mathrm{\bf{h}}\mathrm{log}2-\frac{\tau}{1+w}-\xi_{a}\approx -1-\mathrm{\bf{h}}\mathrm{log}2-\frac{\tau}{1+w}-(-\tau)$. Therefore, setting $\xi_{b}=0$ we get $\tau=\tau_{+}$. For $\tau_{-}$ we consider the bounce towards the left, i.e. for $\xi_{c}$, this refers to $\tau<0$. By putting $\xi_{c}=0$ and taking the approximation $\xi_{a}\approx\tau$, we get the $\tau_{-}$. Note that in both cases we ignored the $\mathrm{\bf{h}}\mathrm{log}2$ terms as small.
\\
%Technical Lemma 1%
\begin{lemma} 
\label{lemma:3.4}
($\textbf{Technical Lemma 1}$)

Let $\pi=(a,b,c)\in S_{3}$, $\textbf{f}=(\mathrm{\bf{h}},w,q,z)\in(0,\infty)^{4}$, $\sigma_{*}\in\lbrace-1,+1\rbrace$. Fix $\delta>0$, $\epsilon_{-}\in(0,-\tau_{-})$, $\epsilon_{+}\in(0,\tau_{+})$ where $\tau_{\pm}=\tau_{\pm}(\textbf{f})$. Set \\
\begin{eqnarray*}
\tau_{0-}&=&\tau_{-}+\epsilon_{-}<0 \\
\tau_{0+}&=&\tau_{+}-\epsilon_{+}>0 \\
\Phi_{0}&=&\Phi_{0}(\pi,\textbf{f},\sigma_{*})\big|_{[\tau_{0-},\tau_{0+}]} \\
\mathcal{E}&=&\mathcal{E}(\sigma_{*};\tau_{0-},\tau_{0+}).
\end{eqnarray*}
Assume $\delta\leq 2^{-4}\mathrm{min}\lbrace 1,w,\epsilon_{-},\frac{\epsilon_{+}}{\tau_{+}\tau_{0+}}\rbrace$ holds.
\end{lemma}

The estimates in \cite{RT} for $\beta$'s and their products still hold and are
\begin{eqnarray*}
\vert\beta_{b}\beta_{a}\vert&\leq&2\mathrm{exp}\left(-\frac{1}{4\mathrm{\bf{h}}}\right) \\
\vert\beta_{c}\beta_{a}\vert&\leq&2\mathrm{exp}\left(-\frac{1}{2\mathrm{\bf{h}}}\epsilon_{-}\right) \\
\vert\beta_{b}\vert^{2}&\leq&2^{4}\mathrm{exp}\left(-\frac{1}{\mathrm{\bf{h}}}\mathrm{min}\lbrace\epsilon_{-},\frac{\epsilon_{+}}{\tau_{+}}\rbrace\right) \\
\vert\beta_{c}\vert^{2}&\leq&2^{4}\mathrm{exp}\left(-\frac{2}{\mathrm{\bf{h}}}\epsilon_{-}\right)
\end{eqnarray*}
Therefore, we only need an estimate for $\gamma^{2}[\Phi]$.
\\
We have  
\[\vert\gamma^{2}[\Phi]\vert\leq \rho^{2}[\Phi]\big| 2\mathrm{exp}\left(-\frac{1}{4\mathrm{\bf{h}}}\right)\big|^{1/3}\big|4\text{exp}\left(-\frac{1}{\mathrm{\bf{h}}}\epsilon_{-}\right)\big|^{1/3}\leq2z\mathrm{exp}\left(-\frac{1}{12\mathrm{\bf{h}}}(1+4\epsilon_{-})\right)\]
\\
\\
The new total estimate is then
\[\mathrm{max}\lbrace\vert\beta_{b}\vert^{2},\vert\beta_{c}\vert^{2},\vert\beta_{b}\beta_{a}\vert,\vert\beta_{c}\beta_{a}\vert,\vert\gamma\vert^{2}\rbrace\leq\mathrm{max}\lbrace 2^{4},2z\rbrace \mathrm{exp}\left(-\frac{1}{12\mathrm{\bf{h}}}\mathrm{min}\lbrace1,\epsilon_{-},\frac{\epsilon_{+}}{\tau_{+}}, 1+4\epsilon_{-}\rbrace\right)\]
Since $\epsilon_{-}>0$ per definition, we have $\epsilon_{-}<1+4\epsilon_{-}$ always. Also, $1+4\epsilon_{-}\geq3\epsilon_{-}$. Therefore,
\[\mathrm{max}\lbrace\vert\beta_{b}\vert^{2},\vert\beta_{c}\vert^{2},\vert\beta_{b}\beta_{a}\vert,\vert\beta_{c}\beta_{a}\vert,\vert\gamma\vert^{2}\rbrace\leq2^{4}\mathrm{max}\lbrace 1,z\rbrace \mathrm{exp}\left(-\frac{1}{4\mathrm{\bf{h}}}\mathrm{min}\lbrace1,\epsilon_{-},\frac{\epsilon_{+}}{\tau_{+}}\rbrace\right).\]

\begin{lemma} 
\label{lemma:3.5}
Recall Definitions \ref{def:3.1} and \ref{def:3.2} .  $\forall$ ($a,b,c$)$\in S_{3}$ we have
\begin{eqnarray*}
\frak a_{a}\left[\Phi,\mathrm{\bf{h}},Z,B_{a}\right]&=&+\beta_{b}^{2}+\beta_{c}^{2}-2\beta_{b}\beta_{c}+\frac{2}{3}\gamma^{2} \\
\frak a_{b}\left[\Phi,\mathrm{\bf{h}},Z,B_{a}\right]&=&-\beta_{b}^{2}+\beta_{c}^{2}-2\beta_{a}\beta_{c}+\frac{2}{3}\gamma^{2} \\
\frak a_{c}\left[\Phi,\mathrm{\bf{h}},Z,B_{a}\right]&=&+\beta_{b}^{2}-\beta_{c}^{2}-2\beta_{a}\beta_{b}+\frac{2}{3}\gamma^{2}
\end{eqnarray*}
\end{lemma}

\begin{remark} 
Lemma \ref{lemma:3.5} gives the difference between $\frak a\left[\Phi,\mathrm{\bf{h}},Z\right]=0$ (the actual field) and $\frak a\left[\Phi,\mathrm{\bf{h}},B_{a}\right]=0$ (the reference field). The bounds for the terms are given in Technical Lemma 1, and tend exponentially to zero as $\mathrm{\bf{h}}\rightarrow 0$. Note, however, that $\rho$ was assumed to be some (finite) constant (but different for every field $\Phi$).
\end{remark}

%Definition 3.11%
\begin{definition} 
\label{def:3.11}
$\Phi=\alpha\oplus\beta\oplus\gamma\in\mathbb{R}^{3}\oplus\mathbb{R}^{3}\oplus\mathbb{R}$, $\beta_{i}\neq 0$, $\mathrm{\bf{h}}>0$. Define four real numbers by
\begin{eqnarray*}
\mathrm{\bf{I}}_{1}\left[\Phi,\mathrm{\bf{h}},\pi\right]&=&-\frac{1}{\mathrm{\bf{h}}}\frak a_{a}\left[\Phi,\mathrm{\bf{h}},Z,B_{a}\right]\mathrm{tanh}\phi_{a}\left[\Phi\right] \\
\mathrm{\bf{I}}_{2}\left[\Phi,\mathrm{\bf{h}},\pi\right]&=&(A_{a}\left[\Phi\right])^{-2}\frak a_{a}\left[\Phi,\mathrm{\bf{h}},Z,B_{a}\right](1-\mathrm{tanh}\phi_{a}\left[\Phi\right]) \\
\mathrm{\bf{I}}_{(3,\textbf{p})}\left[\Phi,\mathrm{\bf{h}},\pi\right]&=&\frac{1}{\mathrm{\bf{h}}}\frak a_{\textbf{p}}\left[\Phi,\mathrm{\bf{h}},Z,B_{a}\right]+\frac{1}{\mathrm{\bf{h}}}\frak a_{a}\left[\Phi,\mathrm{\bf{h}},Z,B_{a}\right]
\end{eqnarray*}
with $p\in\lbrace\textbf{b,c}\rbrace$.
\end{definition}

%Technical Lemma 2%

\begin{lemma}
\label{lemma:3.6}
($\textbf{Technical Lemma 2}$)
Recall $\rho[\Phi_{0}]=\sqrt{z}$. In the context of Definition \ref{def:3.11} the following estimates hold:
\begin{eqnarray*}
\vert\mathrm{\bf{I}}_{S}\left[\Phi\right]\vert&\leq&2^{12}\mathrm{max}\lbrace1,z\rbrace\mathrm{max}\lbrace1,\frac{1}{\mathrm{\bf{h}}},\frac{1}{\mathrm{\bf{h}}}\vert\tau\vert\rbrace\mathrm{exp}\left(-\frac{1}{4\mathrm{\bf{h}}}\mathrm{min}\lbrace1,\epsilon_{-},\frac{\epsilon_{+}}{\tau_{+}}\rbrace\right) \\
\vert\mathrm{\bf{I}}_{S}\left[\Phi\right]-\mathrm{\bf{I}}_{S}\left[\Psi\right]\vert&\leq&2^{18}\mathrm{max}\lbrace1,z\rbrace\left(\mathrm{max}\lbrace1,\frac{1}{\mathrm{\bf{h}}},\frac{1}{\mathrm{\bf{h}}}\vert\tau\vert\rbrace\right)^{2}\mathrm{exp}\left(-\frac{1}{4\mathrm{\bf{h}}}\mathrm{min}\lbrace1,\epsilon_{-},\frac{\epsilon_{+}}{\tau_{+}}\rbrace\right) d_{\mathcal{E}}(\Phi,\Psi)
\end{eqnarray*}
for all $S\in\lbrace1,2,(3,b),(3,c)\rbrace$.
\end{lemma}

%Proof%
\noindent\underline{Proof}. 
Set $M:=\text{exp}\left(-\frac{1}{4\mathrm{\bf{h}}}\mathrm{min}\lbrace1,\epsilon_{-},\frac{\epsilon_{+}}{\tau_{+}}\rbrace\right)$, $M_{1}:=\mathrm{max}\lbrace1,\frac{1}{\mathrm{\bf{h}}},\frac{1}{\mathrm{\bf{h}}}\vert\tau\vert\rbrace$, and $M_{2}:=\mathrm{max}\lbrace1, z\rbrace$.
For $x\geq a$ we have the inequality
\\
\begin{eqnarray*}
\vert \text{e}^{x}-\text{e}^{a}\vert&=&\vert \text{e}^{a}\vert\cdot\vert\text{e}^{x-a}-1\vert \\
&=&\vert\text{e}^{a}\vert\cdot\vert(x-a)+\frac{(x-a)^{2}}{2!}+...\vert \\
&=&\vert\text{e}^{a}\vert\cdot\vert x-a\vert\cdot\vert 1+\frac{x-a}{2!}+\frac{(x-a)^{2}}{3!}+...\vert \\
&\leq&\vert\text{e}^{a}\vert\cdot\vert x-a\vert\cdot\vert\text{e}^{x-a}\vert
\end{eqnarray*}
\\
It implies that $\vert\text{e}^{x}-\text{e}^{y}\vert\leq\mathrm{max}\lbrace\text{e}^{x},\text{e}^{y}\rbrace\vert x-y\vert$, $\forall x,y$. \\\\
Recall $\text{log}\vert\beta_{i}\vert=\frac{1}{\mathrm{\bf{h}}}\xi_{i}+\text{log}2$. \\
\\
\emph{Estimates}. Lemmas \ref{lemma:3.4} and \ref{lemma:3.5} imply $\vert\frak a_{i}[\Phi,\mathrm{\bf{h}},Z,B_{a}]\vert\leq2^{7}MM_{2}$, for $i=1,2,3$, and $\phi_{a}[\Phi]\leq2^{2}M_{1}$, and $(A_{a}[\Phi])^{-2}\leq2^{2}$. Therefore, for the $\vert I_{S}\vert$ estimate we have $\vert I_{1}\vert\leq2^{7}MM_{1}M_{2}$ and $\vert I_{2}\vert\leq2^{12}MM_{1}M_{2}$ and $\vert I_{3}\vert\leq2^{8}MM_{1}M_{2}$.
\\
Further, using Definition \ref{def:3.3}, Definition \ref{def:3.6}, we have:

\begin{equation*}
\vert(\beta_{a}\beta_{b}\beta_{c})[\Phi]\vert^{1/3}=2\text{exp}\left(-\frac{1}{12\mathrm{\bf{h}}}(1+4\epsilon_{-})\right)\leq2M
\end{equation*}

and

\begin{eqnarray*}
\vert\beta_{p}[\Phi]\beta_{a}[\Phi]-\beta_{p}[\Psi]\beta_{a}[\Psi]\vert&\leq&\frac{1}{\mathrm{\bf{h}}}\mathrm{max}\lbrace\vert\beta_{p}[\Phi]\beta_{a}[\Phi]\vert, \vert\beta_{p}[\Psi]\beta_{a}[\Psi]\vert\rbrace \vert\xi_{a,p}[\Phi]-\xi_{a,p}[\Psi]\vert \\
&\leq&\frac{1}{\mathrm{\bf{h}}}2^{4}M_{2} Md_{\mathcal{E}}(\Phi,\Psi) \\
&\leq&2^{4}MM_{1}M_{2}d_{\mathcal{E}}(\Phi,\Psi)
\end{eqnarray*}

and

\begin{eqnarray*}
\vert\xi_{a}[\Phi]-\xi_{a}[\Psi]\vert&\leq&\mathrm{\bf{h}}\vert\text{log}A_{a}[\Phi]-\text{log}A_{a}[\Psi]\vert+\mathrm{\bf{h}}\vert\text{log}\:\text{cosh}\phi_{a}[\Phi]-\text{log}\:\text{cosh}\phi_{a}[\Psi]\vert \\
&\leq&\mathrm{\bf{h}}\vert\text{log}A_{a}[\Phi]-\text{log}A_{a}[\Psi]\vert+\mathrm{\bf{h}}\vert\phi_{a}[\Phi]-\phi_{a}[\Psi]\vert \\
&\leq&\mathrm{\bf{h}}\mathrm{max}\lbrace\frac{1}{A_{a}[\Phi]},\frac{1}{A_{a}[\Psi]}\rbrace\vert A_{a}[\Phi]-A_{a}[\Psi]\vert+\mathrm{\bf{h}}\vert\phi_{a}[\Phi]-\phi_{a}[\Psi]\vert \\
&\leq&\frac{\mathrm{\bf{h}}}{2}d_{\mathcal{E}}(\Phi,\Psi)+\mathrm{\bf{h}}2^{2}M_{1}d_{\mathcal{E}}(\Phi,\Psi) \\
&\leq&2^{3}\mathrm{\bf{h}}M_{1}d_{\mathcal{E}}(\Phi,\Psi)
\end{eqnarray*}

and

\begin{eqnarray*}
\vert\beta_{p}[\Phi]-\beta_{p}[\Psi]\vert&\leq&\frac{1}{\mathrm{\bf{h}}}\mathrm{max}\lbrace\vert\beta_{p}[\Phi]\vert,\vert\beta_{p}[\Psi]\vert\rbrace\vert\xi_{p}[\Phi]-\xi_{p}[\Psi]\vert \\
&\leq&\frac{1}{\mathrm{\bf{h}}}2^{2}(M_{2} M)^{1/2}\left(\vert\xi_{a,p}[\Phi]-\xi_{a,p}[\Psi]\vert+\vert\xi_{a}[\Phi]-\xi_{a}[\Psi]\vert\right) \\
&\leq&\frac{1}{\mathrm{\bf{h}}}2^{2}(M_{2} M)^{1/2}\left(d_{\mathcal{E}}(\Phi,\Psi)+\vert\xi_{a}[\Phi]-\xi_{a}[\Psi]\vert\right) \\
&\leq&2^{2}(MM_{2})^{1/2}\left(\frac{1}{\mathrm{\bf{h}}}+2^{3}M_{1}\right)d_{\mathcal{E}}(\Phi,\Psi) \\
&\leq&2^{6}(MM_{2})^{1/2}M_{1}d_{\mathcal{E}}(\Phi,\Psi)
\end{eqnarray*}

and

\begin{eqnarray*}
\vert\phi_{a}[\Phi]-\phi_{a}[\Psi]\vert&\leq&\frac{1}{\mathrm{\bf{h}}}\vert A_{a}[\Phi]-A_{a}[\Psi]\vert\vert\tau\vert+\frac{1}{\mathrm{\bf{h}}}\vert A_{a}[\Phi]\theta_{a}[\Phi]-A_{a}[\Psi]\theta_{a}[\Psi]\vert \\
&\leq&\frac{1}{\mathrm{\bf{h}}}(1+\vert\tau\vert)\vert A_{a}[\Phi]-A_{a}[\Psi]\vert+\frac{1}{\mathrm{\bf{h}}}2\vert\theta_{a}[\Phi]-\theta_{a}[\Psi]\vert \\
&\leq&\frac{1}{\mathrm{\bf{h}}}(3+\vert\tau\vert)d_{\mathcal{E}}(\Phi,\Psi) \\
&\leq&2^{2}M_{1}d_{\mathcal{E}}(\Phi,\Psi)
\end{eqnarray*}

and

\begin{eqnarray*}
\vert\rho^{2}[\Phi](\beta_{a}\beta_{b}\beta_{c})^{1/3}[\Phi]-\rho^{2}[\Psi](\beta_{a}\beta_{b}\beta_{c})^{1/3}[\Psi]\vert&\leq&\vert\rho^{2}[\Phi](\beta_{a}\beta_{b}\beta_{c})^{\frac{1}{3}}[\Phi]-\rho^{2}[\Psi](\beta_{a}\beta_{b}\beta_{c})^{\frac{1}{3}}[\Phi]\vert \\
&+&\vert\rho^{2}[\Psi](\beta_{a}\beta_{b}\beta_{c})^{\frac{1}{3}}[\Phi]-\rho^{2}[\Psi](\beta_{a}\beta_{b}\beta_{c})^{\frac{1}{3}}[\Psi]\vert \\
&\leq&\vert\rho^{2}[\Phi]-\rho^{2}[\Psi]\vert\cdot\vert(\beta_{a}\beta_{b}\beta_{c})^{\frac{1}{3}}[\Phi]\vert \\
&+&\vert\rho^{2}[\Psi]\vert\cdot\vert(\beta_{a}\beta_{b}\beta_{c})^{\frac{1}{3}}[\Phi]-(\beta_{a}\beta_{b}\beta_{c})^{\frac{1}{3}}[\Psi]\vert \\
&\leq&\vert(\beta_{a}\beta_{b}\beta_{c})^{\frac{1}{3}}[\Phi]\vert d_{\mathcal{E}}(\Phi,\Psi) \\
&+&\left(\rho^{2}[\Psi]\mathrm{max}\lbrace\vert(\beta_{a}\beta_{b}\beta_{c})^{\frac{1}{3}}[\Phi]\vert,\vert(\beta_{a}\beta_{b}\beta_{c})^{\frac{1}{3}}[\Psi]\vert\rbrace\right)\times \\
&\times&\left(\bigg|\frac{\xi_{a}[\Phi]+\xi_{b}[\Phi]+\xi_{c}[\Phi]-\xi_{a}[\Psi]-\xi_{b}[\Psi]-\xi_{c}[\Psi]}{3\mathrm{\bf{h}}}\bigg|\right) \\
&\leq&2M d_{\mathcal{E}}(\Phi,\Psi)+\frac{2}{3\mathrm{\bf{h}}}\rho^{2}[\Psi]M(d_{\mathcal{E}}(\Phi,\Psi) \\
&+&\vert\xi_{a,c}[\Phi]-\xi_{a,c}[\Psi]+\xi_{a}[\Psi]-\xi_{a}[\Phi]\vert) \\
&\leq&2M d_{\mathcal{E}}(\Phi,\Psi)+\frac{2}{3\mathrm{\bf{h}}}\rho^{2}[\Psi]M(2d_{\mathcal{E}}(\Phi,\Psi)+\vert\xi_{a}[\Phi]-\xi_{a}[\Psi]\vert) \\
&\leq&2M d_{\mathcal{E}}(\Phi,\Psi)+\frac{2}{3\mathrm{\bf{h}}}M_{2}M(2d_{\mathcal{E}}(\Phi,\Psi)+2^{3}\mathrm{\bf{h}}M_{1}d_{\mathcal{E}}(\Phi,\Psi)) \\
&\leq&2MM_{1}d_{\mathcal{E}}(\Phi,\Psi)+2^{3}MM_{1}M_{2}d_{\mathcal{E}}(\Phi,\Psi) \\
&\leq&2(1+2^{2}M_{2})MM_{1}d_{\mathcal{E}}(\Phi,\Psi) \\
&\leq&2^{4}MM_{1}M_{2}d_{\mathcal{E}}(\Phi,\Psi)
\end{eqnarray*}
\\
\\
The estimates above imply (using $\vert\beta_{p}^{2}[\Phi]-\beta_{p}^{2}[\Psi]\vert\leq2\mathrm{max}\lbrace\vert\beta_{p}[\Phi]\vert,\vert\beta_{p}[\Psi]\vert\rbrace\vert\beta_{p}[\Phi]-\beta_{p}[\Psi]\vert$):
\\
\begin{eqnarray*}
\vert\frak a_{b}[\Phi,\mathrm{\bf{h}},Z,B_{a}]-\frak a_{b}[\Psi,\mathrm{\bf{h}},Z,B_{a}]\vert&\leq&\vert\beta_{b}^{2}[\Phi]-\beta_{b}^{2}[\Psi]\vert+\vert\beta_{c}^{2}[\Phi]-\beta_{c}^{2}[\Psi]\vert+2\vert\beta_{a}[\Phi]\beta_{c}[\Phi]-\beta_{a}[\Psi]\beta_{c}[\Psi]\vert \\
&+&\frac{2}{3}\vert\gamma^{2}[\Phi]-\gamma^{2}[\Psi]\vert \\
&\leq&2^{10}MM_{1}M_{2}d_{\mathcal{E}}(\Phi,\Psi)+2^{5}MM_{1}M_{2}d_{\mathcal{E}}(\Phi,\Psi)+\frac{2^{5}}{3}MM_{1}M_{2}d_{\mathcal{E}}(\Phi,\Psi) \\
&\leq&2^{11}MM_{1}M_{2}d_{\mathcal{E}}(\Phi,\Psi)
\end{eqnarray*}
\\
Analogous estimate holds for $\vert\frak a_{c}[\Phi,\mathrm{\bf{h}},Z,B_{a}]-\frak a_{c}[\Psi,\mathrm{\bf{h}},Z,B_{a}]\vert$.
\\
\\
Further,
\\
\begin{eqnarray*}
\vert\beta_{p}[\Phi]\beta_{q}[\Phi]-\beta_{p}[\Psi]\beta_{q}[\Psi]\vert&\leq&\vert\beta_{q}[\Psi]\vert\cdot\vert\beta_{p}[\Phi]-\beta_{p}[\Psi]\vert+\vert\beta_{p}[\Phi]\vert\cdot\vert\beta_{q}[\Phi]-\beta_{q}[\Psi]\vert \\
&\leq&2^{2}\sqrt{MM_{2}}\cdot 2^{6}M_{1}\sqrt{MM_{2}}d_{\mathcal{E}}(\Phi,\Psi)+2^{2}\sqrt{MM_{2}}\cdot 2^{6}M_{1}\sqrt{MM_{2}}d_{\mathcal{E}}(\Phi,\Psi) \\
&\leq&2^{9}MM_{1}M_{2}d_{\mathcal{E}}(\Phi,\Psi)
\end{eqnarray*}
\\
and therefore
\\
\begin{eqnarray*}
\vert\frak a_{a}[\Phi,\mathrm{\bf{h}},Z,B_{a}]-\frak a_{a}[\Psi,\mathrm{\bf{h}},Z,B_{a}]\vert&\leq&\vert\beta_{b}^{2}[\Phi]-\beta_{b}^{2}[\Psi]\vert+\vert\beta_{c}^{2}[\Phi]-\beta_{c}^{2}[\Psi]\vert+2\vert\beta_{b}[\Phi]\beta_{c}[\Phi]-\beta_{b}[\Psi]\beta_{c}[\Psi]\vert \\
&+&\frac{2}{3}\vert\gamma^{2}[\Phi]-\gamma^{2}[\Psi]\vert \\
&\leq&(2^{3}\sqrt{MM_{2}}\cdot 2^{6}M_{1}\sqrt{MM_{2}}d_{\mathcal{E}}(\Phi,\Psi))\cdot 2+2\cdot2^{9}MM_{1}M_{2}d_{\mathcal{E}}(\Phi,\Psi)\\
&+&\frac{2}{3}2^{4}MM_{1}M_{2}d_{\mathcal{E}}(\Phi,\Psi) \\
&\leq&2^{12}MM_{1}M_{2}d_{\mathcal{E}}(\Phi,\Psi).
\end{eqnarray*}
\\
These estimates imply
\\
\begin{eqnarray*}
\vert I_{1}[\Phi]-I_{1}[\Psi]\vert&=&\big| -\frac{1}{\mathrm{\bf{h}}}\frak a_{a}[\Phi,\mathrm{\bf{h}},Z,B_{a}]\text{tanh}\phi_{a}[\Phi]+\frac{1}{\mathrm{\bf{h}}}\frak a_{a}[\Psi,\mathrm{\bf{h}},Z,B_{a}]\text{tanh}\phi_{a}[\Psi]\big| \\
&\leq&M_{1}\vert-\frak a_{a}[\Phi,\mathrm{\bf{h}},Z,B_{a}]\text{tanh}\phi_{a}[\Phi]+\frak a_{a}[\Psi,\mathrm{\bf{h}},Z,B_{a}]\text{tanh}\phi_{a}[\Psi]\vert \\
&\leq&M_{1}\vert\frak a_{a}[\Phi,\mathrm{\bf{h}},Z,B_{a}]\vert\vert\text{tanh}\phi_{a}[\Phi]-\text{tanh}\phi_{a}[\Psi]\vert+M_{1}\vert\frak a_{a}[\Phi,\mathrm{\bf{h}},Z,B_{a}]-\frak a_{a}[\Psi,\mathrm{\bf{h}},Z,B_{a}]\vert \\
&\leq&2^{13}MM_{1}^{2}M_{2}d_{\mathcal{E}}(\Phi,\Psi)
\end{eqnarray*}
\\
and
\\
\begin{eqnarray*}
\vert I_{(3,p)}[\Phi]-I_{(3,p)}[\Psi]\vert&=&\big| \frac{1}{\mathrm{\bf{h}}}\frak a_{p}[\Phi,\mathrm{\bf{h}},Z,B_{a}]+\frac{1}{\mathrm{\bf{h}}}\frak a_{a}[\Phi,\mathrm{\bf{h}},Z,B_{a}]-\frac{1}{\mathrm{\bf{h}}}\frak a_{p}[\Psi,\mathrm{\bf{h}},Z,B_{a}]-\frac{1}{\mathrm{\bf{h}}}\frak a_{a}[\Psi,\mathrm{\bf{h}},Z,B_{a}]\big| \\
&\leq&M_{1}\vert\frak a_{p}[\Phi,\mathrm{\bf{h}},Z,B_{a}]-\frak a_{p}[\Psi,\mathrm{\bf{h}},Z,B_{a}]\vert+M_{1}\vert\frak a_{a}[\Phi,\mathrm{\bf{h}},Z,B_{a}]-\frak a_{a}[\Psi,\mathrm{\bf{h}},Z,B_{a}]\vert \\
&\leq&2^{13}MM_{1}^{2}M_{2}d_{\mathcal{E}}(\Phi,\Psi)
\end{eqnarray*}
\\
and
\\
\begin{eqnarray*}
\vert I_{2}[\Phi]-I_{2}[\Psi]\vert&\leq&\bigg|\frac{1}{A^{2}_{a}[\Phi]}\frak a_{a}[\Phi,\mathrm{\bf{h}},Z,B_{a}]\bigg|\cdot\vert\phi_{a}[\Phi]\text{tanh}\phi_{a}[\Phi]-\phi_{a}[\Psi]\text{tanh}\phi_{a}[\Psi]\vert \\
&+&\bigg|\frac{1}{A^{2}_{a}[\Phi]}\frak a_{a}[\Phi,\mathrm{\bf{h}},Z,B_{a}]-\frac{1}{A^{2}_{a}[\Psi]}\frak a_{a}[\Psi,\mathrm{\bf{h}},Z,B_{a}]\bigg|\cdot\vert1-\phi_{a}[\Psi]\text{tanh}\phi_{a}[\Psi]\vert \\
&\leq&2^{18}MM_{1}^{2}M_{2}d_{\mathcal{E}}(\Phi,\Psi)
\end{eqnarray*}
\\
The last step followed from (using $\big|\frac{1}{x^{2}}-\frac{1}{y^{2}}\big|\leq\frac{1}{\vert x^{2}y^{2}\vert}2\mathrm{max}\lbrace\vert x\vert,\vert y\vert\rbrace\vert x-y\vert$):
\\
\begin{eqnarray*}
\vert\phi_{a}[\Phi]\text{tanh}\phi_{a}[\Phi]-\phi_{a}[\Psi]\text{tanh}\phi_{a}[\Psi]\vert&\leq&\vert\phi_{a}[\Phi]\vert\vert\phi_{a}[\Phi]-\phi_{a}[\Psi]\vert+\vert\text{tanh}\phi_{a}[\Psi]\vert\vert\phi_{a}[\Phi]-\phi_{a}[\Psi]\vert \\
&\leq&(2^{2}+1)2^{2}M_{1}^{2}d_{\mathcal{E}}(\Phi,\Psi) \\
&\leq&2^{5}M_{1}^{2}d_{\mathcal{E}}(\Phi,\Psi)
\end{eqnarray*}
\\
and
\\
\begin{eqnarray*}
\bigg|\frac{1}{A^{2}_{a}[\Phi]}\frak a_{a}[\Phi,\mathrm{\bf{h}},Z,B_{a}]-\frac{1}{A^{2}_{a}[\Psi]}\frak a_{a}[\Psi,\mathrm{\bf{h}},Z,B_{a}]\bigg|&\leq&\bigg|\frac{1}{A_{a}^{2}[\Phi]}(\frak a_{a}[\Phi,\mathrm{\bf{h}},Z,B_{a}]-\frak a_{a}[\Psi,\mathrm{\bf{h}},Z,B_{a}])\bigg| \\
&+&\bigg|\frak a_{a}[\Psi,\mathrm{\bf{h}},Z,B_{a}]\left(\frac{1}{A_{a}^{2}[\Phi]}-\frac{1}{A_{a}^{2}[\Psi]}\right)\bigg| \\
&\leq&2^{2}\vert\frak a_{a}[\Phi,\mathrm{\bf{h}},Z,B_{a}]-\frak a_{a}[\Psi,\mathrm{\bf{h}},Z,B_{a}]\vert \\
&+&2^{11}MM_{2}d_{\mathcal{E}}(\Phi,\Psi) \\
&\leq&2^{14}MM_{1}M_{2}d_{\mathcal{E}}(\Phi,\Psi)
\end{eqnarray*}
\\
and
\\
\begin{eqnarray*}
\vert 1-\phi_{a}[\Psi]\text{tanh}\phi_{a}[\Psi]\vert&\leq&1+\vert\phi_{a}[\Psi]\vert\vert\text{tanh}\phi_{a}[\Psi]\vert \\
&\leq&2^{3}M_{1}
\end{eqnarray*}

%\end{lemma}
%State vectors%
\begin{definition}
\label{def:3.12}
($\textbf{State Vectors}$).$\forall\pi=(a,b,c)\in S_{3}$, $\mathrm{\bf{f}}=(\mathrm{\bf{h}},w,q,z)\in(0,\infty)^{2}\times\mathbb{R}\times(0,\infty)$, $\sigma_{*}\in\lbrace-1,+1\rbrace^{3}$, the field $\Phi_{*}=\Phi_{*}(\pi,\mathrm{\bf{f}},\sigma_{*})$ is given by
\begin{eqnarray*}
\alpha_{a}\left[\Phi_{*}\right]&=&-1 \\
\alpha_{b}\left[\Phi_{*}\right]&=&\frac{w}{1+w} \\
\alpha_{c}\left[\Phi_{*}\right]&=&-w-4(1+w)\gamma^{2}-\mu \\
\beta_{a}\left[\Phi_{*},\mathrm{\bf{h}}\right]&=&-\frac{1+w}{1+2w}(1+\mathrm{\bf{h}}\mathrm{log}2) \\
\beta_{b}\left[\Phi_{*},\mathrm{\bf{h}}\right]&=&-\frac{1+w}{1+2w}(1+\mathrm{\bf{h}}\mathrm{log}2) \\
\beta_{c}\left[\Phi_{*},\mathrm{\bf{h}}\right]&=&-(1+w)q-\frac{w(1+w)}{1+2w}-\frac{1+3w+w^{2}}{1+2w}\mathrm{\bf{h}}\mathrm{log}2 
\end{eqnarray*}
where $\mu=(1+w)(\beta_{1}^{2}+\beta_{2}^{2}+\beta_{3}^{2}-2\beta_{2}\beta_{3}-2\beta_{3}\beta_{1}-2\beta_{1}\beta_{2})$ is determined uniquely by requiring that the new constraint, i.e. new equation (1.1c) holds (see end of Section 1).
\\
\\
$\rho[\Phi_{*}]=\rho[\Phi_{0}]=\sqrt{z}$.
\end{definition}

\begin{definition} 
\label{def:3.13}
$\forall\pi=(a,b,c)\in S_{3}$, $\sigma_{*}\in\lbrace+1,-1\rbrace^{3}$ let $\mathcal{H}(\pi,\sigma_{*})\subset\mathcal{D}(\sigma_{*})$ be the set of all vectors $\Phi=\alpha\oplus\beta\oplus\gamma\in\mathcal{D}(\sigma_{*})$ with
\begin{equation}
\vert\beta_{a}\vert=\vert\beta_{b}\vert  \:\:\:\:\:\:\:\:  \sum_{(i,j,k)\in\mathcal{C}}(\alpha_{j}\alpha_{k}-(\beta_{i})^{2}+2\beta_{j}\beta_{k})-4\gamma^{2}=0
\end{equation}
\begin{equation}
\label{eq:pos_cond}
0<\alpha_{b}<-\alpha_{a}  \:\:\:\:\:\:\:\:\:\:\:\:  (\alpha_{b}+\vert\alpha_{a}\vert)\mathrm{log}\vert\beta_{a}/\alpha_{a}\vert<\alpha_{b}\mathrm{log}2
\end{equation}
\end{definition}

%LEMMA 3.7%
\begin{lemma} 
\label{lemma:3.7}
Let $\pi=(a,b,c)\in S_{3}$ and $\sigma_{*}\in\lbrace+1,-1\rbrace^{3}$. The set $\mathcal{H}(\pi,\sigma_{*})\subset\mathcal{D}(\sigma_{*})$ is a smooth 4-dim submanifold. The map
\begin{equation*}
(0,\infty)^{3}\times\mathbb{R}\times(0,\infty)\rightarrow\mathcal{H}(\pi,\sigma_{*})
\end{equation*}
\begin{equation}
(\lambda,\mathrm{\bf{h}},w,q,z)\mapsto\lambda\Phi_{*}(\pi,(\mathrm{\bf{h}},w,q,z),\sigma_{*})
\end{equation}
is a diffeomorphism. Its inverse is given by 
\\
\begin{eqnarray}
w&=&-\frac{\alpha_{b}}{\alpha_{a}+\alpha_{b}} \\
\lambda&=&-\alpha_{a} \\
\frac{1}{\mathrm{\bf{h}}}&=&-\frac{1+2w}{1+w}\mathrm{log}\vert\beta_{a}/\alpha_{a}\vert+\frac{w}{1+w}\mathrm{log}2 \\
q&=&-\frac{1}{1+w}\mathrm{\bf{h}}\mathrm{log}\vert\beta_{c}/\alpha_{a}\vert-\frac{w}{1+2w}(1+\mathrm{\bf{h}}\mathrm{log}2) \\
z&=&-\frac{\rho^{2}}{\alpha_{a}}
\end{eqnarray}
\\
Note: $z$ is well-defined by (\ref{eq:pos_cond}).
\end{lemma}

\begin{definition} 
\label{def:3.14}
$\forall\mathrm{\bf{f}}=(\mathrm{\bf{h}},w,q,z)\in(0,\infty)^{4}$ set
\begin{eqnarray*}
\tau_{1-}(\mathrm{\bf{f}})&=&
\begin{cases}
-\frac{1+w}{3+w}q-\frac{1}{3+w}\mathrm{\bf{h}}\mathrm{log}2\:\:\:\:\mathrm{if}\:q\leq 1\\
-\frac{1+w}{3+2w}-\frac{1+w}{3+2w}\mathrm{\bf{h}}\mathrm{log}2\:\:\:\mathrm{if}\:q>1
\end{cases}
\\
\tau_{1+}(\mathrm{\bf{f}})&=&(1+\mathrm{\bf{h}}\mathrm{log}2)\frac{1+w}{1+2w}
\end{eqnarray*}
Note that $\tau_{1-}(\mathrm{\bf{f}})<0$ and $\tau_{1+}(\mathrm{\bf{f}})>0$.
\end{definition}

\emph{Motivation for Definition \ref{def:3.14}}. It gives us the approximate time of crossing for $\beta$'s. For example, if $\beta_{a}$ bounces at $\tau=0$, then $\beta_{b}$ is going to make the next bounce (to the right) at $\tau_{1+}>0$.

\begin{definition} 
\label{def:3.15}
$\textbf{(Initial Data)}$ $\forall\pi=(a,b,c)\in S_{3}$, $\forall\mathrm{\bf{f}}=(\mathrm{\bf{h}},w,q,z)\in(0,\infty)^{4}$, $\forall\sigma_{*}\in\lbrace-1,+1\rbrace^{3}$ let 
\\
$\Phi_{1}=\Phi_{1}(\pi,\mathrm{\bf{f}},\sigma_{*}):\mathbb{R}\rightarrow\mathcal{D}(\sigma_{*})$ be given by
\begin{eqnarray*}
A_{a}\left[\Phi_{1}\right](\tau)&=&A_{a}\left[\Phi_{*}\right], \:\:\:\:\:\:\:\theta_{a}\left[\Phi_{1},\mathrm{\bf{h}}\right](\tau)=\theta_{a}\left[\Phi_{*},\mathrm{\bf{h}}\right] \\
\alpha_{p,a}\left[\Phi_{1}\right](\tau)&=&\alpha_{p,a}\left[\Phi_{*}\right], \:\: \xi_{p,a}\left[\Phi_{1},\mathrm{\bf{h}}\right](\tau)=\xi_{p,a}\left[\Phi_{*},\mathrm{\bf{h}}\right]+(\tau-\tau_{1+})\alpha_{p,a}\left[\Phi_{*}\right] \\
\rho\left[\Phi_{1}\right]&=&\rho\left[\Phi_{*}\right]\:=\:\sqrt{z}
\end{eqnarray*}
for all $\tau\in\mathbb{R}$ and $p\in\lbrace b,c\rbrace$. Here, $\tau_{1+}=\tau_{1+}(\mathrm{\bf{f}})$ and $\Phi_{*}=\Phi_{*}(\pi,\mathrm{\bf{f}},\sigma_{*})$.
\end{definition}

%Lemma 3.8%
\begin{lemma} 
\label{lemma:3.8}
Stays the same as in \cite{RT} since $\rho[\Phi_{0}]=\rho[\Phi_{1}]=\sqrt{z}$. That is, $\forall\pi\in(a,b,c)\in S_{3}$, $\mathrm{\bf{f}}=(\mathrm{\bf{h}},w,q,z)\in(0,\infty)^{4}$, $\sigma_{*}\in\lbrace-1, +1\rbrace^{3}$, set $\Phi_{0}=\Phi_{0}(\pi,\mathrm{\bf{f}},\sigma_{*})$, $\tau_{1+}=\tau_{1+}(\mathrm{\bf{f}})$. Then
\\
\\
(a) $\vert\beta_{a}\left[\Phi_{1}\right](\tau_{1+})\vert=\vert\beta_{b}\left[\Phi_{1}\right](\tau_{1+})\vert$ \\
(b) $c\left[\Phi_{1},\mathrm{\bf{h}},Z\right](\tau_{1+})=0$ \\
(c) $\not d _{\mathcal{D}}(\Phi_{0}(\tau_{1+}),\Phi_{1}(\tau_{1+}))\leq 2^{7}\mathrm{max}\lbrace1+w,\mathrm{\bf{h}}\rbrace\mathrm{exp}\left(-\frac{1}{2\mathrm{\bf{h}}}\mathrm{min}\lbrace1,w+q\rbrace\right)$ \\
(d) $d_{\mathcal{D}}(\Phi_{0}(\tau),\Phi_{1}(\tau))\leq (1+\vert\tau-\tau_{1+}\vert)d_{\mathcal{D}}(\Phi_{0}(\tau_{1+}),\Phi_{1}(\tau_{1+})),\forall \tau\in\mathbb{R}$
\end{lemma}

\begin{definition}
\label{def:3.16}
$\textbf{(Approximate Transfer Maps)}$ Introduce three maps 
\\
\begin{eqnarray*}
\mathcal{P}_{L}: S_{3}\times (0,\infty)^{4}&\rightarrow& S_{3} \\
\mathcal{Q}_{L}:\:\:\:\:\:\:\:\:\:\:(0,\infty)^{4}&\rightarrow& (0,\infty)^{2}\times\mathbb{R}\times(0,\infty) \\
\lambda_{L}: \:\:\:\:\:\:\:\:\:\:(0,\infty)^{4}&\rightarrow&(0,\infty)
\end{eqnarray*}
where $\mathrm{\bf{f}}=(\mathrm{\bf{h}},w,q,z)$, $q_{L}=\mathrm{num}1_{L}/\mathrm{den}_{L}$, $\mathrm{\bf{h}}_{L}=\mathrm{num}2_{L}/\mathrm{den}_{L}$, and:
\\
\begin{itemize}
\item if\:$q\leq1$:
\\
\begin{eqnarray*}
(a',b',c')&=&(c,a,b) \\
w_{L}&=&\frac{1}{1+w} \\
\lambda_{L}&=&2+w \\
\mathrm{num}1_{L}&=&(1+w)(1-q)-\mathrm{\bf{h}}\mathrm{log}2+\mathrm{\bf{h}}w\mathrm{log}(2+w) \\
\mathrm{num}2_{L}&=&\mathrm{\bf{h}}(2+w) \\
\mathrm{den}_{L}&=&(1+w)(q-\mathrm{\bf{h}}\mathrm{log}2)+\mathrm{\bf{h}}(3+w)\mathrm{log}(2+w) \\
z_{L}&=&\frac{1}{2+w}z
\end{eqnarray*}
\item if\:$q>1$
\\
\begin{eqnarray*}
(a',b',c')&=&(b,a,c) \\
w_{L}&=&1+w \\
\lambda_{L}&=&\frac{2+w}{1+w} \\
\mathrm{num}1_{L}&=&(1+w)(q-1-\mathrm{\bf{h}}\mathrm{log}2)-\mathrm{\bf{h}}w\mathrm{log}\frac{2+w}{1+w} \\
\mathrm{num}2_{L}&=&\mathrm{\bf{h}}(2+w) \\
\mathrm{den}_{L}&=&(1+w)-\mathrm{\bf{h}}\mathrm{log}2+\mathrm{\bf{h}}(3+2w)\mathrm{log}\frac{2+w}{1+w} \\
z_{L}&=&\frac{1+w}{2+w}z
\end{eqnarray*}
\end{itemize}
\end{definition}
\vskip 2mm
%APPROXIMATE TRANSFER MAPS%
\noindent\emph{Motivation for Definition \ref{def:3.16}}
\\
Introduce the scaling parameter $\lambda_{L}$. Recall $\rho[\Phi_{0}]=\rho[\Phi_{*}]=\sqrt{z}$. We have 
\begin{equation}
\label{eq:transf_maps}
\Phi_{0}(\pi,f,\sigma)\mid _{\tau_{1-}}=\lambda_{L}\Phi_{*}(\pi',f_{L}, \sigma),
\end{equation}
 with $f=(\mathrm{\bf{h}},w,q,z)$, $\pi=(a,b,c)$ and $\pi'=(a',b',c')$. Therefore, the rescaling for $\sqrt{z}$ will be $\lambda_{L}\cdot\lambda_{L}^{-\frac{1}{6}\cdot3}=\sqrt\lambda_{L}$. To get the $f_{L}=(\mathrm{\bf{h}}_{L},w_{L},q_{L},z_{L})$ we calculate:
\\
\\
$\underline{\mathrm{Case 1}}$: $q>1$. Then $b'=a$, $a'=b$ and (\ref{eq:transf_maps}) is equivalent to
\begin{eqnarray*}
\alpha_{a}\left[\Phi_{0}\right]&=&\lambda_{L}\alpha_{b'}\left[\Phi_{*}\right](\mathrm{\bf{h}}_{L},w_{L},q_{L}) \\
\alpha_{b}\left[\Phi_{0}\right]&=&\lambda_{L}\alpha_{a'}\left[\Phi_{*}\right](\mathrm{\bf{h}}_{L},w_{L},q_{L}) \\
\frac{1}{\mathrm{\bf{h}}}\xi_{a}\left[\Phi_{0}\right]&=&\frac{1}{\mathrm{\bf{h}}_{L}}\xi_{b'}\left[\Phi_{*}\right](\mathrm{\bf{h}}_{L},w_{L},q_{L})+\mathrm{log}\lambda_{L} \\
\frac{1}{\mathrm{\bf{h}}}\xi_{b}\left[\Phi_{0}\right]&=&\frac{1}{\mathrm{\bf{h}}_{L}}\xi_{a'}\left[\Phi_{*}\right](\mathrm{\bf{h}}_{L},w_{L},q_{L})+\mathrm{log}\lambda_{L} \\
\rho\left[\Phi_{0}\right]&=&\sqrt\lambda_{L}\rho\left[\Phi_{*}\right]
\end{eqnarray*}
Solving for $f_{L}$ we get the same expressions for $\mathrm{\bf{h}}_{L}$, $w_{L}$ and $q_{L}$ as in the vacuum case and the energy density for the fluid goes as $z_{L}=\frac{1+w}{2+w}z$.
\\
$\underline{\mathrm{Case 2}}$: $q\leq 1$. Then $b'=a$, $c'=b$ and (\ref{eq:transf_maps}) is equivalent to
\begin{eqnarray*}
\alpha_{a}\left[\Phi_{0}\right]&=&\lambda_{L}\alpha_{b'}\left[\Phi_{*}\right](\mathrm{\bf{h}}_{L},w_{L},q_{L}) \\
\alpha_{b}\left[\Phi_{0}\right]&=&\lambda_{L}\alpha_{c'}\left[\Phi_{*}\right](\mathrm{\bf{h}}_{L},w_{L},q_{L}) \\
\frac{1}{\mathrm{\bf{h}}}\xi_{a}\left[\Phi_{0}\right]&=&\frac{1}{\mathrm{\bf{h}}_{L}}\xi_{b'}\left[\Phi_{*}\right](\mathrm{\bf{h}}_{L},w_{L},q_{L})+\mathrm{log}\lambda_{L} \\
\frac{1}{\mathrm{\bf{h}}}\xi_{b}\left[\Phi_{0}\right]&=&\frac{1}{\mathrm{\bf{h}}_{L}}\xi_{c'}\left[\Phi_{*}\right](\mathrm{\bf{h}}_{L},w_{L},q_{L})+\mathrm{log}\lambda_{L} \\
\rho\left[\Phi_{0}\right]&=&\sqrt\lambda_{L}\rho\left[\Phi_{*}\right]
\end{eqnarray*}
for which we get $z_{L}=\frac{1}{2+w}z$.

%LEMMA 3.9%
\begin{lemma} 
\label{lemma:3.9}
The identities for $\lambda_{L}$, $w_{L}$, $\mathrm{\bf{h}}_{L}$ and $q_{L}$ stay the same as in \cite{RT} (verified by direct substitution using Definition \ref{def:3.14}). Additionally, we have
\\
\begin{equation}
\label{eq:lemma3.9}
\frac{z}{z_{L}}=1-\alpha_{a,a'}[\Phi_{0}](\tau_{1-})=1-\alpha_{a,a'}[\Phi_{0}](\tau)
\end{equation}
\\
with $\Phi_{0}=\Phi_{0}(\pi,\mathrm{\bf{f}},\sigma_{*})$ and $\tau\in\mathbb{R}$.
\end{lemma}
\begin{definition} 
\label{def:3.17}
$\forall\mathrm{\bf{f}}=(\mathrm{\bf{h}},w,q,z)\in(0,\infty)^{4}$ define
\\
\begin{equation}
\label{eq:tau_star_def}
\tau_{*}(\mathrm{\bf{f}})=\begin{cases}
\frac{q}{1+w}\:,\:\mathrm{if \:q\leq 1}\\ 
1\:,\:\mathrm{if \:q>1}. 
\end{cases}
\end{equation}
\end{definition}

\begin{definition} 
\label{def:3.18}
$\forall\mathrm{\bf{f}}=(\mathrm{\bf{h}},w,q,z)\in(0,\infty)^{4}$ set 
\\
\begin{equation}
\label{def_of_K}
\mathrm{\bf{K(f)}}=2^{40}\left(\frac{1}{\mathrm{\bf{h}}}\right)^{2}\mathrm{max}\lbrace\left(\frac{1}{w}\right)^{2},w^{3}\rbrace\mathrm{max}\lbrace\left(\frac{1}{q}\right)^{2},q\rbrace\mathrm{exp}\left(-\frac{1}{\mathrm{\bf{h}}}2^{-7}\tau_{*}(\mathrm{\bf{f}})\right)\mathrm{max}\lbrace 1,z\rbrace^{2} 
\end{equation}
\end{definition}

%Definition 3.19%
\begin{definition} 
\label{def:3.19}
Let $\mathcal{F}$ be the open set of all $\mathrm{\bf{f}}=(\mathrm{\bf{h}},w,q,z)\in(0,\infty)^{4}$ for which $q\neq1$, $\mathrm{\bf{K}}(\mathrm{\bf{f}})<1$, $\mathrm{\bf{h}}<2^{-7}\tau_{*}(\mathrm{\bf{f}})$.
\end{definition}

%PROPOSITION 3.3%
\begin{proposition} 
\label{prop:3.3}
$\pi=(a,b,c)\in S_{3}$, $\sigma_{*}\in\lbrace -1,+1\rbrace^{3}$, $\mathrm{\bf{f}}=(\mathrm{\bf{h}},w,q,z)$. There are $\textbf{unique}$ maps 
\begin{eqnarray*}
\Pi=\Pi\left[\pi,\sigma_{*}\right]&:&\mathcal{F}\rightarrow (0,\infty)^{2}\times\mathbb{R}\times (0,\infty) \\
\Lambda=\Lambda\left[\pi,\sigma_{*}\right]&:&\mathcal{F}\rightarrow \left[1,\infty\right) \\
\tau_{2-}=\tau_{2-}\left[\pi,\sigma_{*}\right]&:&\mathcal{F}\rightarrow (-\infty,0)
\end{eqnarray*}
so that $\forall\mathrm{\bf{f}}=(\mathrm{\bf{h}},w,q,z)\in\mathcal{F}$ we have 
\\
\begin{enumerate}
\item $\vert\vert\Pi(\mathrm{\bf{f}})-Q_{L}(\mathrm{\bf{f}})\vert\vert_{\mathbb{R}^{3}}\leq\mathrm{\bf{K}}(\mathrm{\bf{f}})$ 
\item $\vert\Lambda(\mathrm{\bf{f}})-\lambda_{L}(\mathrm{\bf{f}})\vert\leq\mathrm{\bf{K}}(\mathrm{\bf{f}})$ 
\item $\tau_{-}(\mathrm{\bf{f}})<\tau_{2-}(\mathrm{\bf{f}})<\frac{1}{2}\tau_{1-}(\mathrm{\bf{f}})$ and $\vert\tau_{2-}(\mathrm{\bf{f}})-\tau_{1-}(\mathrm{\bf{f}})\vert\leq\mathrm{\bf{K}}(\mathrm{\bf{f}})$ 
\item $\Pi,\Lambda$ and $\tau_{2-}$ are continuous 
\item if we set $\tau_{2-}=\tau_{2-}(\mathrm{\bf{f}})$, $\tau_{2+}=\tau_{1+}(\mathrm{\bf{f}})$, $\pi'=(a',b',c')=\mathcal{P}_{L}(\pi,\mathrm{\bf{f}})$, $\lambda=\Lambda(\mathrm{\bf{f}})$ and $\mathrm{\bf{f}}'=(\mathrm{\bf{h}}',w',q',z')=\Pi(\mathrm{\bf{f}})$, then $\frac{1}{2}\leq\tau_{2+}-\tau_{2-}\leq3$ and there is a smooth field
\begin{equation*}
\Phi=\alpha\oplus\beta\oplus\gamma\in\mathcal{E}=\mathcal{E}(\sigma_{*};\tau_{2-},\tau_{2+})
\end{equation*}
that satisfies
\begin{itemize}
\item $(\frak{a,b},c,\frak{d})[\Phi,\mathrm{\bf{h}},Z]=0$ on $[\tau_{2-},\tau_{2+}]$ (i.e. solution to the primary system of equations in Section 1)
\item $\Phi(\tau_{2+})=\Phi_{*}(\pi,\mathrm{\bf{f}},\sigma_{*})$ and $\Phi(\tau_{2-})=\lambda\Phi_{*}(\pi',\mathrm{\bf{f}}',\sigma_{*})$, in particular $\Phi(\tau_{2+})\in\mathcal{H}(\pi,\sigma_{*})$ and $\Phi(\tau_{2-})\in\mathcal{H}(\pi',\sigma_{*})$
\item $\vert\beta_{a}[\Phi](\tau)\vert\geq\vert\beta_{a'}[\Phi](\tau)\vert$ for all $\tau\in[\tau_{2-},\frac{1}{2}\tau_{1-}(\mathrm{\bf{f}})]$ with equality if and only of $\tau=\tau_{2-}$
\item $d_{\mathcal{E},(\pi,\mathrm{\bf{h}})}(\Phi,\Phi_{0})\leq\mathrm{\bf{K}}(\mathrm{\bf{f}})$, where $\Phi_{0}=\Phi_{0}(\pi,\mathrm{\bf{f}},\sigma_{*})\big|_{[\tau_{2-},\tau_{2+}]}$
\item $\mathrm{sup}_{\tau\in[\tau_{2-},\tau_{2+}]}\mathrm{max}\lbrace\alpha_{b,c}[\Phi],\alpha_{c,a}[\Phi],\alpha_{a,b}[\Phi]\rbrace(\tau)\leq-2^{-2}\mathrm{min}\lbrace w^{2},w^{-1}\rbrace$ \\ \\
\end{itemize}
\end{enumerate}
\end{proposition}
%PROOF%
\noindent\underline{$\textbf{Proof}$}. The logic of the proof is the same as for the corresponding proposition in \cite{RT}. We start by introducing the parameter vector $\ell=(\ell_{1},...,\ell_{8})\in\mathbb{R}^{8}$.
\\
$\pi=(a,b,c)\in S_{3}$, $\sigma_{*}\lbrace -1,+1\rbrace^{3}$, $\mathrm{\bf{f}}=(\mathrm{\bf{h}},w,q,z)\in (0,\infty)^{4}$, $q\neq 1$, $\mathrm{\bf{h}}\leq 1$. 
\\
For any $s=(s_{1},...,s_{8})\in\mathbb{R}^{8}$ define
\begin{eqnarray*}
\mathrm{\bf{X}}(s)&=&\mathrm{\bf{X}}(s_{1},...,s_{8}) \\
&=&2^{s_{1}}\left(\frac{1}{\mathrm{\bf{h}}}\right)^{s_{2}}\times 
\begin{cases}
\frac{1}{w}^{s_{3}}\:\:\mathrm{if}\:w\leq1 \\
w^{s_{4}}\:\:\mathrm{if}\:w>1
\end{cases}
\times
\begin{cases}
\frac{1}{q}^{s_{5}}\:\:\mathrm{if}\:q\leq1 \\
q^{s_{6}}\:\:\mathrm{if}\:q>1
\end{cases}
\times \mathrm{exp}\left(\frac{1}{\mathrm{\bf{h}}}s_{7}\tau_{*}\right)\times \mathrm{max}\lbrace 1,z\rbrace^{s_{8}}
\end{eqnarray*}
\\
%Properties of X(s)%
$Properties\:of\:\mathrm{\bf{X}}(s)$ (stay unchanged):
\begin{itemize}
\item $\mathrm{\bf{X}}(s)\mathrm{\bf{X}}(s')=\mathrm{\bf{X}}(s+s')$;
\item $\mathrm{\bf{X}}(0,...,0)=1$;
\item $\mathrm{\bf{X}}(-1,0,0,-1,-1,0,0,0)=\frac{q}{2w}\leq\tau_{*}$;
\item $\mathrm{\bf{X}}(s)$ is positive, non-decreasing in each argument. \\
\end{itemize}
%Basic smallness assumptions%
$Basic$ $smallness$ $assumptions$. Introduce parameter vector $\ell=(\ell_{1},...,\ell_{9})\in\mathbb{R}^{9}$ with $(\ell_{1},...,\ell_{8})\geq (0,0,0,0,0,0,-\infty,0)$ and $\ell_{9}\geq 0$. $(\bullet)_{1}$
\\
Basic assumptions on $\mathrm{\bf{f}}=(\mathrm{\bf{h}},w,q,z)$ are 
\\
\begin{equation}
\label{eq:f_assumptions}
q\neq 1\:\:\:\:\:\:\:\: \mathrm{\bf{K}}:=\mathrm{\bf{X}}(\ell)<1\:\:\:\:\:\:\:\: \mathrm{\bf{h}}< 2^{-\ell_{9}}\tau_{*}
\end{equation}
\\
We are interested in the ratio $\mathrm{\bf{h}}/\tau_{*}$. The component of $\ell$ giving a bound for it is $\ell_{9}$.
\\
\\
%Abbreviations%
$Abbreviations$. See \cite{RT}.
\\ \\
%Preliminaries 1%
$Preliminaries\:1$. Introduce $\epsilon_{-}$ and $\epsilon_{+}$ by $\tau_{0-}=\tau_{-}+\epsilon_{-}$ and $\tau_{0+}=\tau_{+}-\epsilon_{+}$. Recall Lemma \ref{lemma:3.4}. By using the definitions for $\tau_{+}$, $\tau_{1+}$, $\tau_{-}$ and $\tau_{1-}$, we find
\[\epsilon_{+}=\tau_{+}-\tau_{0+}=\tau_{+}-\tau_{1+}=\frac{1+w}{1+2w}\left(1+\frac{1}{w}-\mathrm{\bf{h}}\mathrm{log}2\right)\]
\[\epsilon_{-}=\tau_{0-}-\tau_{-}=\frac{1}{2}(\tau_{1-}-\tau_{-})=\begin{cases} 
\frac{1}{2(3+w)}\left(\frac{1+w}{2+w}q-\mathrm{\bf{h}}\mathrm{log}2\right)\:\:\mathrm{if}\:q<1 \\
\frac{1+w}{2(3+2w)}\left(\frac{1+w}{2+w}-\mathrm{\bf{h}}\mathrm{log}2\right)\:\:\mathrm{if}\:q>1.
\end{cases}
\]
\\
\\
Require $\ell_{9}\geq 2$ $(\bullet)_{2}$. Then $\mathrm{\bf{h}}\mathrm{log}2\leq\mathrm{\bf{h}}$ and $\mathrm{\bf{h}}\overset{(\ref{eq:f_assumptions}),(\bullet)_{2}}<2^{-2}\tau_{*}\overset{(\ref{eq:tau_star_def})}\leq2^{-2}\mathrm{min}\lbrace 1,q\rbrace$. Recall $\tau_{+}=1+\frac{1}{w}$. Then we have 
\\
\begin{eqnarray*}
\frac{\epsilon_{+}}{\tau_{+}}&=&\frac{1+w}{1+2w}\left(1+\frac{1}{w}-\mathrm{\bf{h}}\mathrm{log}2\right)\cdot\frac{w}{1+w} \\
&=&\frac{1+w}{1+2w}-\frac{w}{1+2w}\mathrm{\bf{h}}\mathrm{log}2 \\
&\overset{(\bullet)_{2}}\geq&\frac{1}{2}-\frac{1}{2}\cdot\frac{1}{4}=\frac{3}{8}>2^{-2}
\end{eqnarray*}
\\
On the other hand, $\frac{\epsilon_{+}}{\tau_{+}}\leq\frac{1+w}{1+2w}\leq1$. Therefore, 
\\
\begin{equation}
\label{e+_tau+_bounds}
2^{-2}\leq\frac{\epsilon_{+}}{\tau_{+}}\leq1
\end{equation}
\\
Analogous calculations (for $q<1, \frac{\epsilon_{-}}{\tau_{*}}=\frac{1+w}{q}\frac{1}{2(3+w)}\left(\frac{1+w}{2+w}q-\mathrm{\bf{h}}\mathrm{log}2\right)\leq\frac{(1+w)^{2}}{2(1+w)(2+w)}\leq 2^{-1}$, etc.) lead to
\\
\begin{eqnarray}
\label{e-_taustar_bounds}
2^{-5}\leq\frac{\epsilon_{-}}{\tau_{*}}\leq2^{-1}
\end{eqnarray}
\\
We have
\\
\begin{equation}
\label{tau_overview}
-1<\tau_{-}<\tau_{0-}<\tau_{1-}<0<\frac{1}{2}<\tau_{0+}=\tau_{1+}=\tau_{2+}<\mathrm{min}\lbrace2,\tau_{+}\rbrace
\end{equation}
\\
Define $\delta:=2^{-9}\mathrm{min}\lbrace1,w\rbrace\tau_{*}$. Recall $Properties\:of\:\mathrm{\bf{X}}(s)$. Then we have
\\
\begin{align}
\label{delta_ineq}
\delta&=\mathrm{\bf{X}}(-9,0,-1,0,0,0,0,0)\tau_{*} \nonumber \\
&\geq\mathrm{\bf{X}}(-9,0,-1,0,0,0,0,0)\mathrm{\bf{X}}(-1,0,0,-1,-1,0,0,0) \nonumber \\
&=\mathrm{\bf{X}}(-10,0,-1,-1,-1,0,0,0)
\end{align}
\\
Therefore, from (\ref{e-_taustar_bounds}) follows that $\tau_{*}\leq 2^{5}\epsilon_{-}$ and $\delta\leq 2^{-4}\mathrm{min}\lbrace1,w,\epsilon_{-},\frac{\epsilon_{+}}{\tau_{+}\tau_{0+}}\rbrace$. This implies that the assumption of Lemma \ref{lemma:3.4} is satisfied.
\\
\\
%Preliminaries 2%
$Preliminaries\:2$. Require $\ell_{8}\geq7$ $(\bullet)_{3}$. Recall Lemmas \ref{lemma:3.2} and \ref{lemma:3.8}. Then we have
\\
\begin{eqnarray*}
d_{\mathcal{E}}(\Phi_{0},\Phi_{1})&\overset{Lemma\ref{lemma:3.8}, d)}\leq& 2^{2}d_{\mathcal{D}}(\Phi_{0}(\tau_{1+}),\Phi_{*})  \\
&\overset{Lemma\ref{lemma:3.2},b)and(\bullet)_{3}}\leq& 2^{11}\not d_{\mathcal{D}}(\Phi_{0}(\tau_{1+}),\Phi_{*})  \\
&\overset{Lemma\ref{lemma:3.8},c)}\leq&2^{18}(1+w)\mathrm{exp}\left(-\frac{1}{2\mathrm{\bf{h}}}\mathrm{min}\lbrace1,q\rbrace\right)  \\
&\overset{(\ref{eq:tau_star_def})}\leq&2^{18}(1+w)\mathrm{exp}\left(-\frac{1}{2\mathrm{\bf{h}}}\tau_{*}\right)  \\
&\leq&\mathrm{\bf{X}}(19,0,0,1,0,0,-2^{-1},0) \\
&\overset{(\ref{delta_ineq})}\leq&2^{-2}\delta\mathrm{\bf{X}}(31,0,1,2,1,0,-2^{-1},0)
\end{eqnarray*}
\\
Note on the second inequality: the assumption $(\bullet)_{3}$ plays the crucial role here. In particular, we need to make sure that the assumptions of Lemma \ref{lemma:3.2} are satisfied, that is, 
\\
\begin{itemize}
\item $\mathrm{\bf{h}}\leq 1$ $\longrightarrow$ ok by assumption $(\bullet)_{3}$;
\item constants $C,D\geq 1$ such that $C^{-1}\leq A_{a}[X]\leq C$ and $D^{-1}\leq\mathrm{\bf{h}}\vert\phi_{a}[X]\vert\leq D$ $\longrightarrow$ ok by putting $C=D=2$ (recall $A_{a}[\Phi_{0}](\tau_{1+})=1$ and $\mathrm{\bf{h}}\phi_{a}[\Phi_{0}](\tau_{1+})=\tau_{1+}\in[\frac{1}{2},2]$, by Definition \ref{def:3.14}). We want to use statement b) from Lemma \ref{lemma:3.2}. Indeed, $\mathrm{exp}\left(-\frac{1}{\mathrm{\bf{h}}}2^{-2}2^{-1}\right)\leq 2^{-12}$ is satisfied for $l_{8}$ at least 7. That is the motivation for  the assumption $(\bullet)_{3}$. \\
\end{itemize}
Recall the estimate for $d_{\mathcal{E}}(\Phi_{0},\Phi_{1})$. Therefore, the assumption $(\ell_{1},...,\ell_{8})\geq (31,0,1,2,1,0,-2^{-1},0)$ $(\bullet)_{4}$ implies immediately that
\begin{equation}
\Phi_{1}\in B_{\mathcal{E}}[2^{-2}\delta, \Phi_{0}]
\end{equation}
\\
%Fixed point construction%
$Construction\:of\:\Phi$. Fixed point construction is technically carried out on the interval $[\tau_{0-},\tau_{0+}]$. Define a map $P:B_{\mathcal{E}}[\delta,\Phi_{0}]\rightarrow B_{\mathcal{E}}[\delta,\Phi_{0}], \Psi\mapsto P(\Psi)$ by
\\
\begin{eqnarray}
\label{def_of_P}
A_{a}\left[P(\Psi)\right](\tau)-A_{a}\left[\Phi_{1}\right](\tau)&=&\int_{\tau_{0+}}^{\tau} \mathrm{d}\tau'\mathrm{\bf{I}}_{1}\left[\Psi,\mathrm{\bf{h}},\pi\right](\tau') \\
\label{def_of_P_1}
\theta_{a}\left[P(\Psi),\mathrm{\bf{h}}\right](\tau)-\theta_{a}\left[\Phi_{1},\mathrm{\bf{h}}\right](\tau)&=&\int_{\tau_{0+}}^{\tau} \mathrm{d}\tau'\mathrm{\bf{I}}_{2}\left[\Psi,\mathrm{\bf{h}},\pi\right](\tau') 
\\
\label{def_of_P_2}
\alpha_{p,a}\left[P(\Psi)\right](\tau)-\alpha_{p,a}\left[\Phi_{1}\right](\tau)&=&\int_{\tau_{0+}}^{\tau}\mathrm{d}\tau'\mathrm{\bf{I}}_{(3,p)}\left[\Psi,\mathrm{\bf{h}},\pi\right](\tau') \\
\label{def_of_P_3}
\xi_{p,a}\left[P(\Psi),\mathrm{\bf{h}}\right](\tau)-\xi_{p,a}\left[\Phi_{1},\mathrm{\bf{h}}\right](\tau)&=&\int_{\tau_{0+}}^{\tau}\mathrm{d}\tau'' \int_{\tau_{0+}}^{\tau''}\mathrm{d}\tau'\mathrm{\bf{I}}_{(3,p)}\left[\Psi,\mathrm{\bf{h}},\pi\right](\tau') \\
\label{def_of_P_end}
\rho[P(\Psi)](\tau)-\rho[\Phi_{1}](\tau)&=&0
\end{eqnarray}
\\
for all $p\in\lbrace b,c\rbrace$ and $\tau\in\left[\tau_{0-},\tau_{0+}\right]$. We now want to make sure that $P$ is well defined. Let $\Psi\in B_{\mathcal{E}}[\delta,\Phi_{0}]$. The requirement $(\ell_{1},...,\ell_{8})\geq (29,2,1,1,1,0,-2^{-7},1)$ $(\bullet)_{5}$ makes sure that it's the case, since we have
\\
\begin{eqnarray*}
d_{\mathcal{E}}(P(\Psi),\Phi_{0})&\leq& d_{\mathcal{E}}(P(\Psi),\Phi_{1})+d_{\mathcal{E}}(\Phi_{1},\Phi_{0}) \\
&\leq&\mathrm{\bf{X}}(16,1,0,0,0,0,-2^{-7},1)+2^{-2}\delta \\
&\leq&(\mathrm{\bf{X}}(26,1,1,1,1,0,-2^{-7},1)+2^{-2})\delta \\
&\overset{(\bullet)_{5}}\leq&\delta
\end{eqnarray*}
\\
(Note: although in the definition of $P$ we have the difference in $\rho$'s and in the definition of $d_{\mathcal{E}}$ in $\rho^{2}$'s, the estimate above is still ok for $(\ref{def_of_P_end})$ since $\vert\rho^{2}[\Phi]-\rho^{2}[\Psi]\vert\leq\vert\rho[\Phi]-\rho[\Psi]\vert$ always holds. To be precise, $\vert\rho^{2}[\Phi]-\rho^{2}[\Psi]\vert\leq2\mathrm{max}\lbrace\vert\rho[\Phi]\vert, \vert\rho[\Psi]\vert\rbrace\vert\rho[\Phi]-\rho[\Psi]\vert$). Further, recall Lemma \ref{lemma:3.6}. We have
\\
\begin{equation}
\label{abs_I_estimate}
\vert \mathrm{\bf{I}}_{S}[\Psi,\mathrm{\bf{h}},\pi]\vert\leq\mathrm{\bf{X}}(13,1,0,0,0,0,-2^{-7},1)\overset{(\ref{delta_ineq})}\leq 2^{-6}\delta\mathrm{\bf{X}}(29,1,1,1,1,0,-2^{-7},1)\overset{(\bullet)_{5},(\ref{eq:f_assumptions})}\leq2^{-6}\delta
\end{equation} 
\\
\begin{equation}
\label{abs_I_diff_estimate}
\vert\mathrm{\bf{I}}_{S}\left[\Psi\right]-\mathrm{\bf{I}}_{S}\left[\Psi'\right]\vert\leq2^{-5}\mathrm{\bf{X}}(24,2,0,0,0,0,-2^{-7},1)d_{\mathcal{E}}(\Psi,\Psi')\overset{(\bullet)_{5}}\leq 2^{-5}d_{\mathcal{E}}(\Psi,\Psi')
\end{equation}
\\
We now want to apply Banach Fixed Point Theorem to $P(\Psi)$ and, therefore, show that $P$ admits a unique fixed point on $B_{\mathcal{E}}[\frac{1}{2}\delta, \Phi_{0}]$. For that, we need the following assumptions to be satisfied:
\\
\begin{itemize}
\item $P$ is defined on a non-empty complete metric space;
\item $P$ is a contraction, i.e. $\exists const. L<1$ such that $\forall\Psi,\Psi'\in B_{\mathcal{E}}[\frac{1}{2}\delta,\Phi_{0}]$: $d_{\mathcal{E}} (P(\Psi),P(\Psi'))\leq L d_{\mathcal{E}}(\Psi,\Psi')$ \\
\end{itemize}
The first requirement is satisfied by Definitions \ref{def:3.5} and \ref{def:3.6}. We now show that $P$ is a contraction. Recall $\tau_{0+}=\tau_{1+}=\tau_{2+}<\mathrm{min}\lbrace2,\tau_{+}\rbrace$. Therefore, 
\\
\begin{equation}
\label{sup_estimate}
\mathrm{sup}_{\tau\in[\tau_{0-},\tau_{0+}]}\vert\tau-\tau_{0+}\vert\leq\mathrm{sup}_{\tau\in[\tau_{0-},\tau_{0+}]}(\vert\tau\vert+\vert\tau_{0+}\vert)\leq\mathrm{sup}_{\tau\in[\tau_{0-},\tau_{0+}]}\vert\tau\vert+\mathrm{sup}\vert\tau_{0+}\vert\leq 4
\end{equation}
\\
Recall the definition of $P(\Psi)$ and the right hand sides (RHS) of equations (\ref{def_of_P})-(\ref{def_of_P_end}). By (\ref{abs_I_estimate}) and (\ref{sup_estimate}) each of the RHS satisfies $\leq 2^{-2}\delta$. Therefore, $P(\Psi)\in B_{\mathcal{E}}[\frac{1}{2}\delta,\Phi_{0}]$ and is a contraction with $L\leq\frac{1}{2}$.
\\
So, by Banach Fixed Point Theorem it follows that $P$ has a unique fixed point
\\
\begin{equation}
\label{fixed_point}
\boxed{\Phi\in B_{\mathcal{E}}[\frac{1}{2}\delta,\Phi_{0}]}
\end{equation}
\\
\\
\emph{Proof that the fixed point satisfies $(\frak a,\frak b, c, \frak d)[\Phi,\mathrm{\bf{h}},Z]=0$}. Fixed point is smooth. Observe that $\Phi(\tau_{0+})=\Phi_{1}(\tau_{0+})\overset{(\ref{tau_overview})}=\Phi_{1}(\tau_{1+})\overset{Definition \ref{def:3.15}}=\Phi_{*}$. Further, $c[\Phi,\mathrm{\bf{h}},Z](\tau_{0+})\overset{(\ref{tau_overview})}=c[\Phi,\mathrm{\bf{h}},Z](\tau_{1+})\overset{Lemma \ref{lemma:3.8},b}=0$. 
\\
\\
Set $\Psi=P(\Psi)=\Phi$ in (\ref{def_of_P})-(\ref{def_of_P_end}) and differentiate with respect to $\tau$. Recall Definition \ref{def:3.11}. We have:
\\
\underline{for (\ref{def_of_P}) and (\ref{def_of_P_1})}: The result of differentiation both of them can be written in the matrix form (so that we can use Lemma \ref{lemma:3.1})
\\
\begin{equation*}
\frac{d}{d\tau}\left(\begin{array}{c}A_{a} \\\theta_{a}\end{array}\right)=\frac{1}{(A_{a})^{2}}\left(\begin{array}{cc}\frac{1}{\mathrm{\bf{h}}}(A_{a})^{2}\mathrm{tanh}\phi_{a} & \frac{1}{\mathrm{\bf{h}}}(A_{a})^{2}\mathrm{sech}\phi_{a} \\\phi_{a}\mathrm{tanh}\phi_{a}-1 & \mathrm{sinh}\phi_{a}+\phi_{a}\mathrm{sech}\phi_{a}\end{array}\right) \left(\begin{array}{c}\frak a_{a}[\Phi,\mathrm{\bf{h}},B_{a}]-\frak a_{a}[\Phi,\mathrm{\bf{h}},Z] \\-(\sigma_{*})_{a}\frak b_{a}[\Phi,\mathrm{\bf{h}},B_{a}]+(\sigma_{*})_{a}\frak b_{a}[\Phi,\mathrm{\bf{h}},Z]\end{array}\right)
\end{equation*}
\\
Comparing it with the equation in Lemma \ref{lemma:3.1}, we conclude that $\boxed{\frak a_{a}[\Phi,\mathrm{\bf{h}},Z]=\frak b_{a}[\Phi,\mathrm{\bf{h}},Z]=0}$ must hold.
\\
\underline{for (\ref{def_of_P_2})}: The result of differentiation is
\\
\begin{equation}
\label{eq:help1}
\frac{d}{d\tau}\alpha_{p,a}[\Phi]=\frac{1}{\mathrm{\bf{h}}}\frak a_{p}[\Phi,\mathrm{\bf{h}},Z,B_{a}]+\frac{1}{\mathrm{\bf{h}}}\frak a_{a}[\Phi,\mathrm{\bf{h}},Z,B_{a}]
\end{equation}
\\
Recall equation (\ref{eq:frak_a}), which implies (no energy density terms appear because $B_{a}$ has $m=0$) 
\begin{equation}
\label{eq:help2}
\frak a_{p}[\Phi,\mathrm{\bf{h}},B_{a}]+\frak a_{a}[\Phi,\mathrm{\bf{h}},B_{a}]=-\mathrm{\bf{h}}\frac{d}{d\tau}\alpha_{p,a}[\Phi]
\end{equation}
Using the earlier obtained result $\frak a_{a}[\Phi,\mathrm{\bf{h}},Z]=0$ we calculate
\\
\begin{equation*}
\frak a_{p}[\Phi,\mathrm{\bf{h}},B_{a}]+\frak a_{a}[\Phi,\mathrm{\bf{h}},B_{a}]\overset{(\ref{eq:help2})}=-\mathrm{\bf{h}}\frac{d}{d\tau}\alpha_{p,a}[\Phi]\overset{(\ref{eq:help1})}=-(\frak a_{p}[\Phi,\mathrm{\bf{h}},Z]-\frak a_{p}[\Phi,\mathrm{\bf{h}},B_{a}]+\underbrace{\frak a_{a}[\Phi,\mathrm{\bf{h}},Z]}_{=0}-\frak a_{a}[\Phi,\mathrm{\bf{h}},B_{a}])
\end{equation*}
\\
which implies $\boxed{\frak a_{p}[\Phi,\mathrm{\bf{h}},Z]=0}$.
\\
\underline{for (\ref{def_of_P_3})}: The result of differentiation and (\ref{eq:help1})  give $\frac{d}{d\tau}\xi_{p,a}[\Phi,\mathrm{\bf{h}}]=\alpha_{p,a}[\Phi]$. The explicit calculation gives
\\
\begin{eqnarray*}
\alpha_{p}[\Phi]+\alpha_{a}[\Phi]&=&\frac{d}{d\tau}(\xi_{p}[\Phi]+\xi_{a}[\Phi]) \\
&\overset{Definition \ref{def:3.3}}=&\frac{d}{d\tau}(\mathrm{\bf{h}}\mathrm{log}\vert\frac{1}{2}\beta_{p}[\Phi]\vert)+\frac{d}{d\tau}(\mathrm{\bf{h}}\mathrm{log}\vert\frac{1}{2}\beta_{a}[\Phi]\vert) \\
&=&\mathrm{\bf{h}}\left(\frac{1}{\vert\beta_{p}\vert}\frac{d}{d\tau}\vert\beta_{p}\vert+\frac{1}{\vert\beta_{a}\vert}\frac{d}{d\tau}\vert\beta_{a}\vert\right) 
\end{eqnarray*}
\\
which is equivalent to 
\\
\begin{equation*}
-\beta_{a}[\Phi]\left(-\mathrm{\bf{h}}\frac{d}{d\tau}\beta_{p}[\Phi]+\alpha_{p}[\Phi]\beta_{p}[\Phi]\right)-\beta_{p}[\Phi]\underbrace{\left(-\mathrm{\bf{h}}\frac{d}{d\tau}\beta_{a}[\Phi]+\alpha_{a}[\Phi]\beta_{a}[\Phi]\right)}_{=\frak b_{a}[\Phi,\mathrm{\bf{h}},Z]=0}=0
\end{equation*}
\\
Therefore, $\boxed{\frak b_{p}[\Phi,\mathrm{\bf{h}},Z]=0}$.
\\
\underline{for (\ref{def_of_P_end})}: The result of differentiation gives $\frac{d}{d\tau}\rho[\Phi]=0$, which by Definition \ref{def:3.3} is equivalent to $\boxed{\frak d [\Phi,\mathrm{\bf{h}},Z]=0}$.
\\
Now, Proposition \ref{prop:3.2} and $\frak d[\Phi,\mathrm{\bf{h}},Z]=0$ imply that $\boxed{c[\Phi,\mathrm{\bf{h}},Z]=0}$ on $[\tau_{0-},\tau_{0+}]$.
\\
\\
\emph{Estimates on $\Phi$}. We have $P(\Phi)=\Phi$ by the fixed point equation. Further,
\\
\begin{eqnarray*}
d_{\mathcal{E}}(\Phi_{0},\Phi)&\leq&d_{\mathcal{E}}(\Phi_{0},\Phi_{1})+d_{\mathcal{E}}(\Phi_{1},\Phi) \\
&=&d_{\mathcal{E}}(\Phi_{0},\Phi_{1})+d_{\mathcal{E}}(\Phi_{1},P(\Phi)) \\
&\overset{Preliminaries 2,(\ref{abs_I_estimate})}\leq&\mathrm{\bf{X}}(19,0,0,1,0,0,-2^{-1},0)+\mathrm{\bf{X}}(17,1,0,0,0,0,-2^{-7},1) \\
&\leq&\mathrm{\bf{X}}(20,1,0,1,0,0,-2^{-7},1)
\end{eqnarray*}
\\
The result above motivates to require $(\ell_{1},...,\ell_{8})\geq (20,1,0,1,0,0,-2^{-7},1)$ $(\bullet)_{6}$ so that we can have (recall equation (\ref{eq:f_assumptions})) $d_{\mathcal{E}}(\Phi_{0},\Phi)\leq\mathrm{\bf{K}}$. We now want to make an estimate on the metric space $(\mathcal{D}(\sigma_{*}),\not{d}_{\mathcal{D}(\sigma_{*}),\mathrm{\bf{h}}})$. That is, we want an estimate for $\not{d}_{\mathcal{D}}(\Phi_{0}(\tau),\Phi(\tau))$ for $\tau$ in a sub-interval of $[\tau_{0-},\tau_{0+}]$. The desired bound for $\not{d}_{\mathcal{D}}(\Phi_{0}(\tau),\Phi(\tau))$ comes from Lemma \ref{lemma:3.2} a). Define $\mathcal{J}:=[\tau_{0-},\frac{1}{2}\tau_{1-}]\subset [\tau_{0-},\tau_{0+}]$. Morally, $\mathcal{J}$ represents the interval that allowed us "extra space" for the fixed point construction. The latter was technically carried out on $[\tau_{0-},\tau_{0+}]$, but the idea was "from crossing to crossing". To the right, we had exactly $\tau_{0+}=\tau_{1+}$, but on the left $\tau_{0-}<\tau_{1-}$. That difference is represented by $\mathcal{J}$.
\\
Setting $C=2$ and $D=12\mathrm{max}\lbrace 1,\frac{1}{q}\rbrace$ the assumptions of Lemma \ref{lemma:3.2} are satisfied. Observe that $C^{2}D\leq\mathrm{\bf{X}}(6,0,0,0,1,0,0,0)$ and we estimate
\\
\begin{equation}
\label{eq:def_of_M}
\not{d}_{\mathcal{D}}(\Phi_{0}(\tau),\Phi(\tau))\leq 2^{3}C^{2}D d_{\mathcal{E}}(\Phi_{0},\Phi)\leq\mathrm{\bf{X}}(29,1,0,1,1,0,-2^{-7},1)=:\mathrm{\bf{M}}
\end{equation}
\\
\\
%Construction of tau 2-%
$Construction\:of\:\tau_{2-}$. Recall that $a'=c$ if $q<1$ and $a'=b$ if $q>1$. Recall Lemma \ref{lemma:3.9}. For all $\tau\in\mathcal{J}$ we have
\\
\begin{equation}
\label{eq:xi_ref_diff}
(\xi_{a}[\Phi_{0},\mathrm{\bf{h}}]-\xi_{a'}[\Phi_{0},\mathrm{\bf{h}}])F=\tau-\tau_{1-}-\underbrace{2\mathrm{\bf{h}}\mathrm{log}(1+\mathrm{e}^{2\tau/\mathrm{\bf{h}}})F}_{=:T_{1}}
\end{equation}
\\
By adding and subtracting $\xi_{a}[\Phi,\mathrm{\bf{h}}]$ and $\xi_{a'}[\Phi,\mathrm{\bf{h}}]$ to the LHS of (\ref{eq:xi_ref_diff}) we get
\\
\begin{equation}
\label{eq:xi_actual_diff}
(\xi_{a}[\Phi,\mathrm{\bf{h}}]-\xi_{a'}[\Phi,\mathrm{\bf{h}}])F=\tau-\tau_{1-}-T_{2}
\end{equation}
\\
with $T_{2}:=T_{1}-(\xi_{a}[\Phi,\mathrm{\bf{h}}]-\xi_{a}[\Phi_{0},\mathrm{\bf{h}}])F+(\xi_{a'}[\Phi,\mathrm{\bf{h}}]-\xi_{a'}[\Phi_{0},\mathrm{\bf{h}}])F$. Both $T_{1}$ and $T_{2}$ are functions of $\tau\in\mathcal{J}$. For all such $\tau$ we now want to make an estimate for the RHS of (\ref{eq:xi_actual_diff}). That is, how $\vert\tau-\tau_{1-}\vert$ compares to $\vert T_{2}\vert$. We have 
\\
\begin{equation*}
0<T_{1}\overset{Def.}=2\mathrm{\bf{h}}\mathrm{log}(1+\mathrm{e}^{2\tau/\mathrm{\bf{h}}})F\overset{\mathrm{log}(1+x)\leq x}\leq2\mathrm{\bf{h}}\mathrm{e}^{2\tau/\mathrm{\bf{h}}}F\overset{\tau\leq\frac{1}{2}\tau_{1-}}\leq 2\mathrm{\bf{h}}\mathrm{e}^{\tau_{1-}/\mathrm{\bf{h}}}F\leq\mathrm{\bf{M}}F
\end{equation*}
\\
Therefore,
\\
\begin{eqnarray*}
\vert T_{2}\vert&=&\vert T_{1}-(\xi_{a}[\Phi,\mathrm{\bf{h}}]-\xi_{a}[\Phi_{0},\mathrm{\bf{h}}])F+(\xi_{a'}[\Phi,\mathrm{\bf{h}}]-\xi_{a'}[\Phi_{0},\mathrm{\bf{h}}])F\vert \\
&\leq&\vert T_{1}\vert+\vert\xi_{a}[\Phi,\mathrm{\bf{h}}]-\xi_{a}[\Phi_{0},\mathrm{\bf{h}}]\vert F+\vert\xi_{a'}[\Phi,\mathrm{\bf{h}}]-\xi_{a'}[\Phi_{0},\mathrm{\bf{h}}]\vert F \\
&\leq&\mathrm{\bf{M}}F+2\not{d}_{\mathcal{D}}(\Phi,\Phi_{0})F \\
&\leq&3\mathrm{\bf{M}}F \\
&\overset{F\leq\frac{1}{2}by\:Def.}\leq&\frac{3}{2}\mathrm{\bf{M}}
\end{eqnarray*}
\\
Set
\\
\begin{equation*}
\mathrm{dist}_{\mathbb{R}}(\tau_{1-},\mathbb{R}\setminus\mathcal{J})=\mathrm{min}\lbrace\frac{1}{2}\vert\tau_{1-}\vert,\epsilon_{-}\rbrace\overset{(\ref{e-_taustar_bounds})}\geq 2^{-5}\tau_{*}\geq\mathrm{\bf{X}}(-6,0,0,-1,-1,0,0,0)
\end{equation*}
\\
Then
\\
\begin{eqnarray*}
\vert T_{2}\vert&\leq&\frac{3}{2}\mathrm{\bf{X}}(29,1,0,1,1,0,-2^{-7},1) \\
&=&\frac{3}{2}\mathrm{\bf{X}}(-6,0,0,-1,-1,0,0,0)\mathrm{\bf{X}}(35,1,0,2,2,0,-2^{-7},1) \\
&\leq&\mathrm{\bf{X}}(37,1,0,2,2,0,-2^{-7},1)\mathrm{dist}_{\mathbb{R}}(\tau_{1-},\mathbb{R}\setminus\mathcal{J})
\end{eqnarray*}
\\
Require $(\ell_{1},...,\ell_{8})\geq(37,1,0,2,2,0,-2^{-7},1)$ $(\bullet)_{7}$. Therefore, $\vert T_{2}\vert\leq\mathrm{dist}_{\mathbb{R}}(\tau_{1-},\mathbb{R}\setminus\mathcal{J})$.
\\
\\
Set 
\\
\begin{equation}
\label{eq:def_tau2-}
\tau_{2-}=\mathrm{sup}\lbrace\tau\in\mathcal{J}|\xi_{a}[\Phi,\mathrm{\bf{h}}](\tau)\leq\xi_{a'}[\Phi,\mathrm{\bf{h}}](\tau)\rbrace
\end{equation}
\\
The rest stays the same. The condition $(\ell_{1},...,\ell_{8})\geq(31,1,0,1,1,0,-2^{-7},1)$ $(\bullet)_{8}$ implies $\vert\tau_{2-}-\tau_{1-}\vert\leq\frac{3}{2}\mathrm{\bf{M}}\overset{(\bullet)_{8}}\leq\mathrm{\bf{K}}$.
\\
\\
$Estimates\:on\:\Phi_{0}$. Here, we want to estimate how big were the terms that we neglected for $\alpha$ and $\xi$. For example, we assumed that $\xi_{a}(\tau)\approx\tau$, where the "$\approx$" meant neglecting the $\mathcal{O}(\mathrm{exp}(-2\vert\tau\vert))$ term for the limiting cases $\tau\gg\mathrm{\bf{h}}$ and $\tau\ll \mathrm{\bf{h}}$. We now make it quantitative. For all $\tau\in\mathcal{J}$ we have
\begin{eqnarray*}
\vert\alpha_{a}[\Phi_{0}](\tau)-1\vert&=&\vert\mathrm{tanh}\frac{1}{\mathrm{\bf{h}}}\vert\tau\vert-1\vert\leq 2\mathrm{exp}\left(-\frac{2}{\mathrm{\bf{h}}}\vert\tau\vert\right) \\
\vert\xi_{a}[\Phi_{0},\mathrm{\bf{h}}](\tau)-\tau\vert&=&\big|\mathrm{\bf{h}}\mathrm{log}(2\mathrm{cosh}\frac{1}{\mathrm{\bf{h}}}\vert\tau\vert)-\vert\tau\vert\big|\leq\mathrm{\bf{h}}\mathrm{exp}\left(-\frac{2}{\mathrm{\bf{h}}}\vert\tau\vert\right) \\
\mathrm{exp}\left(-\frac{2}{\mathrm{\bf{h}}}\vert\tau\vert\right)&\leq&\mathrm{exp}\left(-\frac{1}{\mathrm{\bf{h}}}\vert\tau_{1-}\vert\right)\leq\mathrm{exp}\left(-\frac{1}{\mathrm{\bf{h}}}2^{-2}\tau_{*}\right)\leq 2^{-29}\mathrm{\bf{M}}
\end{eqnarray*}
\\
\\
%Construction of lambda%
\emph{Construction of $\lambda$}. Set
\\
\begin{equation}
\label{eq:lambda}
\lambda=\alpha_{a'}[\Phi](\tau_{2-})
\end{equation}
\\
Recall Lemma \ref{lemma:3.9}. Then,
\\
\begin{eqnarray*}
\vert\lambda-\lambda_{L}\vert&=&\vert\alpha_{a'}[\Phi](\tau_{2-})-(1-\alpha_{a,a'}[\Phi_{0}](\tau_{1-}))\vert \\
&\leq&\vert\alpha_{a'}[\Phi](\tau_{2-})-\alpha_{a'}[\Phi_{0}](\tau_{2-})\vert+\vert\alpha_{a'}[\Phi_{0}](\tau_{2-})-\alpha_{a'}[\Phi_{0}](\tau_{1-})\vert+\vert\alpha_{a'}[\Phi_{0}](\tau_{1-})+(1-\alpha_{a,a'}[\Phi_{0}](\tau_{1-}))\vert \\
&\leq&\vert\alpha_{a'}[\Phi](\tau_{2-})-\alpha_{a'}[\Phi_{0}](\tau_{2-})\vert+\left(\vert\alpha_{a}[\Phi_{0}](\tau_{2-})-1\vert+\vert\alpha_{a}[\Phi_{0}](\tau_{1-})-1\vert\right)+\vert1-\alpha_{a}[\Phi_{0}](\tau_{1-})\vert \\
&\leq&d_{\mathcal{E}}(\Phi,\Phi_{0})+6\cdot 2^{-29}\mathrm{\bf{M}} \\
&\overset{(\ref{eq:def_of_M})}\leq&2\mathrm{\bf{M}}
\end{eqnarray*}
\\
Require $(\ell_{1},...,\ell_{8})\geq(32,1,0,2,1,0,-2^{-7},1)$ $(\bullet)_{9}$. If it holds, then $\lambda\geq\lambda_{L}-(1+w)^{-1}\geq 1$.
\\
\\
%Construction of z'%
$Construction\:of\:z'$. Recall equation (\ref{eq:lemma3.9}). Set
\\
\begin{equation}
\label{eq:z'}
z'=-\frac{\rho^{2}[\Phi](\tau_{2-})}{\alpha_{a'}[\Phi](\tau_{2-})}
\end{equation}
\\
Then,
\\
\begin{eqnarray*}
\vert z'-z_{L}\vert&=&\bigg| -\frac{\rho^{2}[\Phi]}{\alpha_{a'}[\Phi](\tau_{2-})}-\frac{z}{1-\alpha_{a,a'}[\Phi_{0}](\tau_{1-})}\bigg| \\
&=&\bigg| \frac{\rho^{2}[\Phi](\tau_{2-})}{\lambda}-\frac{\rho^{2}[\Phi_{0}](\tau_{1-})}{\lambda_{L}}\bigg| \\
&=&\bigg|\frac{\rho^{2}[\Phi](\tau_{2-})\lambda_{L}-\rho^{2}[\Phi_{0}](\tau_{1-})\lambda}{\lambda\lambda_{L}}\bigg| \\
&\leq& \frac{1}{\vert\lambda\lambda_{L}\vert}\left(\rho^{2}[\Phi](\tau_{2-})\vert\lambda-\lambda_{L}\vert+\vert\lambda\vert\vert\rho^{2}[\Phi](\tau_{2-})-\rho^{2}[\Phi_{0}](\tau_{1-})\vert\right) \\
&\leq& \frac{1}{\vert\lambda\lambda_{L}\vert}\left(\rho^{2}[\Phi](\tau_{2-})\vert\lambda-\lambda_{L}\vert+\vert\lambda\vert\vert\rho^{2}[\Phi](\tau_{2-})-\rho^{2}[\Phi_{0}](\tau_{2-})\vert\right) \\
&\leq& \frac{1}{\vert\lambda\lambda_{L}\vert}2z\mathrm{\bf{M}}+\frac{1}{\vert\lambda_{L}\vert}\mathrm{\bf{M}} \\
&\leq& 2z\mathrm{\bf{M}}+\mathrm{\bf{M}} \\
&=&\mathrm{\bf{X}}(30,1,0,1,1,0,-2^{-7},2)+\mathrm{\bf{X}}(29,1,0,1,1,0,-2^{-7},1) \\
&\leq&\mathrm{\bf{X}}(31,1,0,1,1,0,-2^{-7},2)
\end{eqnarray*}
\\
\\
where $\lambda\geq\lambda_{L}-\frac{1}{1+w}\geq 1$ was used, which also implies $\vert\lambda\lambda_{L}\vert\neq 0$, $\frac{1}{\vert\lambda_{L}\vert}\leq\frac{1}{1+\frac{1}{1+w}}< 1$, and $\vert\lambda\lambda_{L}\vert\geq 1+\frac{1}{1+w}$. 
\\
Require $(\ell_{1},...,\ell_{8})\geq(31,1,0,1,1,0,-2^{-7},2)$ $(\bullet)_{10}$. Then, $\vert z'-z_{L}\vert\leq 1$.
\\
\\
%Construction of w'%
\emph{Construction of $w'$}. (stays the same as in \cite{RT}) Require $(\ell_{1},...,\ell_{8})\geq(32,1,0,2,1,0,-2^{-7},1)$ $(\bullet)_{11}$. Set
\\
\begin{equation}
\label{eq:w'}
w'=\frac{\alpha_{a}[\Phi](\tau_{2-})}{-\alpha_{a,a'}[\Phi](\tau_{2-})}>0
\end{equation}
\\
To check that $w'$ is well-defined, i.e. denominator is non-zero and $w'>0$, recall \emph{Estimates on $\Phi_{0}$} and (\ref{eq:def_of_M}). We have

\begin{eqnarray*}
\vert\alpha_{a,a'}[\Phi](\tau)-\alpha_{a,a'}[\Phi_{0}](\tau)\vert&\leq&2\mathrm{\bf{M}} \\ 
\vert\alpha_{a}[\Phi](\tau)-1\vert&=&\vert\alpha_{a}[\Phi](\tau)-\alpha_{a}[\Phi_{0}](\tau)+\alpha_{a}[\Phi_{0}](\tau)-1\vert\leq\vert\alpha_{a}[\Phi](\tau)-\alpha_{a}[\Phi_{0}](\tau)\vert+\vert\alpha_{a}[\Phi_{0}](\tau)-1\vert\leq 2\mathrm{\bf{M}}
\end{eqnarray*}

Further, 4$\mathrm{\bf{M}}\leq\mathrm{\bf{X}}(-1,0,0,-1,0,0,0,0)\leq\frac{1}{1+w}\overset{(\ref{ref_field_alpha_b}),(\ref{ref_field_alpha_c})}\leq\vert\alpha_{a,a'}[\Phi_{0}](\tau)\vert$. Recall $(\bullet)_{9}$, then $4\mathrm{\bf{M}}<1$. We estimate
\\
\begin{eqnarray*}
\vert\alpha_{a,a'}[\Phi](\tau)-\alpha_{a,a'}[\Phi_{0}](\tau)\vert&\leq&\frac{1}{2}\vert\alpha_{a,a'}[\Phi_{0}](\tau)\vert \\
\vert\alpha_{a}[\Phi](\tau)-1\vert&\leq&\frac{1}{2}
\end{eqnarray*}
\\
The last inequality implies $\frac{1}{2}\leq\alpha_{a}[\Phi](\tau)\leq\frac{3}{2}$ for all $\tau\in\mathcal{J}$, in particular, $\alpha_{a}[\Phi](\tau_{2-})>0$. From the second to last inequality we get $\alpha_{a,a'}[\Phi](\tau_{2-})\leq\frac{1}{2}\alpha_{a,a'}[\Phi_{0}](\tau_{2-})<0$. The last inequality follows from (\ref{ref_field_alpha_b}),(\ref{ref_field_alpha_c}). Therefore, $w'$ is well-defined.
\\
\\
Recall Lemma \ref{lemma:3.9}. We estimate
\\
\begin{eqnarray*}
\vert w'-w_{L}\vert&=&\bigg|\frac{\alpha_{a}[\Phi](\tau_{2-})}{-\alpha_{a,a'}[\Phi](\tau_{2-})}-\left(-\frac{1}{\alpha_{a,a'}[\Phi_{0}](\tau_{2-})}\right)\bigg| \\
&\leq&\bigg|\frac{\alpha_{a}[\Phi](\tau_{2-})}{-\alpha_{a,a'}[\Phi](\tau_{2-})}-\frac{\alpha_{a}[\Phi_{0}](\tau_{2-})}{-\alpha_{a,a'}[\Phi](\tau_{2-})}\bigg| \\
&+&\bigg|\frac{\alpha_{a}[\Phi_{0}](\tau_{2-})}{-\alpha_{a,a'}[\Phi](\tau_{2-})}-\frac{\alpha_{a}[\Phi_{0}](\tau_{2-})}{-\alpha_{a,a'}[\Phi_{0}](\tau_{2-})}\bigg|+\bigg|\frac{\alpha_{a}[\Phi_{0}](\tau_{2-})}{-\alpha_{a,a'}[\Phi_{0}](\tau_{2-})}-\frac{1}{\alpha_{a,a'}[\Phi_{0}](\tau_{2-})}\bigg| \\
&\leq&\frac{1}{\vert\alpha_{a,a'}[\Phi](\tau_{2-})\vert}\vert\alpha_{a}[\Phi](\tau_{2-})-\alpha_{a}[\Phi_{0}](\tau_{2-})\vert+\bigg|\frac{\alpha_{a}[\Phi_{0}](\tau_{2-})}{-\alpha_{a,a'}[\Phi](\tau_{2-})}-\frac{\alpha_{a}[\Phi_{0}](\tau_{2-})}{-\alpha_{a,a'}[\Phi_{0}](\tau_{2-})}\bigg|+ \\
&+&\bigg|\frac{\alpha_{a}[\Phi_{0}](\tau_{2-})}{-\alpha_{a,a'}[\Phi_{0}](\tau_{2-})}-\frac{1}{\alpha_{a,a'}[\Phi_{0}](\tau_{2-})}\bigg| \\
&\leq&2w_{L}\mathrm{\bf{M}}+4w^{2}_{L}\mathrm{\bf{M}}+w_{L}\mathrm{\bf{M}} \\
&\leq&2^{3}(1+w)^{2}\mathrm{\bf{M}} \\
&\leq&\frac{1}{2}\mathrm{\bf{X}}(6,0,0,2,0,0,0,0)\mathrm{\bf{M}}
\end{eqnarray*}
\\
Require $(\ell_{1},...,\ell_{8})\geq(35,1,0,3,1,0,-2^{-7},1)$ $(\bullet)_{12}$. Therefore, $\vert w'-w_{L}\vert\leq\frac{1}{2}\mathrm{\bf{K}}\leq\frac{1}{2}$.
\\
\\
\emph{Construction of $\mathrm{\bf{h}}'$}. Stays the same as in \cite{RT}. Inequalities $(\bullet)_{13},(\bullet)_{14},(\bullet)_{15}$ will be added. Main result:

\begin{equation}
\label{eq:h'}
\mathrm{\bf{h}'}=\frac{\mathrm{\bf{h}}}{\mu}>0
\end{equation}

where $\mu=\frac{1+2w'}{1+w'}(-\xi_{a}[\Phi,\mathrm{\bf{h}}](\tau_{2-})+\mathrm{\bf{h}}\mathrm{log}\lambda)-\mathrm{\bf{h}}\mathrm{log}2$. Further, $\vert \mathrm{\bf{h}'}-\mathrm{\bf{h}}_{L}\vert\leq\frac{1}{2}$.
\\
\\
\emph{Construction of $q'$}. Stays the same as in \cite{RT}. Inequality $(\bullet)_{16}$ will be added. Main result:

\begin{equation}
\label{eq:q'}
q'=\frac{1}{1+w'}\left(\mathrm{\bf{h}'}\mathrm{log}\lambda-\frac{\mathrm{\bf{h}'}}{\mathrm{\bf{h}}}\xi_{c'}[\Phi,\mathrm{\bf{h}}](\tau_{2-})-\frac{w'(1+w')}{1+2w'}-\frac{1+3w'+(w')^{2}}{1+2w'}\mathrm{\bf{h}'}\mathrm{log}2\right)
\end{equation}
\\
\\
\emph{The maximum of $\alpha_{b,c},\alpha_{c,a},\alpha_{a,b}$}. Recall \emph{Estimates on $\Phi$}. We had $d_{\mathcal{E}}(\Phi,\Phi_{0})\leq\mathrm{\bf{X}}(20,1,0,1,0,0,-2^{-7},1)$. Recall Definition \ref{def:3.6}. We have
\\
\begin{eqnarray*}
\alpha_{a,p}[\Phi]&\leq&\alpha_{a,p}[\Phi_{0}]+\alpha_{a}[\Phi]+\alpha_{p}[\Phi]-\alpha_{a}[\Phi_{0}]-\alpha_{p}[\Phi_{0}] \\
&\overset{(\ref{ref_field_alpha_b}),(\ref{ref_field_alpha_c})}\leq&-\frac{1}{1+w}+d_{\mathcal{D}}(\Phi_{0},\Phi) \\
&\leq&-\frac{1}{1+w}+\mathrm{\bf{X}}(20,1,0,1,0,0,-2^{-7},1)
\end{eqnarray*}
\\
and
\\
\begin{eqnarray*}
\alpha_{b,c}[\Phi]&\leq&\alpha_{a,b}[\Phi]+\alpha_{a,c}[\Phi]-2\alpha_{a}[\Phi] \\
&\overset{Definition \ref{def:3.3}}\leq&\alpha_{a,b}[\Phi]+\alpha_{a,c}[\Phi]+2A_{a}[\Phi] \\
&\leq&2^{2}d_{\mathcal{D}}(\Phi_{0},\Phi)+\alpha_{a,b}[\Phi_{0}]+\alpha_{a,c}[\Phi_{0}]+2A_{a}[\Phi_{0}] \\
&\leq&2^{2}\mathrm{\bf{X}}(20,1,0,1,0,0,-2^{-7},1)+\alpha_{a,b}[\Phi_{0}]+\alpha_{a,c}[\Phi_{0}]+2A_{a}[\Phi_{0}]
\end{eqnarray*}
\\
Now, require $(\ell_{1},...,\ell_{8})\geq(24,1,2,2,0,0,-2^{-7},1)$ $(\bullet)_{17}$. Then $\mathrm{\bf{X}}(20,1,0,1,0,0,-2^{-7},1)\leq 2^{-3}(1+w)^{-1}\mathrm{min}\lbrace w^{2},1\rbrace$ and $d_{\mathcal{D}}(\Phi_{0}(\tau),\Phi(\tau))\leq2^{-3}(1+w)^{-1}\mathrm{min}\lbrace w^{2},1\rbrace$ for all $\tau\in [\tau_{0-},\tau_{0+}]$. This implies that
\\
\begin{equation*}
\alpha_{a,p}[\Phi]\leq -\frac{1}{1+w}+2^{-3}\frac{1}{1+w}\mathrm{min}\lbrace w^{2},1\rbrace\leq -2^{-1}(1+w)^{-1}
\end{equation*}
\\
and
\\
\begin{equation*}
\alpha_{b,c}[\Phi]\overset{(\ref{ref_field_A}),(\ref{ref_field_alpha_b}),(\ref{ref_field_alpha_c})}\leq -2^{-1}(1+w)^{-1}w^{2}
\end{equation*}
\\
for all $\tau\in[\tau_{0-},\tau_{0+}]$ and $p=\lbrace b,c\rbrace$.
\\
\\
\emph{Definitions of maps $\Pi,\Lambda$ and $\tau_{2-}$}. Set $(\ell_{1},...,\ell_{8})=(40,2,2,3,2,1,-2^{-7},2)$ and $\ell_{9}=7$. With this choice, all inequalities $(\bullet)$ hold. The constant $\mathrm{\bf{K}}$ defined by (\ref{eq:f_assumptions}) coincides with $\mathrm{\bf{K}}(\mathrm{\bf{f}})$ defined in (\ref{def_of_K}). Vector $\mathrm{\bf{f}}=(\mathrm{\bf{h}},w,q,z)\in(0,\infty)^{4}$ satisfies (\ref{eq:f_assumptions}) if and only if $\mathrm{\bf{f}}\in\mathcal{F}$. Therefore, we can set
\\
\begin{align}
\Pi[\pi,\sigma_{*}]&:& &\mathcal{F}& &\rightarrow& &(0,\infty)^{2}\times \mathbb{R}\times (0,\infty) & \:\:&\mathrm{\bf{f}}& &\mapsto& &\mathrm{right\:hand\:side\:of (\ref{eq:h'}),(\ref{eq:w'}),(\ref{eq:q'}),(\ref{eq:z'})}& \nonumber \\
\Lambda[\pi,\sigma_{*}]&:& &\mathcal{F}& &\rightarrow& &[1,\infty)& \:\:&\mathrm{\bf{f}}& &\mapsto& &\mathrm{right\:hand\:side\:of (\ref{eq:lambda})}& \nonumber \\
\tau_{2-}[\pi,\sigma_{*}]&:& &\mathcal{F}& &\rightarrow& &(\infty,0)& \:\:&\mathrm{\bf{f}}& &\mapsto& &\mathrm{right\:hand\:side\:of (\ref{eq:def_tau2-})}& \nonumber 
\end{align}
\\
Equation $\Phi(\tau_{2-})=\lambda\Phi_{*}(\pi',\mathrm{\bf{f}}',\sigma_{*})$ with $\mathrm{\bf{f}}'=(\mathrm{\bf{h}}',w',q',z')$ is equivalent to (recall $b'=a$)
\\
\begin{eqnarray*}
\label{eq:scaling_1}
\alpha_{a'}[\Phi](\tau_{2-})&=&-\lambda \\
\label{eq:scaling_2}
\alpha_{a}[\Phi](\tau_{2-})&=&\lambda\frac{w'}{1+w'} \\
\label{eq:scaling_3}
\alpha_{c'}[\Phi](\tau_{2-})&=&\lambda(-w'-\mu'-4(1+w')\gamma^{2}) \\
\label{eq:scaling_4}
\frac{1}{\mathrm{\bf{h}}}\xi_{a'}[\Phi,\mathrm{\bf{h}}](\tau_{2-})&=&\mathrm{log}\lambda+\frac{1}{\mathrm{\bf{h}}'}\left(-\frac{1+w'}{1+2w'}(1+\mathrm{\bf{h}}'\mathrm{log}2)\right) \\
\label{eq:scaling_5}
\frac{1}{\mathrm{\bf{h}}}\xi_{a}[\Phi,\mathrm{\bf{h}}](\tau_{2-})&=&\mathrm{log}\lambda+\frac{1}{\mathrm{\bf{h}}'}\left(-\frac{1+w'}{1+2w'}(1+\mathrm{\bf{h}}'\mathrm{log}2)\right) \\
\label{eq:scaling_6}
\frac{1}{\mathrm{\bf{h}}}\xi_{c'}[\Phi,\mathrm{\bf{h}}](\tau_{2-})&=&\mathrm{log}\lambda+\frac{1}{\mathrm{\bf{h}}'}\left(-(1+w')q'-\frac{w'(1+w')}{1+2w'}-\frac{1+3w'+(w')^{2}}{1+2w'}\mathrm{\bf{h}}'\mathrm{log}2\right) \\
\label{eq:scaling_7}
\rho^{2}[\Phi](\tau_{2-})&=&\lambda z'
\end{eqnarray*}
\\
with $\mu'=(1+w')(\beta_{1}^{2}+\beta_{2}^{2}+\beta_{3}^{2}-2\beta_{2}\beta_{3}-2\beta_{3}\beta_{1}-2\beta_{1}\beta_{2})\big|_{\beta=\beta[\Phi_{*}(\pi',\mathrm{\bf{f}}',\sigma_{*})]}$ and $\gamma^{2}=\rho^{2}\vert\beta_{1}\beta_{2}\beta_{3}\vert^{1/3}\big|_{\beta=\beta[\Phi_{*}(\pi',\mathrm{\bf{f}}',\sigma_{*})]}$
\\
\\
\emph{Continuity of maps $\Pi,\Lambda$ and $\tau_{2-}$}. Fix $\mathrm{\bf{f}}^{\Psi}=(\mathrm{\bf{h}}^{\Psi},w^{\Psi},q^{\Psi},z^{\Psi})\in\mathcal{F}$. Let $r>0$ and let $\mathrm{\bf{f}}^{\Upsilon}=(\mathrm{\bf{h}}^{\Upsilon},w^{\Upsilon},q^{\Upsilon},z^{\Upsilon})\in\mathcal{F}$ with $\vert\vert\mathrm{\bf{f}}^{\Psi}-\mathrm{\bf{f}}^{\Upsilon}\vert\vert_{\mathbb{R}}\leq r$. Introduce notation $\mathrm{\bf{f}}^{B}\in\mathcal{F}$ with $B=\Psi,\Upsilon$. Following this convention, the contraction mapping fixed points are denoted $\Phi^{B}\in\mathcal{E}^{B}$. We also write $\Phi^{\Psi}=\Psi$ and $\Phi^{\Upsilon}=\Upsilon$.
\\
Suppose $r\leq\frac{1}{2}\vert q^{\Psi}-1\vert$. Then $0\neq\mathrm{sgn}(q^{\Psi}-1)=\mathrm{sgn}(q^{\Upsilon}-1)$.
\\
\\
Now, if we want to compare the solutions $\Psi$ and $\Upsilon$, we have to apply the symmetry transformation from Proposition \ref{prop:3.1} to make sure they solve the same equation. That is, we introduce a new field of the form "$\Xi=\Upsilon\circ\chi$". Recall that in the Proposition \ref{prop:3.1} we had the general form $\chi(\tau)=p\tau+q$. Here, introduce $\chi:\mathbb{R}\rightarrow\mathbb{R}$ by
\begin{equation}
\label{def_of_chi}
\chi(\tau)=\frac{\mathrm{\bf{h}}^{\Upsilon}}{\mathrm{\bf{h}}^{\Psi}}(\tau-\tau_{0+}^{\Psi})+\tau_{0+}^{\Upsilon}
\end{equation}
The motivation for such form of $\chi$ is as follows. The fixed point construction was carried out on the interval $[\tau_{0-},\tau_{0+}]$. Here, introduce $\mathcal{I}^{B}=[\tau_{0-}^{B},\tau_{0+}^{B}]$ with $B=\Psi,\Upsilon$, and $\mathcal{I}^{\Xi}=[\chi^{-1}(\tau_{0-}^{\Upsilon}),\tau_{0+}^{\Psi}]$. With these definitions, we have $\chi(\mathcal{I}^{\Xi})=\mathcal{I}^{\Upsilon}$, and by Proposition \ref{prop:3.1} the field $\Xi=\Upsilon\circ(\chi\big|_{\mathcal{I}^{\Xi}})$ satisfies $(\frak a,\frak b,c,\frak d)[\Xi,\mathrm{\bf{h}}^{\Psi},Z]=0$ on $\mathcal{I}^{\Xi}$.
\\
\\
Define $\mathcal{I}=\mathcal{I}^{\Psi}\cap\mathcal{I}^{\Xi}$. Recall the definition of $\mathcal{J}$ and set $\mathcal{J}^{B}=[\tau_{0-}^{B},\frac{1}{2}\tau_{1-}^{B}]\subset\mathcal{I}^{B}$. Further, $\vert\tau_{2-}^{B}-\tau_{1-}^{B}\vert\leq\frac{1}{2}\mathrm{dist}_{\mathbb{R}}(\tau_{1-}^{B},\mathbb{R}\setminus\mathcal{J}^{B})$.
\\
\\
Define $\mathcal{J}=\mathcal{J}^{\Psi}\cap\mathcal{J}^{\Xi}\subset\mathcal{I}$, with $\mathcal{J}^{\Xi}=\chi^{-1}(\mathcal{J}^{\Upsilon})$. For $r>0$ sufficiently small, we want to show
\\
\begin{itemize}
\item $\tau_{2-}^{\Psi}\in\mathcal{J}$
\item $\chi^{-1}(\tau_{2-}^{\Upsilon})\in\mathcal{J}$\\
\end{itemize}
To prove the first one, we need to show that $\tau_{2-}^{\Psi}\in\mathcal{J}^{\Psi}$ and $\tau_{2-}^{\Psi}\in\mathcal{J}^{\Xi}$. Clearly, $\tau_{2-}^{\Psi}\in\mathcal{J}^{\Psi}$. The inclusion $\tau_{2-}^{\Psi}\in\mathcal{J}^{\Xi}$ is equivalent to $\chi(\tau_{2-}^{\Psi})\in\mathcal{J}^{\Upsilon}$ by definition of $\chi(\tau)$. So, the goal is to show that $\vert\chi(\tau_{2-}^{\Psi})-\tau_{1-}^{\Upsilon}\vert\leq\frac{1}{2}\mathrm{dist}_{\mathbb{R}}(\tau_{1-}^{\Upsilon},\mathbb{R}\setminus\mathcal{J}^{\Upsilon})$. We have
\\
\begin{eqnarray*}
\vert\chi(\tau_{2-}^{\Psi})-\tau_{1-}^{\Upsilon}\vert&\leq&\vert\chi(\tau_{2-}^{\Psi})-\chi(\tau_{1-}^{\Psi})\vert+\vert\chi(\tau_{1-}^{\Psi})-\tau_{1-}^{\Upsilon}\vert \\
&\leq&\frac{\mathrm{\bf{h}}^{\Upsilon}}{\mathrm{\bf{h}}^{\Psi}}\frac{1}{2}\mathrm{dist}_{\mathbb{R}}(\tau_{1-}^{\Psi},\mathbb{R}\setminus\mathcal{J}^{\Psi})+\vert\chi(\tau_{1-}^{\Psi})-\tau_{1-}^{\Upsilon}\vert \\
&\overset{\mathrm{for}\:\mathrm{\bf{f}}^{\Upsilon}=\mathrm{\bf{f}}^{\Psi}}=&\frac{1}{2}\mathrm{dist}_{\mathbb{R}}(\tau_{1-}^{\Upsilon},\mathbb{R}\setminus\mathcal{J}^{\Upsilon}) 
\end{eqnarray*}
\\
Note that the RHS of the second inequality is a continuous function of $\mathrm{\bf{f}}^{\Upsilon}$.
\\
\\
Introduce abbreviations $\mathcal{D}=\mathcal{D}^{\Psi}=\mathcal{D}^{\Upsilon}=\mathcal{D}(\sigma_{*})$, $\mathcal{E}=\mathcal{E}(\sigma_{*};\mathcal{I})$, $\Phi_{0}=\Phi_{0}(\pi,\mathrm{\bf{f}}^{\Psi},\sigma_{*})\big|_{\mathcal{I}}=\Phi_{0}^{\Psi}\big|_{\mathcal{I}}$, $d_{\mathcal{X}}=d_{\mathcal{X},(\pi,\mathrm{\bf{h}}^{\Psi})}$ for $\mathcal{X}=\mathcal{E},\mathcal{D}$, $d_{\mathcal{X}^{B}}=d_{\mathcal{X}^{B},(\pi,\mathrm{\bf{h}}^{B})}$ for $B=\Psi,\Upsilon$.
\\
\\
Recall (\ref{fixed_point}). We have $d_{\mathcal{E}^{B}}(B,\Phi_{0}^{B})\leq\frac{1}{2}\delta^{B}$. If $r>0$ is sufficiently small then we have
\\
\begin{equation*}
d_{\mathcal{E}}(\Psi\big|_{\mathcal{I}},\Phi_{0})\leq d_{\mathcal{E}^{\Psi}}(\Psi,\Phi_{0}^{\Psi})\leq\frac{1}{2}\delta^{\Psi}
\end{equation*}
\\
and
\\
\begin{eqnarray*}
d_{\mathcal{E}}(\Xi\big|_{\mathcal{I}},\Phi_{0})&\overset{\Xi=\Upsilon\circ\chi|_{\mathcal{I}}}\leq&d_{\mathcal{E}}(\Upsilon\circ\chi|_{\mathcal{I}},\Phi_{0}^{\Upsilon}\circ\chi|_{\mathcal{I}})+d_{\mathcal{E}}(\Phi_{0}^{\Upsilon}\circ\chi|_{\mathcal{I}},\Phi_{0}) \\
&\overset{(\ref{def_of_chi})}\leq&\mathrm{max}\lbrace1,\frac{\mathrm{\bf{h}}^{\Psi}}{\mathrm{\bf{h}}^{\Upsilon}}\rbrace d_{\mathcal{E}^{\Upsilon}}(\Upsilon,\Phi_{0}^{\Upsilon})+d_{\mathcal{E}}(\Phi_{0}^{\Upsilon}\circ\chi|_{\mathcal{I}},\Phi_{0}) \\
&\leq&\mathrm{max}\lbrace1,\frac{\mathrm{\bf{h}}^{\Psi}}{\mathrm{\bf{h}}^{\Upsilon}}\rbrace \frac{1}{2}\delta^{\Upsilon}+d_{\mathcal{E}}(\Phi_{0}^{\Upsilon}\circ\chi|_{\mathcal{I}},\Phi_{0})
\end{eqnarray*}
\\
Note that the RHS of the last inequality is a continuous function of $\mathrm{\bf{f}}^{\Upsilon}$ and is equal to $\frac{1}{2}\delta^{\Psi}$ when $\mathrm{\bf{f}}^{\Upsilon}=\mathrm{\bf{f}}^{\Psi}$. Therefore, we have $d_{\mathcal{E}}(\Psi\big|_{\mathcal{I}},\Phi_{0})\leq\delta^{\Psi}$ and $d_{\mathcal{E}}(\Xi\big|_{\mathcal{I}},\Phi_{0})\leq\delta^{\Psi}$.
\\
\\
Now, recall equations (\ref{def_of_P})-(\ref{def_of_P_end}). Recall Definition \ref{def:3.15}. Both $X=\Psi\big|_{\mathcal{I}}$ and $X=\Xi\big|_{\mathcal{I}}$ satisfy $(\frak a,\frak b, c, \frak d)[X,\mathrm{\bf{h}}^{\Psi},Z]=0$ on $\mathcal{I}$. For all $p\in\lbrace b,c\rbrace$ and $\tau\in\mathcal{I}$, we have
\\
\begin{eqnarray*}
\label{def_of_P_cont}
A_{a}[X](\tau)&=&A_{a}[X](\tau_{0+}^{\Psi})+\int_{\tau_{0+}^{\Psi}}^{\tau} \mathrm{d}\tau'\mathrm{\bf{I}}_{1}\left[\Psi,\mathrm{\bf{h}},\pi\right](\tau') \\
\label{def_of_P_1_cont}
\theta_{a}[X](\tau)&=&\theta_{a}[X,\mathrm{\bf{h}}^{\Psi}](\tau_{0+}^{\Psi})+\int_{\tau_{0+}^{\Psi}}^{\tau} \mathrm{d}\tau'\mathrm{\bf{I}}_{2}\left[\Psi,\mathrm{\bf{h}},\pi\right](\tau') 
\\
\label{def_of_P_2_cont}
\alpha_{p,a}[X](\tau)&=&\alpha_{p,a}[X](\tau_{0+}^{\Psi})+\int_{\tau_{0+}^{\Psi}}^{\tau}\mathrm{d}\tau'\mathrm{\bf{I}}_{(3,p)}\left[\Psi,\mathrm{\bf{h}},\pi\right](\tau') \\
\label{def_of_P_3_cont}
\xi_{p,a}[X](\tau)&=&\xi_{p,a}[X,\mathrm{\bf{h}}^{\Psi}](\tau_{0+}^{\Psi})+\alpha_{p,a}[X](\tau_{0+}^{\Psi})(\tau-\tau_{0+}^{\Psi})+\int_{\tau_{0+}^{\Psi}}^{\tau}\mathrm{d}\tau'' \int_{\tau_{0+}}^{\tau''}\mathrm{d}\tau'\mathrm{\bf{I}}_{(3,p)}\left[\Psi,\mathrm{\bf{h}},\pi\right](\tau') \\
\label{def_of_P_end_cont}
\rho[X](\tau)&=&\rho[X](\tau_{0+}^{\Psi})
\end{eqnarray*}
\\
Recall (\ref{abs_I_diff_estimate}) and $\mathrm{sup}_{\tau\in\mathcal{I}}\vert\tau-\tau_{0+}^{\Psi}\vert\leq 4$. Then we have $d_{\mathcal{E}}(\Psi\big|_{\mathcal{I}},\Xi\big|_{\mathcal{I}})\leq 2^{3}d_{\mathcal{D}}(\Psi(\tau_{0+}^{\Psi}),\Xi(\tau_{0+}^{\Psi}))+2^{-1}d_{\mathcal{E}}(\Psi\big|_{\mathcal{I}},\Xi\big|_{\mathcal{I}})$. Therefore,
\\
\begin{equation}
\label{dist_cont}
d_{\mathcal{E}}(\Psi\big|_{\mathcal{I}},\Xi\big|_{\mathcal{I}})\leq2^{4}d_{\mathcal{D}}(\Psi(\tau_{0+}^{\Psi}),\Xi(\tau_{0+}^{\Psi}))=2^{4}d_{\mathcal{D}}(\Phi_{*}(\pi,\mathrm{\bf{f}}^{\Psi},\sigma_{*}),\Phi_{*}(\pi,\mathrm{\bf{f}}^{\Upsilon},\sigma_{*}))\xrightarrow{\mathrm{\bf{f}}^{\Upsilon}\rightarrow\mathrm{\bf{f}}^{\Psi}} 0
\end{equation}
\\
Further, we have
\\
\begin{eqnarray*}
d_{\mathcal{D}}(\lambda^{\Psi}\Phi_{*}(\pi',\mathrm{\bf{f}}^{\Psi},\sigma_{*}),\lambda^{\Upsilon}\Phi_{*}(\pi',\mathrm{\bf{f}}^{\Upsilon},\sigma_{*}))&=&d_{\mathcal{D}}(\Psi(\tau_{2-}^{\Psi}),\Xi(\chi^{-1}(\tau_{2-}^{\Upsilon}))) \\
&\leq&d_{\mathcal{D}}(\Psi(\tau_{2-}^{\Psi}),\Psi(\chi^{-1}(\tau_{2-}^{\Upsilon})))+d_{\mathcal{D}}(\Psi(\chi^{-1}(\tau_{2-}^{\Upsilon})),\Xi(\chi^{-1}(\tau_{2-}^{\Upsilon}))) \\
&\overset{(\ref{dist_cont})}\leq&d_{\mathcal{D}}(\Psi(\tau_{2-}^{\Psi}),\Psi(\chi^{-1}(\tau_{2-}^{\Upsilon})))+2^{4}d_{\mathcal{D}}(\Phi_{*}(\pi,\mathrm{\bf{f}}^{\Psi},\sigma_{*}),\Phi_{*}(\pi,\mathrm{\bf{f}}^{\Upsilon},\sigma_{*}))
\end{eqnarray*}
\\
Now, to prove that $\Pi,\Lambda$ and $\tau_{2-}$ are continuous, it suffices to show that $\chi^{-1}(\tau_{2-}^{\Upsilon})\rightarrow\tau_{2-}^{\Psi}$ as $\mathrm{\bf{f}}^{\Upsilon}\rightarrow\mathrm{\bf{f}}^{\Psi}$. 
\\
\\
Recall that $\alpha_{a}[\Psi](\tau)\geq\frac{1}{2}$ and $\alpha_{a,a'}[\Psi](\tau)\leq 0$ for all $\tau\in\mathcal{J}\subset\mathcal{J}^{\Psi}$. Therefore, $\forall\tau\in\mathcal{J}$ we have
\\
\begin{equation*}
\frac{d}{d\tau}(\xi_{a}[\Psi,\mathrm{\bf{h}}^{\Psi}]-\xi_{a'}[\Psi,\mathrm{\bf{h}}^{\Psi}])\overset{\mathrm{Definition \ref{def:3.3}}}=\alpha_{a}[\Psi]-\alpha_{a'}[\Psi]=2\alpha_{a}[\Psi]-\alpha_{a,a'}[\Psi]\geq 1
\end{equation*}
\\
and it follows that $\vert\xi_{a}[\Psi,\mathrm{\bf{h}}^{\Psi}](\tau)-\xi_{a'}[\Psi,\mathrm{\bf{h}}^{\Psi}](\tau)\vert\geq\vert\tau-\tau_{2-}^{\Psi}\vert$, $\tau\in\mathcal{J}$. Set $\tau=\chi^{-1}(\tau_{2-}^{\Upsilon})\in\mathcal{J}$. Recall Definition \ref{def:3.6}. Then
\\
\begin{eqnarray*}
\vert\chi^{-1}(\tau_{2-}^{\Upsilon})-\tau_{2-}^{\Psi}\vert&\leq&\vert\xi_{a}[\Psi,\mathrm{\bf{h}}^{\Psi}](\chi^{-1}(\tau_{2-}^{\Upsilon}))-\xi_{a'}[\Psi,\mathrm{\bf{h}}^{\Psi}](\chi^{-1}(\tau_{2-}^{\Upsilon}))\vert \\
&\leq&2\not{d}_{\mathcal{D},\mathrm{\bf{h}}^{\Psi}}(\Psi(\chi^{-1}(\tau_{2-}^{\Upsilon})),\Xi(\chi^{-1}(\tau_{2-}^{\Upsilon})))
\end{eqnarray*}
\\
Since $\mathrm{\bf{f}}^{\Upsilon}\rightarrow\mathrm{\bf{f}}^{\Psi}$ implies $d_{\mathcal{E}}(\Psi\big|_{\mathcal{I}},\Xi\big|_{\mathcal{I}})\rightarrow 0$, we have $\vert\chi^{-1}(\tau_{2-}^{\Upsilon})-\tau_{2-}^{\Psi}\vert\rightarrow 0$ as required.
\\
\\
\emph{Uniqueness of $\Pi,\Lambda$ and $\tau_{2-}$}. The argument stays exactly the same as for the vacuum case in \cite{RT}.
%SECTION 4%
\newpage
\section{Era-to-era and epoch-to-epoch maps}
\begin{definition}
\label{def:4.1}
($\mathrm{\bf{Epoch-to-epoch map}}$) Set $\mathcal{Q}_{R}:(0,\infty)\setminus\mathbb{Q}\rightarrow (0,\infty)\setminus\mathbb{Q}$
\\
\[w\mapsto\mathcal{Q}_{R}(w)=
\begin{cases}
\frac{1}{w}-1 \:\:if\:w<1 \\
w-1 \:\:if\:w>1
\end{cases}\]
For every $w\in(0,\infty)\setminus\mathbb{Q}$, set
\begin{equation}
\label{eq:EpochToEpoch}
\mathcal{\bf{Q}}_{R} \lbrace w\rbrace (q,\mathrm{\bf{h}},z)=\left(\frac{\mathrm{num1}}{\mathrm{den}}, \frac{\mathrm{num2}}{\mathrm{den}},\mathrm{num3}\right)
\end{equation}
where, if $w<1$
\\
\begin{eqnarray*}
\mathrm{num1}&=&1+w+\mathrm{\bf{h}}\mathrm{log}2-\mathrm{\bf{h}}(1+2w)\mathrm{log}\left(1+\frac{1}{w}\right) \\
\mathrm{num2}&=&\mathrm{\bf{h}} \\
\mathrm{den}&=&(1+w)(1+q+\mathrm{\bf{h}}\mathrm{log}2)-\mathrm{\bf{h}}(2+w)\mathrm{log}\left(1+\frac{1}{w}\right) \\
\mathrm{num3}&=&\left(1+\frac{1}{w}\right)z
\end{eqnarray*}
\\
and, if $w>1$,
\\
\begin{eqnarray*}
\mathrm{num1}&=&(1+w)(1+q+\mathrm{\bf{h}}\mathrm{log}2)-\mathrm{\bf{h}}(2+w)\mathrm{log}\left(1+\frac{1}{w}\right) \\
\mathrm{num2}&=&\mathrm{\bf{h}}w \\
\mathrm{den}&=&1+w+\mathrm{\bf{h}}\mathrm{log}2-\mathrm{\bf{h}}(1+2w)\mathrm{log}\left(1+\frac{1}{w}\right) \\
\mathrm{num3}&=&\left(1+\frac{1}{w}\right)z
\end{eqnarray*}
\\
Further, $\forall w\in(0,\infty)\setminus\mathbb{Q}$ set
\\
\begin{eqnarray*}
\mathcal{Q}^{n}_{R}&=&(\mathcal{Q}_{R}\circ ...\circ\mathcal{Q}_{R})(w) \\
\mathcal{\bf{Q}}^{n}_{R}\lbrace w\rbrace&=&\mathcal{\bf{Q}}_{R}\lbrace\mathcal{Q}^{n-1}_{R}(w)\rbrace\circ ...\circ\mathcal{\bf{Q}}_{R}\lbrace\mathcal{Q}^{2}_{R}(w)\rbrace\circ\mathcal{\bf{Q}}_{R}\lbrace\mathcal{Q}_{R}(w)\rbrace\circ\mathcal{\bf{Q}}_{R}\lbrace w\rbrace
\end{eqnarray*}
\end{definition}

\begin{definition} 
\label{def:4.2}
The floor function is $\mathbb{R}\ni x\mapsto\lfloor x\rfloor=\mathrm{max}\lbrace n\in\mathbb{Z}\vert n\leq x \rbrace$.
\end{definition}

\begin{definition}
\label{def:4.3}
($\textbf{Era-to-era map}$). Define $\mathcal{E}_{R}:(0,1)\setminus\mathbb{Q}\rightarrow (0,1)\setminus\mathbb{Q}$ by $\mathcal{E}_{R}(w)=\mathcal{Q}^{\lfloor 1/w\rfloor}_{R}(w)$. For every $w\in (0,1)\setminus\mathbb{Q}$, denote by $\mathcal{\bf{E}}_{R}\lbrace w\rbrace$ the pair of rational functions over $\mathbb{R}$ given by $\mathcal{\bf{E}}_{R}\lbrace w\rbrace=\mathcal{\bf{Q}}^{\lfloor 1/w \rfloor}_{R\lbrace w\rbrace}$. Finally, for all $w\in (0,1)\setminus\mathbb{Q}$ and all integers $n\geq 0$, set
\\
\begin{eqnarray*}
\mathcal{E}^{n}(w)&=&(\mathcal{E}_{R}\circ ...\circ\mathcal{E}_{R})(w) \\
\mathcal{\bf{E}}^{n}_{R}\lbrace w\rbrace&=&\mathcal{\bf{E}}_{R}\lbrace\mathcal{E}^{n-1}_{R}(w)\rbrace\circ ...\circ\mathcal{\bf{E}}_{R}\lbrace\mathcal{E}^{2}_{R}(w)\rbrace\circ\mathcal{\bf{E}}_{R}\lbrace\mathcal{E}_{R}(w)\rbrace\circ\mathcal{\bf{E}}_{R}\lbrace w\rbrace
\end{eqnarray*}
\end{definition}
%LEMMA 4.1%
\begin{lemma}
\label{lemma:4.1}
For all integers $m,n\geq 0$,
\\
\begin{itemize}
\item $\mathcal{\bf{Q}}^{m+n}_{R}\lbrace w\rbrace=\mathcal{\bf{Q}}^{m}_{R}\lbrace\mathcal{Q}^{n}_{R}(w)\rbrace\circ\mathcal{\bf{Q}}^{n}_{R}\lbrace w\rbrace$ for $w\in (0,\infty)\setminus\mathbb{Q}$
\item $\mathcal{\bf{E}}^{m+n}_{R}\lbrace w\rbrace=\mathcal{\bf{E}}^{m}_{R}\lbrace\mathcal{E}^{n}_{R}(w)\rbrace\circ\mathcal{\bf{E}}^{n}_{R}\lbrace w\rbrace$ for $w\in (0,1)\setminus\mathbb{Q}$ \\
\end{itemize}
\end{lemma}
\underline{Proof}: Consider the equation for $\mathcal{\bf{Q}}^{m+n}_{R}\lbrace w\rbrace$. We have
\\
\begin{eqnarray*}
\mathrm{LHS}&=&\mathcal{\bf{Q}}^{m+n}_{R}\lbrace w \rbrace \\
&=& \mathcal{\bf{Q}}_{R} \lbrace \mathcal{Q}^{m+n-1}_{R}(w) \rbrace \circ ... \circ\mathcal{\bf{Q}}_{R}\lbrace \mathcal{Q}_{R}(w)\rbrace\circ\mathcal{\bf{Q}}_{R}\lbrace w \rbrace 
\end{eqnarray*}
\\
and
\\
\begin{eqnarray*}
\mathrm{RHS}&=&\mathcal{\bf{Q}}^{m}_{R}\lbrace\mathcal{Q}^{n}_{R}(w)\rbrace\circ\mathcal{\bf{Q}}^{n}_{R}\lbrace w\rbrace \\
&=&\mathcal{\bf{Q}}^{m}_{R}\lbrace\mathcal{Q}^{n}_{R}(w)\rbrace\circ(\mathcal{\bf{Q}}_{R}\lbrace\mathcal{Q}^{n-1}_{R}(w)\rbrace\circ ...\circ\mathcal{\bf{Q}}_{R}\lbrace\mathcal{Q}^{2}_{R}(w)\rbrace\circ\mathcal{\bf{Q}}_{R}\lbrace\mathcal{Q}_{R}(w)\rbrace\circ\mathcal{\bf{Q}}_{R}\lbrace w\rbrace) \\
&=&\mathcal{\bf{Q}}_{R}\lbrace\mathcal{Q}^{m-1}_{R}(\mathcal{Q}^{n}_{R}(w))\rbrace\circ ...\circ\mathcal{\bf{Q}}_{R}\lbrace\mathcal{Q}_{R}(\mathcal{Q}^{n}_{R}(w))\rbrace\circ\mathcal{\bf{Q}}_{R}\lbrace \mathcal{Q}^{n}_{R}(w)\rbrace\circ\mathcal{\bf{Q}}_{R}\lbrace \mathcal{Q}^{n-1}_{R}(w)\rbrace\circ ... \circ\mathcal{\bf{Q}}_{R}\lbrace w \rbrace \\
&=&\mathcal{\bf{Q}}_{R}\lbrace\mathcal{Q}^{m+n-1}_{R}(w)\rbrace\circ ...\circ \mathcal{\bf{Q}}_{R}\lbrace \mathcal{Q}^{1+n}_{R}(w)\rbrace\circ \mathcal{\bf{Q}}_{R}\lbrace \mathcal{Q}^{n}_{R}(w)\rbrace\circ \mathcal{\bf{Q}}_{R}\lbrace \mathcal{Q}^{n-1}_{R}(w)\rbrace\circ ... \circ \mathcal{\bf{Q}}_{R}\lbrace w \rbrace
\end{eqnarray*}
\\
Analogous statement for $\mathcal{\bf{E}}^{m+n}_{R}\lbrace w \rbrace$. $\square$
\\
\\
%PROPORSITION 4.1%
\begin{proposition} 
\label{prop:4.1}
Let $w\in(0,1)\setminus\mathbb{Q}$. Then, for every integer $1\leq r\leq\lfloor \frac{1}{w}\rfloor$,
\\
\[\mathcal{\bf{Q}}^{r}_{R}\lbrace w \rbrace (q,\mathrm{\bf{h}},z)=\left(\frac{\mathrm{num1}_{r}}{\mathrm{den}_{r}},\frac{\mathrm{num2}_{r}}{\mathrm{den}_{r}},\mathrm{num3}_{r}\right)\]
where
\\
\begin{eqnarray*}
\mathrm{num1}_{r}&=&(1+w)(r+rq-q)+\mathrm{\bf{h}}A_{1}(w,r) \\
\mathrm{num2}_{r}&=&\mathrm{\bf{h}}(w+1-wr) \\
\mathrm{den}_{r}&=&(1+w)(1+q)+\mathrm{\bf{h}}A_{2}(w,r) \\
\mathrm{num3}_{r}&=&\frac{1}{1-(r-1)w}\left(1+\frac{1}{w}\right)z
\end{eqnarray*}
\\
and where
\\
\begin{eqnarray*}
A_{1}(w,r)&=&(2r-1+wr-w)\mathrm{log}2-(2r-1+wr+w)\mathrm{log}\left(1+\frac{1}{w}\right) \\
&+&\sum_{k=1}^{r-1}(1+2k-2k^{2}w-w)\mathrm{log}\left(1+\frac{w}{1-kw}\right)+r\sum_{k=1}^{r-1}((2k-1)w-2)\mathrm{log}\left(1+\frac{w}{1-kw}\right) \\
A_{2}(w,r)&=&(1+wr)\mathrm{log}2-(2+w)\mathrm{log}\left(1+\frac{1}{w}\right)+r\sum_{k=1}^{r-1}((2k-1)w-2)\mathrm{log}\left(1+\frac{w}{1-kw}\right)
\end{eqnarray*}
\\
Furthermore, $\mathcal{E}_{R}(w)=\frac{1}{w}-\lfloor \frac{1}{w} \rfloor$, that  is, $\mathcal{E}_{R}$ is a Gauss map, and 
\\
\begin{equation}
\label{eq2:Prop4.1}
\mathcal{\bf{E}}_{R}\lbrace w \rbrace (q,\mathrm{\bf{h}},z)=\left(\frac{\mathrm{num1}_{\lfloor 1/w \rfloor}}{\mathrm{den}_{\lfloor 1/w \rfloor}}, \frac{\mathrm{num2}_{\lfloor 1/w \rfloor}}{\mathrm{den}_{\lfloor 1/w \rfloor}}, \mathrm{num3}_{\lfloor 1/w \rfloor}\right)
\end{equation}
\end{proposition}

\underline{Proof (sketch)}: Let $w\in (0,1)\setminus\mathbb{Q}$. We use the proof by induction over $1\leq r\leq\lfloor \frac{1}{w}\rfloor$. The base case $r=1$ gives by direct substitution exactly $\mathcal{\bf{Q}}_{R}\lbrace w \rbrace (q,\mathrm{\bf{h}})$ from Def.4.1. The induction step is the identity
\\
\begin{equation}
\label{eq:inductionStep}
\mathcal{\bf{Q}}_{R}\lbrace\mathcal{Q}^{r-1}_{R}(w)\rbrace\left(\frac{\mathrm{num1}_{r-1}}{\mathrm{den}_{r-1}},\frac{\mathrm{num2}_{r-1}}{\mathrm{den}_{r-1}},\mathrm{num3}_{r-1}\right)=\left(\frac{\mathrm{num1}_{r}}{\mathrm{den}_{r}},\frac{\mathrm{num2}_{r}}{\mathrm{den}_{r}}, \mathrm{num3}_{r}\right)
\end{equation}
\\
for all $2\leq r\leq\lfloor \frac{1}{w} \rfloor$. We have
\\
\begin{eqnarray*}
\lambda\mathrm{num1}_{r}&=&\left(1+\left(\frac{1}{w}-r+1\right)\right)(\mathrm{den}_{r-1}+\mathrm{num1}_{r-1}+\mathrm{num2}_{r-1}\mathrm{log}2) \\
&-&\mathrm{num2}_{r-1}\left(2+\left(\frac{1}{w}-r+1\right)\right)\mathrm{log}\left(1+\frac{w}{1-(r-1)w}\right) \\
\lambda\mathrm{num2}_{r}&=&\mathrm{num2}_{r-1}\left(\frac{1}{w}-r+1\right) \\
\lambda\mathrm{den}_{r}&=&\mathrm{den}_{r-1}\left(1+\left(\frac{1}{w}-r+1\right)\right)+\mathrm{num2}_{r-1}\mathrm{log}2 \\
&-&\mathrm{num2}_{r-1}\left(1+2\left(\frac{1}{w}-r+1\right)\right)\mathrm{log}\left(1+\frac{w}{1-(r-1)w}\right) 
\end{eqnarray*}
\\
where $\lambda=2+\frac{1}{w}-r>2$.
\\
Further, by equation (\ref{eq:EpochToEpoch}) we have $\left(1+\frac{1}{\frac{1}{w}-r+1}\right)\mathrm{num3}_{r-1}=\mathrm{num3}_{r}$. Therefore, $\lambda\mathrm{num3}_{r-1}=\left(\frac{1}{w}-r+1\right)\mathrm{num3}_{r}$. 
\\
\\
Equation (\ref{eq2:Prop4.1}) follows from the definition of $\mathcal{\bf{E}}_{R}\lbrace w \rbrace$.
$\square$
\\
\\
\begin{lemma}
\label{lemma:4.2}
For every $w\in(0,1)\setminus\mathbb{Q}$ and every integer $r$ with $1\leq r\leq\lfloor \frac{1}{w} \rfloor$, 
\\
\begin{eqnarray*}
0&\leq& A_{1}(w,r)-rA_{2}(w,r)+\mathrm{log}2\:\leq \:6\frac{1}{w} \\
-8\mathrm{log}\left(1+\frac{1}{w}\right)\:&\leq& \:A_{2}(w,r)\leq 0
\end{eqnarray*}
\\
with $A_{1},A_{2}$ as defined in Proposition \ref{prop:4.1}.
\end{lemma}

\underline{Proof}. Straightforward. $\square$
\\
\\
%PROPOSITION 4.2%
\begin{proposition} 
\label{prop:4.2}
For every $w\in(0,1)\setminus\mathbb{Q}$, every $p>0$ and every integer $1\leq r\leq\lfloor\frac{1}{w}\rfloor$, let $(\mu', \nu',\zeta')$ be the triple of rational functions over $\mathbb{R}$ in the triple of abstract variables $(\mu,\nu,\zeta)$ given implicitly by
\\
\begin{equation*}
\left(p'+\frac{\mu'}{\nu'},\frac{1+w'}{\nu'},\zeta'\right)=\mathcal{\bf{Q}}^{r}_{R}\lbrace w \rbrace \left(p+\frac{\mu}{\nu},\frac{1+w}{\nu},\zeta\right)
\end{equation*}
\\
where $w'=\mathcal{Q}^{r}_{R}(w)=\frac{1}{w}-r$ and $p'=r-\frac{p}{1+p}$, that is $(p',0,0)=\mathcal{\bf{Q}}^{r}_{R}\lbrace w \rbrace (p,0,0)$. Then, $\mu'$ is actually a linear polynomial over $\mathbb{R}$ in $\mu$, and $\nu'$ is a linear polynomial over $\mathbb{R}$ in $(\mu,\nu)$, and $\zeta'$ is a linear polynomial over $\mathbb{R}$ in $\zeta$. Explicitly
\\
\begin{equation}
\label{eq:Prop4.2}
\left(\begin{array}{c}\mu' \\\nu' \\\zeta'\end{array}\right)=\frac{1}{w}\left(\begin{array}{ccc}-\frac{1}{1+p} & 0 & 0 \\1 & 1+p & 0 \\0 & 0 & \frac{1+w}{1+w(1-r)}\end{array}\right)\left(\begin{array}{c}\mu \\\nu \\\zeta\end{array}\right)+\frac{1}{w}\left(\begin{array}{c}A_{1}(w,r)-p'A_{2}(w,r) \\A_{2}(w,r) \\0\end{array}\right)
\end{equation}
\\
The first and the second entries of the vector 
\\
\begin{equation*}
\frac{1}{w}\left(\begin{array}{c}A_{1}(w,r)-p'A_{2}(w,r) \\A_{2}(w,r) \\0\end{array}\right)
\end{equation*}
\\
are bounded in absolute value by $\leq 2^{4}\left(\frac{1}{w}\right)^{2}$ and $2^{3}\frac{1}{w}\mathrm{log}\left(1+\frac{1}{w}\right)$, respectively.
\end{proposition}

\begin{remark} 
Observe that $\zeta'$ is well-defined, that is, for all $1\leq r\leq\lfloor\frac{1}{w}\rfloor$ we have $\zeta'>0$ (physical meaning $\rightarrow$ energy density) and $1+w(1-r)\neq 0$.
\end{remark}

\underline{Proof}(of Proposition \ref{prop:4.2}). Equation (\ref{eq:Prop4.2}) follows directly from Proposition \ref{prop:4.1}. The bounds follow from Lemma \ref{lemma:4.2}. $\square$
\\
\\
\begin{definition} 
\label{def:4.4}
For every sequence of strictly positive integers $(k_{n})_{n\geq 0}$, we denote the associated infinite continued fraction by
\\
\begin{equation*}
\langle k_{0}, k_{1}, ...\rangle=\frac{1}{k_{0}+\frac{1}{k_{1}+...}} \:\:\: \in \:\:\:\left(\frac{1}{k_{0}+1},\frac{1}{k_{0}}\right)\setminus\mathbb{Q}
\end{equation*}
\\
Every element of $(0,1)\setminus\mathbb{Q}$ has a unique continued fraction expansion of this form.
\end{definition}

\underline{Goal}: to show that for $\mathrm{\bf{h}}=0$, the era-to-era maps can be realized as a left-shift operator on two-sided sequences of positive integers.
\\
\\
%PROPOSITION 4.3%
\begin{proposition} 
\label{prop:4.3}
Fix $(k_{n})_{n\in\mathbb{Z}}$ and define $(p_{n})_{n\in\mathbb{Z}}$ and $(w_{n})_{n\in\mathbb{Z}}$ by
\\
\begin{eqnarray*}
\frac{1}{1+p_{n}}&=&\langle k_{n},k_{n-1},k_{n-2}...\rangle \\
w_{n}&=&\langle k_{n+1},k_{n+2},k_{n+3}...\rangle
\end{eqnarray*}
\\
Then $w_{n+1}=\mathcal{E}_{R}(w_{n})$ and $(p_{n+1},0,0)=\mathcal{\bf{E}}_{R}\lbrace w_{n}\rbrace (p_{n},0,0)$ for all $n\in \mathbb{Z}$, and $\mathcal{E}^{n}_{R}(w_{0})=w_{n}$, and $\mathcal{\bf{E}}^{n}_{R}\lbrace w_{0}\rbrace (p_{0},0,0)=(p_{n},0,0)$ for all $n\geq 0$.
\end{proposition}

\underline{Proof}. We have
\\
\begin{eqnarray*}
\mathcal{E}_{R}(w_{n})&=&\frac{1}{w_{n}}-\lfloor \frac{1}{w}\rfloor \\
&=&k_{n+1}+\frac{1}{k_{n+2}+\frac{1}{k_{n+3}+...}}-k_{n+1} \\
&=&\frac{1}{k_{n+2}+\frac{1}{k_{n+3}+...}} \\
&=&w_{n+1}
\end{eqnarray*}
\\
and
\\
\begin{eqnarray*}
\mathcal{\bf{E}}_{R}\lbrace w_{n}\rbrace (p_{n},0,0)&=&\left(\lfloor \frac{1}{w_{n}}\rfloor -1+\frac{1}{1+p_{n}},0,0 \right) \\
&=&(k_{n+1}-1+\langle k_{n},k_{n-1},k_{n-2},...\rangle, 0,0) \\
&=&\left(k_{n+1}-1+\frac{1}{k_{n}+\frac{1}{k_{n-1}+\frac{1}{...}}}, 0,0\right) \\
&=&(p_{n+1},0,0).
\end{eqnarray*}
$\square$
\\
\\
\begin{definition} 
\label{def:4.5}
Fix any two-sided sequence $(k_{n})_{n\in\mathbb{Z}}$ of strictly positive integers and define $(p_{n})_{n\in\mathbb{Z}}$ and $(w_{n})_{n\in\mathbb{Z}}$ as in Proposition \ref{prop:4.3}. For every integer $n\geq 0$, let $(\mu_{n},\nu_{n},\zeta_{n})$ be the triple of linear polynomials over $\mathbb{R}$ in the abstract variables $(\mu_{0},\nu_{0},\zeta_{0})$ with coefficients depending only on the fixed sequence $(k_{n})_{n\in\mathbb{Z}}$, given implicitly by
\\
\begin{equation*}
\left(p_{n}+\frac{\mu_{n}}{\nu_{n}},\frac{1+w_{n}}{\nu_{n}},\zeta_{n}\right)=\mathcal{\bf{E}}^{n}_{R}\lbrace w_{0}\rbrace \left(p_{0}+\frac{\mu_{0}}{\nu_{0}},\frac{1+w_{0}}{\nu_{0}},\zeta_{0}\right)
\end{equation*}
\\
or by the equivalent recursive prescription
\\
\begin{equation}
\label{eq:Def.4.5Rec}
\left(p_{n+1}+\frac{\mu_{n+1}}{\nu_{n+1}},\frac{1+w_{n+1}}{\nu_{n+1}},\zeta_{n+1}\right)=\mathcal{\bf{E}}_{R}\lbrace w_{n}\rbrace \left(p_{n}+\frac{\mu_{n}}{\nu_{n}},\frac{1+w_{n}}{\nu_{n}},\zeta_{n}\right)
\end{equation}
\\
By Proposition \ref{prop:4.2}, equation (\ref{eq:Def.4.5Rec}) is $V_{n+1}=X_{n}V_{n}+Y_{n}$, where $V_{n}=(\mu_{n},\nu_{n},\zeta_{n})^{T}$ and
\\
\begin{eqnarray*}
X_{n}&=&\frac{1}{w_{n}}\left(\begin{array}{ccc}-\frac{1}{1+p_{n}} & 0 & 0 \\1 & 1+p_{n} & 0 \\0 & 0 & \frac{1+w_{n}}{1+w_{n}\left(1-\lfloor \frac{1}{w_{n}} \rfloor\right)}\end{array}\right) \\
Y_{n}&=&\frac{1}{w_{n}}\left(\begin{array}{c}A_{1}(w_{n})-p_{n+1}A_{2}(w_{n}) \\A_{2}(w_{n}) \\0\end{array}\right)
\end{eqnarray*}
\\
Here, $A_{1}(w)=A_{1}\left(w,\lfloor \frac{1}{w}\rfloor\right)$ and $A_{2}(w)=A_{2}\left(w,\lfloor \frac{1}{w}\rfloor\right)$.
\end{definition}

%EXAMPLE 4.1%
\underline{Example4.1} Consider Definition \ref{def:4.5} when $k_{n}=1$ for all $n\in\mathbb{Z}$. Then $w_{n}=p_{n}=w=\frac{1}{2}(\sqrt{5}-1)\in(0,1)\setminus\mathbb{Q}$, for all $n\in\mathbb{Z}$. We have $\lfloor\frac{1}{w}\rfloor=1$ and, therefore,
\\
\begin{eqnarray*}
X_{n}&=&\left(\begin{array}{ccc}-1 & 0 & 0 \\1+w & 2+w & 0 \\0 & 0 & 1+w\end{array}\right) \\
Y_{n}&=&\left(\begin{array}{c}-2\mathrm{log}(1+w) \\(2+w)\mathrm{log}2-(6+4w)\mathrm{log}(1+w) \\0\end{array}\right)
\end{eqnarray*}
\\
for all $n\geq0$. It follows that
\\
\begin{eqnarray*}
\mu_{n+1}&=&-\mu_{n}-2\mathrm{log}(1+w) \\
\nu_{n+1}&=&(1+w)\mu_{n}+(2+w)\nu_{n}+(2+w)\mathrm{log}2-(6+4w)\mathrm{log}(1+w) \\
\zeta_{n+1}&=&(1+w)\zeta_{n}
\end{eqnarray*}
\\
and we have $\mu_{n+2}=\mu_{n}$ for all $n\geq0$, which implies $\mu_{2n}=\mu_{0}$ and $\mu_{2n+1}=-\mu_{0}-2\mathrm{log}(1+w)$. We can, therefore, identify unique $\lambda_{1}=\lambda_{1}(\mu_{0})$ and $\lambda_{2}=\lambda_{2}(\mu_{0})$, such that
\\
\begin{eqnarray*}
\nu_{2n}&=&(2+w)^{2n}(\nu_{0}-\lambda_{1})+\lambda_{1} \\
\nu_{2n+1}&=&(2+w)^{2n}(\nu_{1}-\lambda_{2})+\lambda_{2}.
\end{eqnarray*} 
\\
\\
%PROPAGATOR%
\begin{definition}
\label{def:4.6}
($\mathrm{\bf{Propagator}}$). Let $(p_{n})_{n\in\mathbb{Z}}$, $(w_{n})_{n\in\mathbb{Z}}$, $(X_{n})_{n\geq0}$ be as in Definition \ref{def:4.5}. Then for all integers $n\geq m\geq 0$, let $P_{n,m}=X_{n-1}\cdot\cdot\cdot X_{m}$. Explicitly, 
\\
\begin{equation*}
P_{n,m}=\left(\begin{array}{ccc}a_{n-1}\cdot\cdot\cdot a_{m} & 0 & 0 \\\sum^{n-1}_{l=m}x_{l} & c_{n-1}\cdot\cdot\cdot c_{m} & 0 \\0 & 0 & d_{n-1}\cdot\cdot\cdot d_{m}\end{array}\right)
\end{equation*}
\\
where $x_{l}=c_{n-1}\cdot\cdot\cdot c_{l+1}b_{l}a_{l-1}\cdot\cdot\cdot a_{m}$ whenever $n-1\geq l\geq m$, and for all $l\geq 0$,
\\
\[X_{l}=\left(\begin{array}{ccc}a_{l} & 0 & 0 \\b_{l} & c_{l} & 0 \\0 & 0 & d_{l}\end{array}\right) \:\:\:\:\:\:\: a_{l}=\frac{-1}{w_{l}(1+p_{l})} \:\:\:\:\:\:\: b_{l}=\frac{1}{w_{l}} \:\:\:\:\:\:\: c_{l}=\frac{1+p_{l}}{w_{l}} \:\:\:\:\:\:\: d_{l}=\frac{1}{w_{l}}\frac{1+w_{l}}{1+w_{l}\left(1-\lfloor \frac{1}{w_{l}} \rfloor\right)}\]
\\
In this definition, a sequence of dots $\cdot\cdot\cdot$ indicates that indices increase towards the left, one by one. A product of the form $F_{k}\cdot\cdot\cdot F_{j}$ is equal to one if $k=j-1$. In particular, $P_{n,n}=\mathbb{I}_{3}$.
\end{definition}
%LEMMA 4.3%
\begin{lemma}
\label{lemma:4.3}
Recall Definition \ref{def:4.5}. We have
\begin{equation*}
V_{n}=P_{n,0}V_{0}+\sum^{n-1}_{l=0} P_{n,l+1}Y_{l}
\end{equation*}
\end{lemma}
\underline{Proof}: By direct inspection using Definition \ref{def:4.5}. Namely,
\\
\begin{eqnarray*}
V_{n}&=&X_{n-1}V_{n-1}+Y_{n-1} \\
&=& X_{n-1}(X_{n-2}V_{n-2}+Y_{n-2})+Y_{n-1} \\
&=&X_{n-1}X_{n-2}V_{n-2}+X_{n-1}Y_{n-2}+Y_{n-1} \\
&=&X_{n-1}X_{n-2}(X_{n-3}V_{n-3}+Y_{n-3})+X_{n-1}Y_{n-2}+Y_{n-1} \\
&=&X_{n-1}X_{n-2}X_{n-3}V_{n-3}+X_{n-1}X_{n-2}Y_{n-3}+X_{n-1}Y_{n-2}+Y_{n-1} \\
&=& ... \\
&=&P_{n,0}V_{0}+\sum^{\infty}_{l=0}P_{n,l+1}Y_{l}
\end{eqnarray*}
\\
$\square$

%LEMMA 4.4%
\begin{lemma}  
\label{lemma:4.4}
Recall Definition \ref{def:4.6}. For all integers $n\geq m\geq0$, we have
\\
\begin{align}
\frac{1}{2}&\leq& &\frac{w_{n-1}}{w_{m-1}}(-1)^{m+n}a_{n-1}\cdot\cdot\cdot a_{m}& &\leq& 2 \nonumber \\
(1-\delta_{mn})\frac{1}{4}&\leq& &\frac{w_{n-1}}{w^{2}_{m-1}}(w_{n-2}\cdot\cdot\cdot w_{m-1})^{2}\sum_{l=m}^{n-1}x_{l}& &\leq& 2 \nonumber \\
\frac{1}{2}&\leq& &\frac{w_{n-1}}{w_{m-1}}(w_{n-2}\cdot\cdot\cdot w_{m-1})^{2}c_{n-1}\cdot\cdot\cdot c_{m}& &\leq& 2 \nonumber \\
 \frac{1}{2}&< & &\frac{w^{2}_{n-1}}{w^{2}_{m-1}}(w_{n-2}\cdot\cdot\cdot w_{m-1})^{2}d_{n-1}\cdot\cdot\cdot d_{m}& &\leq&2
\end{align}
\end{lemma}

\underline{Proof}: Only for the new product $d_{n-1}\cdot\cdot\cdot d_{m}$. Recall $n\geq m$ so in the sequence $d_{n-1}\cdot\cdot\cdot d_{m}$ the indices are increasing towards the left. For the other three stays exactly the same as for the vacuum. Recall Definition \ref{def:4.6}. 
Recall Proposition \ref{prop:4.3}. We have $w_{n+1}=\mathcal{E}_{R}(w_{n})$. Therefore,
\\
\begin{eqnarray*}
1+w_{l}\left(1-\lfloor \frac{1}{w_{l}} \rfloor\right)&=&1+w_{l}\left(1-\frac{1}{w_{l}}+w_{l+1}\right) \\
&=&w_{l}(1+w_{l+1})
\end{eqnarray*} 
\\
and we get
\\
\begin{eqnarray*}
d_{n-1}\cdot\cdot\cdot d_{m}&=&\frac{1}{w_{n-2}\cdot\cdot\cdot w_{m-1}}\frac{w_{m-1}}{w_{n-1}}\frac{(1+w_{n-1})\cdot\cdot\cdot (1+w_{m})}{w_{n-1}(1+w_{n})w_{n-2}(1+w_{n-1})\cdot\cdot\cdot w_{m}(1+w_{m+1})} \\
&=&\frac{1}{w_{n-2}\cdot\cdot\cdot w_{m-1}}\frac{w_{m-1}}{w_{n-1}}\frac{1+w_{m}}{1+w_{n}}\frac{1}{w_{n-1}w_{n-2}\cdot\cdot\cdot w_{m}} \\
&=&\frac{1}{w_{n-2}\cdot\cdot\cdot w_{m-1}}\frac{w_{m-1}}{w_{n-1}}\frac{1+w_{m}}{1+w_{n}}\frac{w_{m-1}}{w_{n-1}}\frac{1}{w_{n-2}\cdot\cdot\cdot w_{m-1}} \\
&=&\left(\frac{1}{w_{n-2}\cdot\cdot\cdot w_{m-1}}\right)^{2}\frac{w^{2}_{m-1}}{w^{2}_{n-1}}\frac{1+w_{m}}{1+w_{n}}
\end{eqnarray*}
\\
Therefore,
\\
\begin{equation*}
\frac{1}{2}\leq\frac{w^{2}_{n-1}}{w^{2}_{m-1}}(w_{n-2}\cdot\cdot\cdot w_{m-1})^{2} d_{n-1}\cdot\cdot\cdot d_{m}\leq 2
\end{equation*}
$\square$
\\
\\
%PROPOSITION 4.4%
\begin{proposition} 
\label{prop:4.4}
For all $w_{0}\in (0,1)\setminus\mathbb{Q}$ and $q_{0}\in (0,\infty)\setminus\mathbb{Q}$, introduce
\\
\begin{itemize}
\item a two sided sequence of strictly positive integers $(k_{n})_{n\in\mathbb{Z}}$ by
\\
\begin{eqnarray*}
(1+q_{0})^{-1}&=&\langle k_{0},k_{-1}, k_{-2},...\rangle \\
w_{0}&=&\langle k_{1},k_{2},k_{3},...\rangle
\end{eqnarray*}
\\
\item ($Era$ $Pointer$) $J:\mathbb{Z}\rightarrow\mathbb{Z}$ by $J(0)=0$ and $J(n+1)=J(n)+k_{n+1}$ \\
\item ($Era$ $Counter$) $N:\mathbb{Z}\rightarrow\mathbb{Z}$ by $N(0)=0$ and $N(j+1)=N(j)+\chi_{J(\mathbb{Z})}(j)$, with $\chi$-characteristic function \\
\item sequences $(w_{j})_{j\in\mathbb{Z}}$ and $(p_{j})_{j\in\mathbb{Z}}$ by
\\
\begin{eqnarray*}
w_{j}&=&\langle k_{N(j)+1}, k_{N(j)+2}, ...\rangle +J(N(j))-j \\
p_{j}&=&\langle k_{N(j)-1}, k_{N(j)-2}, ...\rangle +k_{N(j)}+j-J(N(j))-1 \\
\end{eqnarray*}
\end{itemize}
\textbf{Part 1}. Then $p_{0}=q_{0}$ and $w_{j},p_{j}>0$ and $\mathcal{Q}_{R}(w_{j})=w_{j+1}$ and $\mathcal{\bf{Q}}_{R} \lbrace w_{j}\rbrace (p_{j},0,0)=(p_{j+1},0,0)$ for all $j\in\mathbb{Z}$, and $\mathcal{Q}^{j}_{R}(w_{0})=w_{j}$ and $\mathcal{\bf{Q}}^{j}_{R}\lbrace w_{0} \rbrace (q_{0},0,0)=(p_{j},0,0)$ for all $j\geq 0$.
\\
\\
\textbf{Part 2}. Introduce $\rho_{+}$ and $\textbf{C}_{0}=\textbf{C}_{0}(w_{0},q_{0})$. Suppose $\textbf{C}_{0}(w_{0},q_{0})<\infty$. Fix $\mathrm{\bf{h}}_{0}$ in the interval $0<\mathrm{\bf{h}}_{0}<\textbf{C}_{0}(w_{0},q_{0})$. \\
Then there are sequences $(q_{j})_{j\geq 0}$, $(\mathrm{\bf{h}}_{j})_{j\geq 0}$, $(z_{j})_{j\geq 0}$ of real numbers such that for every $j\geq 0$
\\
\begin{equation*}
(q_{j+1},\mathrm{\bf{h}}_{j+1},z_{j+1})=\mathcal{\bf{Q}}_{R}\lbrace w_{j} \rbrace (q_{j}, \mathrm{\bf{h}}_{j}, z_{j})
\end{equation*}
\\
or $(q_{j},\mathrm{\bf{h}}_{j},z_{j})=\mathcal{\bf{Q}}^{j}_{R}\lbrace w_{0}\rbrace(q_{0},\mathrm{\bf{h}}_{0},z_{0})$. For all $j\geq 0$,
\\
\begin{itemize}
\item $0<\mathrm{\bf{h}}_{j}\leq 2^{6}\mathrm{\bf{h}}_{0}\rho_{+}^{-2N(j)}$ and
\\
\begin{equation*}
\frac{1}{4}\leq\frac{\mathrm{\bf{h}}_{j}}{\mathrm{\bf{h}}_{0}}\frac{1+w_{0}}{1+w_{j}}\prod_{l=0}^{N(j)-1}\frac{1}{w_{J(l)}w_{J(l-1)}}\leq 4
\end{equation*}
\\
\item $0<z_{j}\leq2^{3}z_{0}\rho_{+}^{-2N(j)}$ 
and
\\
\begin{equation*}
\frac{1}{2}\leq\frac{z_{j}}{z_{0}}\prod_{l=1}^{N(j)-1}w_{J(l)}w_{J(l-1)}\leq 2
\end{equation*}
\\
\item $q_{j}\in(0,\infty)\setminus\mathbb{Z}$ and $\vert q_{j}-p_{j}\vert\leq 2^{12}\mathrm{\bf{h}}_{0}N(j)\rho_{+}^{-2N(j)}k_{N(j)}$ \\
\item $q_{j}\in(0,1)$ if and only if $p_{j}\in(0,1)$ if and only if $j-1\in J(\mathbb{Z})$ \\
\item $\mathrm{max}\lbrace\frac{1}{w_{j}},\frac{1}{q_{j}},\frac{1}{\vert q_{j}-1\vert},q_{j},z_{j}\rbrace\leq 2^{4}\mathrm{max}\lbrace k_{N(j)-2},k_{N(j)-1},k_{N(j)}k_{N(j)+1}\rbrace$ \\
\end{itemize}
\textbf{Part 3}. Let the map $\mathcal{Q}_{L}:(0,\infty)^{4}\rightarrow(0,\infty)^{2}\times\mathbb{R}\times (0,\infty)$ be given as in Definition \ref{def:3.16}. Then the sequences $(\mathrm{\bf{h}}_{j})_{j\geq 0}$, $(w_{j})_{j\geq 0}$, $(q_{j})_{j\geq 0}$, $(z_{j})_{j\geq 0}$ in Part 2 satisfy for all $j\geq 0$:
\\
\begin{equation*}
(\mathrm{\bf{h}}_{j},w_{j},q_{j},z_{j})=\mathcal{Q}_{L}(\mathrm{\bf{h}}_{j+1},w_{j+1},q_{j+1},z_{j+1})
\end{equation*}
\end{proposition}
\underline{Proof of Part 1}. The two basic properties of $J$ and $N$ remain the same as for the vacuum case. That is, for all $j\in\mathbb{Z}$:
\\
\begin{itemize}
\item $N\circ J$ is the identity and, therefore, $J(N(j))=j\Leftrightarrow j\in J(\mathbb{Z})$ \\
\item $J(N(j))\geq j$ and $J(N(j)-1)\leq j-1$ by definition of $Era$ $Counter$. Therefore,
\\
\begin{equation}
j\leq J(N(j))\leq k_{N(j)}+j-1
\end{equation}
\end{itemize}
The second bullet implies that $J(N(j))-j\geq 0$ and, therefore, $w_{j}>0$ $\forall j\in\mathbb{Z}$. Also, it implies that  $k_{N(j)}+j-1-J(N(j))\geq 0$ and, therefore, $p_{j}>0$ $\forall j\in\mathbb{Z}$.
\\
The first bullet implies that $w_{j}\in (0,1)\Leftrightarrow j\in J(\mathbb{Z})$ (follows directly from definition of $w_{j}$). Then we have
\\
\begin{eqnarray*}
\mathcal{Q}_{R}(w_{j})&=&
\begin{cases}
\frac{1}{w_{j}}-1 \:\: \mathrm{if}\:\: j\in J(\mathbb{Z}) \\
w_{j}-1 \:\: \mathrm{if} \:\: j\notin J(\mathbb{Z})
\end{cases}
\\
\mathcal{\bf{Q}}_{R}\lbrace w_{j} \rbrace (p_{j},0,0)&=&
\begin{cases}
\left(\frac{1}{1+p_{j}},0,0\right) \:\: \mathrm{if} \:\: j\in J(\mathbb{Z}) \\
(1+p_{j},0,0) \:\: \mathrm{if} \:\: j\notin J(\mathbb{Z})
\end{cases}
\end{eqnarray*} 
\\
\\
\underline{Proof of Part 2}. We first construct sequences $(q_{j})_{j\geq 0}$, $(\mathrm{\bf{h}}_{j})_{j\geq 0}$, and $(z_{j})_{j\geq 0}$. Then we verify that they have the desired properties. A sequence of dots $\cdot\cdot\cdot$ indicates that indices increase towards the left, one by one. A product of the form $F_{m}\cdot\cdot\cdot F_{n}$ is equal to one if $m=n-1$. 
\\
\\
Define sequences $(w^{*}_{n})_{n\in\mathbb{Z}}$ and $(p^{*}_{n})_{n\in\mathbb{Z}}$ by $w^{*}_{n}=w_{J(n)}\in (0,1)\setminus\mathbb{Q}$ and $p^{*}_{n}=p_{J(n)}\in (0,\infty)\setminus\mathbb{Q}$. Using the results from Part 1, we can write 
\\
\begin{eqnarray*}
\frac{1}{1+p^{*}_{n}}&=&\langle k_{n}, k_{n-1}, k_{n-2}...\rangle \\
w^{*}_{n}&=&\langle k_{n+1}, k_{n+2}, k_{n+3}... \rangle
\end{eqnarray*}
\\
Proposition \ref{prop:4.3} implies that for any such $p^{*}_{n}$ and $w^{*}_{n}$ we have $w^{*}_{n+1}=\mathcal{E}_{R}(w^{*}_{n})$ and $(p^{*}_{n+1},0,0)=\mathcal{\bf{E}}_{R}\lbrace w^{*}_{n} \rbrace (p^{*}_{n},0,0)$.
\\
\\
Recall Definition \ref{def:4.5}. Let $V^{*}_{n}=(\mu^{*}_{n}, \nu^{*}_{n}, \zeta^{*}_{n})^{T}$ be the solution to $V^{*}_{n+1}=X^{*}_{n}V^{*}_{n}+Y^{*}_{n}$ $\forall n\geq 0$ with $\mu^{*}_{0}=0$, $\nu^{*}_{0}=\frac{1+w^{*}_{0}}{\mathrm{\bf{h}}_{0}}>0$, and $\zeta^{*}_{0}=z_{0}$. Further, we have 
\\
\begin{eqnarray*}
X^{*}_{n}&=&\frac{1}{w^{*}_{n}}\left(\begin{array}{ccc}-\frac{1}{1+p^{*}_{n}} & 0 & 0 \\1 & 1+p^{*}_{n} & 0 \\0 & 0 & \frac{1+w^{*}_{n}}{1+w^{*}_{n}\left( 1-\lfloor \frac{1}{w^{*}_{n}}\rfloor\right)}\end{array}\right) \\
Y^{*}_{n}&=&\frac{1}{w^{*}_{n}}\left(\begin{array}{c}A_{1}(w^{*}_{n})-p^{*}_{n+1}A_{2}(w^{*}_{n}) \\A_{2}(w^{*}_{n}) \\0\end{array}\right)
\end{eqnarray*}
\\
Let $V_{j}=(\mu_{j},\nu_{j},\zeta_{j})^{T}$ be given by $V_{0}=V^{*}_{0}$ and $V_{j}=X^{*}_{N(j)-1}V^{*}_{N(j)-1}+Y_{j}$, $\forall j\geq 1$, with 
\\
\begin{equation*}
Y_{j}=\frac{1}{w^{*}_{s}}\left(\begin{array}{c}A_{1}(w^{*}_{s},j-J(s))-p_{j}A_{2}(w^{*}_{s},j-J(s)) \\A_{2}(w^{*}_{s},j-J(s)) \\0\end{array}\right)
\end{equation*}
\\
for $s=N(j)-1$. The functions $A_{1}, A_{2}$ are well-defined because $A_{1}(w)=A_{1}(w,\lfloor\frac{1}{w}\rfloor)$ (as given with only one argument in the expressions for X and Y) and $1\leq j-J(s)\leq  k_{N(j)}=\lfloor \frac{1}{w^{*}_{s}}\rfloor$. Both inequalities are the direct consequence of the second bullet in the Proof of Part 1. The last equality is just the definition of $w^{*}_{s}$.
\\
\\
(two observations about $Y_{j}$ stay the same as for the vacuum case since the third component is zero).
\\
\\
Recall Definition \ref{def:4.6}. Set $P^{*}_{n,m}=X^{*}_{n-1}\cdot\cdot\cdot X^{*}_{m}$ for all $n\geq m\geq 0$. By Lemma \ref{lemma:4.3} we have
\begin{equation*}
V_{j}=X^{*}_{s}(P^{*}_{s,0}V^{*}_{0}+\sum^{s-1}_{l=0}P^{*}_{s,l+1}Y^{*}_{l})+Y_{j}=P^{*}_{s+1,0}V^{*}_{0}+\sum^{s-1}_{l=0}P^{*}_{s+1,l+1}Y^{*}_{l}+Y_{j}
\end{equation*}
which implies
\\
\begin{eqnarray*}
\mu_{j}&=&a^{*}_{s}\cdot\cdot\cdot a^{*}_{0}\mu^{*}_{0}+\sum^{s-1}_{l=0}a^{*}_{s}\cdot\cdot\cdot a^{*}_{l+1}\frac{1}{w^{*}_{l}}(A_{1}(w^{*}_{l})-p^{*}_{l+1}A_{2}(w^{*}_{l}))+\frac{1}{w_{j}}(A_{1}(w_{j})-p_{j+1}A_{2}(w_{j})) \\
&=^{\mu^{*}_{0}=0}&\sum^{s-1}_{l=0}a^{*}_{s}\cdot\cdot\cdot a^{*}_{l+1}\frac{1}{w^{*}_{l}}(A_{1}(w^{*}_{l})-p^{*}_{l+1}A_{2}(w^{*}_{l}))+\frac{1}{w_{j}}(A_{1}(w_{j})-p_{j+1}A_{2}(w_{j})) \\
\nu_{j}&=&\mu^{*}_{0}\sum^{s}_{l=0}x_{l}+c^{*}_{s}\cdot\cdot\cdot c^{*}_{0}\nu^{*}_{0}+\sum^{s-1}_{l=0}\left(\sum^{s}_{k=l+1}x_{k}\frac{1}{w^{*}_{l}}(A_{1}(w^{*}_{l})-p^{*}_{l+1}A_{2}(w^{*}_{l}))+c^{*}_{s}\cdot\cdot\cdot c^{*}_{l+1}\frac{1}{w_{l}}A_{2}(w^{*}_{l})\right) \\
\zeta_{j}&=&d^{*}_{s}\cdot\cdot\cdot d^{*}_{0}\zeta^{*}_{0}
\end{eqnarray*}
\\
Therefore, using Lemma \ref{lemma:4.4} (Version 2)
\\
\begin{equation*}
\vert\zeta_{j}\vert=\vert d^{*}_{s}\cdot\cdot\cdot d^{*}_{0}\vert\vert\zeta^{*}_{0}\vert\leq 2\left(\frac{1}{w^{*}_{s-1}\cdot\cdot\cdot w^{*}_{-1}}\right)^{2}\left(\frac{w^{*}_{-1}}{w^{*}_{s}}\right)^{2} \vert\zeta^{*}_{0}\vert 
\end{equation*}
\\
and
\\
\begin{equation*}
\vert\zeta_{j}\vert=\vert d^{*}_{s}\cdot\cdot\cdot d^{*}_{0}\vert\vert\zeta^{*}_{0}\vert\geq\frac{1}{2}\left(\frac{1}{w^{*}_{s-1}\cdot\cdot\cdot w^{*}_{-1}}\right)^{2}\left(\frac{w^{*}_{-1}}{w^{*}_{s}}\right)^{2} \vert\zeta^{*}_{0}\vert 
\end{equation*}
\\
\\
and the old estimates for $\vert\mu_{j}\vert$ and $\nu_{j}$ stay the same (checked), that is
\\
\begin{eqnarray*}
\vert\mu_{j}\vert&\leq&\frac{2^{5}}{w^{*}_{s}}\sum_{l=0}^{s}\frac{1}{w^{*}_{l}} \\
\nu_{j}&\geq&\frac{1}{2w^{*}_{s}}\left(\frac{1}{w^{*}_{s-1}\cdot\cdot\cdot w^{*}_{-1}}\right)^{2}\left(w^{*}_{-1}\nu_{0}-2^{8}\sum_{l=0}^{s}(w^{*}_{l-1}\cdot\cdot\cdot w^{*}_{-1})^{2}\mathrm{log}\left(1+\frac{1}{w^{*}_{l}}\right)\right) \\
\nu_{j}&\leq&\frac{2}{w^{*}_{s}}\left(\frac{1}{w^{*}_{s-1}\cdot\cdot\cdot w_{-1}^{*}}\right)^{2}\left(w^{*}_{-1}\nu_{0}+2^{6}\sum_{l=0}^{s}(w^{*}_{l-1}\cdot\cdot\cdot w^{*}_{-1})^{2}\mathrm{log}\left(1+\frac{1}{w^{*}_{l}}\right)\right)
\end{eqnarray*}
\\
for all $j\geq1$ and $s=N(j)-1$. All four estimates are also true when $j=0,s=-1$. Now, using the (old) estimate $2^{8}\sum_{l=0}^{s}(w^{*}_{l-1}\cdot\cdot\cdot w^{*}_{-1})^{2}\mathrm{log}(1+1/w^{*}_{l})\leq 2^{-1}w^{*}_{-1}\frac{1}{\mathrm{\bf{h}}_{0}}$ (checked), we have by direct calculation
\\
\begin{equation}
\label{est:1}
\frac{1}{4}\leq\frac{w^{*}_{N(j)-1}}{w^{*}_{-1}}(w^{*}_{N(j)-2}\cdot\cdot\cdot w^{*}_{-1})^{2}\frac{\mathrm{\bf{h}}_{0}}{1+w^{*}_{0}}\nu_{j}\leq 4
\end{equation}
\\
Further, we estimate
\\
\begin{equation}
\label{est:2}
\frac{1}{2}\leq\left(\frac{w^{*}_{N(j)-1}}{w^{*}_{-1}}\right)^{2}(w^{*}_{N(j)-2}\cdot\cdot\cdot w^{*}_{-1})^{2}\frac{1}{\zeta^{*}_{0}}\zeta_{j}\leq2
\end{equation}
\\
\\
Introduce $\mathcal{Z}_{j}=z_{0}\prod_{l=1}^{N(j)-1}\frac{1}{w_{J(l)}w_{J(l-1)}}$. Therefore, we have
\\
\begin{equation*}
\frac{1}{2}\leq\frac{z_{j}}{\mathcal{Z}_{j}}\leq2
\end{equation*}
$\square$.
%SECTION 5%
\section{Abstract semi-global existence theorem}  Stays exactly the same as for the vacuum case. See \cite{RT}.
\\
\\
%MAIN THEOREMS%
\section{Main Theorems}
%\subsection{}
\begin{definition} 
\label{def:6.1}
Let $\vert\vert\cdot\vert\vert$ be the Euclidean distance in $\mathbb{R}^{4}$. For every $\delta >0$ and every 
$\mathrm{\bf{f}}\in\mathbb{R}^{4}$, set $B[\delta,\mathrm{\bf{f}}]=\lbrace\mathrm{\bf{g}}\in\mathbb{R}^{4}|\: \vert\vert\mathrm{\bf{g}}-\mathrm{\bf{f}}\vert\vert\leq\delta\rbrace$
\end{definition}

\begin{definition} 
\label{def:6.2}
Let $\mathcal{F}\subset (0,\infty)^{4}$ be as in Definition \ref{def:3.19}. For all $\zeta\geq 1$ set
\\
\begin{equation*}
B_{\zeta}\mathcal{F}=\lbrace(\delta,\mathrm{\bf{f}})\in[0,\infty)\times\mathcal{F}|\: B[\zeta\delta,\mathrm{\bf{f}}]\subset\mathcal{F}\rbrace \:\:\:\:\:and\:\:\:\:\:B\mathcal{F}=B_{1}\mathcal{F}
\end{equation*}
\end{definition}

%Lemma6.1%
\begin{lemma}
\label{lemma:6.1}
For all $(\delta,\mathcal{F})\in B\mathcal{F}$ set
\\
\begin{align}
W(\delta,\mathrm{\bf{f}})= &\mathrm{max}\lbrace\frac{1}{w-\delta},w+\delta,\frac{1}{q-\delta},\frac{1}{\vert q-1\vert-\delta},q+\delta\rbrace& &\in[1,\infty)& \nonumber \\
W(\mathrm{\bf{f}})= &W(0,\mathrm{\bf{f}})=\mathrm{max}\lbrace\frac{1}{w},w,\frac{1}{q},\frac{1}{\vert q-1\vert},q\rbrace& &\in[1,\infty)& \nonumber
\end{align}
\\
where $\mathrm{\bf{f}}=(\mathrm{\bf{h}},w,q,z)$. Then
\\
\begin{enumerate}
\item $W(\mathrm{\bf{g}})\leq W(\delta,\mathrm{\bf{f}})$ for all $\mathrm{\bf{g}}\in B[\delta,\mathrm{\bf{f}}]$. 
\item If $(\delta,\mathrm{\bf{f}})\in B_{2}\mathcal{F}\subset B\mathcal{F}$ then $W(\delta,\mathrm{\bf{f}})\leq 2W(\mathrm{\bf{f}})$. \\
\end{enumerate}
\end{lemma}
\underline{Proof}: straightforward by direct substitution. $\square$
%NewLemma6.2%
\begin{lemma} 
\label{lemma:6.2}
Let $\mathrm{Err}:B\mathcal{F}\rightarrow[0,\infty)$ be given by
\\
\begin{equation*}
\mathrm{Err}(\delta,\mathrm{\bf{f}})=2^{42}\left(\frac{1}{\mathrm{\bf{h}}-\delta}\right)^{2}W(\delta,\mathrm{\bf{f}})^{5}\mathrm{exp}\left(-\frac{1}{\mathrm{\bf{h}}}2^{-9}W(\delta,\mathrm{\bf{f}})^{-2}\right)\mathrm{max}\lbrace1,z\rbrace^{2}
\end{equation*}
\\
where $\mathrm{\bf{f}}=(\mathrm{\bf{h}},w,q,z)$. Then for all $(\delta,\mathrm{\bf{f}})\in B\mathcal{F}$, we have $\mathrm{\bf{K}}(\mathrm{\bf{g}})\leq\mathrm{Err}(\delta,\mathrm{\bf{f}})$ for all $\mathrm{\bf{g}}\in B[\delta,\mathrm{\bf{f}}]\subset\mathcal{F}$.
\end{lemma}

\underline{Proof}: Let $\mathrm{\bf{g}}=(\mathrm{\bf{h}}',w',q',z')\in B[\delta,\mathrm{\bf{f}}]$. Then, by $\tau_{*}\geq\mathrm{\bf{X}}(-1,0,0,-1,-1,0,0,0)$ in the notation of Proposition \ref{prop:3.3}, we have $\tau_{*}(\mathrm{\bf{g}})\geq\frac{1}{2}W(\mathrm{\bf{g}})^{-2}$. Further, $0<x-\delta\leq x'\leq x+\delta\leq2x$ for $x\in\lbrace\mathrm{\bf{h}},z\rbrace$, which obviously implies $\frac{1}{\mathrm{\bf{h}}'}\geq\frac{1}{2\mathrm{\bf{h}}}$ and $\mathrm{max}\lbrace 1,z'\rbrace^{2}\leq 2^{2}\mathrm{max}\lbrace 1,z \rbrace^{2}$. Recall Definition \ref{def:3.18}, and it follows 
\\
\begin{eqnarray*}
\mathrm{\bf{K}}(\mathrm{\bf{g}})&\leq&2^{42}\left(\frac{1}{\mathrm{\bf{h}}-\delta}\right)^{2}W(\mathrm{\bf{g}})^{5}\mathrm{exp}\left(-\frac{1}{\mathrm{\bf{h}}}2^{-9}W(\mathrm{\bf{g}})^{-2}\right)\mathrm{max}\lbrace 1,z\rbrace^{2} \\
&\overset{Lemma \ref{lemma:6.1},(1)}\leq&2^{42}\left(\frac{1}{\mathrm{\bf{h}}-\delta}\right)^{2}W(\delta,\mathrm{\bf{f}})^{7}\mathrm{exp}\left(-\frac{1}{\mathrm{\bf{h}}}2^{-9}W(\delta,\mathrm{\bf{f}})^{-2}\right)\mathrm{max}\lbrace 1,z\rbrace^{2}
\end{eqnarray*}
$\square$
%NewLemma6.3%
\begin{lemma} 
\label{lemma:6.3}
Let $\mathcal{Q}_{L}$ be as in Definition \ref{def:3.16}. Set $\mathrm{Lip}:B\mathcal{F}\rightarrow[0,\infty)$, $\mathrm{Lip}(\delta,\mathrm{\bf{f}})=2^{13}W(\delta,\mathrm{\bf{f}})^{3}\mathrm{max}\lbrace 1,z\rbrace$. Then $\vert\vert\mathcal{Q}_{L}(\mathrm{\bf{g}})-\mathcal{Q}_{L}(\mathrm{\bf{g}}')\vert\vert\leq\mathrm{Lip}(\delta,\mathrm{\bf{f}})\vert\vert\mathrm{\bf{g}}-\mathrm{\bf{g}}'\vert\vert$ for all $\mathrm{\bf{g}},\mathrm{\bf{g}}'\in B[\delta,\mathrm{\bf{f}}]$.
\end{lemma}

\underline{Proof}. Let $\mathrm{\bf{f}}=(\mathrm{\bf{h}},w,q,z)$. Recall NewLemma B.1 of Appendix B (result summary: the same result as for the old case just multiplied by $\mathrm{max}\lbrace 1,z\rbrace$ due to $\vert dz_{L}/dw\vert$ derivative). The case $\mathrm{\bf{g}}=\mathrm{\bf{g}}'$ is trivial. Suppose $\mathrm{\bf{g}}\neq\mathrm{\bf{g}}'$. Then identify $\mathrm{\bf{f}}_{1}$ and $\mathrm{\bf{f}}_{2}$ in the formulation of NewLemma B.1 with  $\mathrm{\bf{f}}_{1}=(\mathrm{\bf{h}}_{1},w_{1},q_{1},z_{1})=\mathrm{\bf{g}}$ and $\mathrm{\bf{f}}_{2}=(\mathrm{\bf{h}}_{2},w_{2},q_{2},z_{2})=\mathrm{\bf{g}}'$.
\\
Observe $0<\mathrm{\bf{h}}_{i}\leq 1$ for $i=1,2$, by $\mathrm{\bf{g}},\mathrm{\bf{g}}'\in B[\delta,\mathrm{\bf{f}}]\subset\mathcal{F}$.
\\
$\delta<\vert q-1\vert$ $\Rightarrow$ either $q,q_{1},q_{2}<1$ or $q,q_{1},q_{2}>1$.
\\
Using $w_{\mathrm{max}}\leq\mathrm{max}\lbrace W(\mathrm{\bf{g}}),W(\mathrm{\bf{g}}')\rbrace$, $q_{\mathrm{max}}\leq\mathrm{max}\lbrace W(\mathrm{\bf{g}}),W(\mathrm{\bf{g}}')\rbrace$, and $q^{-1}_{\mathrm{min}}=\mathrm{max}\lbrace q^{-1}_{1},q^{-1}_{2}\rbrace\leq\mathrm{max}\lbrace W(\mathrm{\bf{g}}),W(\mathrm{\bf{g}}')\rbrace$, we estimate
\\
\begin{eqnarray*}
2^{12}q^{-2}_{\mathrm{min}}\mathrm{log}(2+w_{\mathrm{max}})\mathrm{max}\lbrace 1,z\rbrace&\leq&2^{12}W(\delta,\mathrm{\bf{f}})^{2}(1+w_{\mathrm{max}})\mathrm{max}\lbrace 1,z\rbrace \\
&\leq&2^{12}W(\delta,\mathrm{\bf{f}})^{2}(1+W(\delta,\mathrm{\bf{f}}))\mathrm{max}\lbrace 1,z\rbrace \\
&\leq&2^{13}W(\delta,\mathrm{\bf{f}})^{3}\mathrm{max}\lbrace 1,z\rbrace
\end{eqnarray*}
$\square$
\\
\\
%Main Theorem 1%
\begin{theorem} 
\label{main_theorem_1}
Recall the definitions of $\mathcal{P}_{L},\mathcal{Q}_{L},\mathcal{F},\Pi$ from Section 3, Definition \ref{def:6.2}, Lemma \ref{lemma:6.1}, Lemma \ref{lemma:6.2} and Lemma \ref{lemma:6.3}. Suppose:
\\
\begin{enumerate}
\item $(\mathrm{\bf{f}}_{j})_{j\geq0}$, with $\mathrm{\bf{f}}_{j}=(\mathrm{\bf{h}}_{j},w_{j},q_{j},z_{j})\in\mathcal{F}$ satisfies $\mathrm{\bf{f}}_{j-1}=\mathcal{Q}_{L}(\mathrm{\bf{f}}_{j})$ for all $j\geq 1$. 
\item The sequence $(\delta_{j})_{j\geq0}$ given by
\\
\begin{equation*}
\delta_{j}=\sum_{l=j+1}^{\infty}\lbrace\prod_{k=j+1}^{l-1}2^{16}W(\mathrm{\bf{f}}_{k})^{3}\mathrm{max}\lbrace 1,z_{k}\rbrace\rbrace 2^{49}\left(\frac{1}{\mathrm{\bf{h}}_{l}}\right)^{2}W(\mathrm{\bf{f}}_{l})^{5}\mathrm{exp}\left(-\frac{1}{\mathrm{\bf{h}}_{l}}2^{-11}W(\mathrm{\bf{f}}_{l})^{-2}\right)\mathrm{max}\lbrace 1,z_{l}\rbrace^{2}
\end{equation*}
\\
satisfies $\delta_{j}<\infty$ and $(\delta_{j},\mathrm{\bf{f}}_{j})\in B_{2}\mathcal{F}$ for all $j\geq 0$.
\item $\pi_{0}\in S_{3}$ and $(\pi_{j})_{j\geq0}$ is the unique sequence in $S_{3}$ that satisfies $\pi_{j-1}=\mathcal{P}_{L}(\pi_{j},\mathrm{\bf{f}}_{j})$ for all $j\geq 1$.
\item $\sigma\in\lbrace -1,+1\rbrace^{3}$. \\ \\
\end{enumerate}
Then, there exists a sequence $(\mathrm{\bf{g}}_{j})_{j\geq 0}$ with $\mathrm{\bf{g}}_{j}\in B[\delta_{j},\mathrm{\bf{f}}_{j}]\subset\mathcal{F}$ such that for all $j\geq 1$:
\\
\begin{equation*}
\mathrm{\bf{g}}_{j-1}=\Pi[\pi_{j},\sigma_{*}](\mathrm{\bf{g}}_{j}) \:\:\:\:\:\:\: and \:\:\:\:\:\:\: \pi_{j-1}=\mathcal{P}_{L}(\pi_{j},\mathrm{\bf{g}}_{j})
\end{equation*}
\end{theorem}
\underline{Proof}: For the proof of this theorem, we use exclusively Proposition 5.1. The abstract object of the latter proposition are in our case identified with
\\
\begin{align}
d&\rightarrow 4 \nonumber \\
\mathcal{F}&\rightarrow\mathcal{F}\: \mathrm{as\:in\:Definition\:\ref{def:3.19}} \nonumber \\
\Pi_{j}&\rightarrow\Pi[\pi_{j},\sigma_{*}], \:\mathrm{see\:Proposition\:\ref{prop:3.3}\:and\:hypotheses\:(3)\:and\:(4)\:of\:Theorem\: \ref{main_theorem_1}} \nonumber \\
\mathcal{Q}_{L}&\rightarrow\mathcal{Q}_{L}\big|_{\mathcal{F}}, \: \mathrm{with}\:\mathcal{Q}_{L}\:\mathrm{as\:in\:Definition\:\ref{def:3.16}} \nonumber \\
\mathrm{Err}&\rightarrow\mathrm{Err}\:\mathrm{as\:in\:Lemma\:\ref{lemma:6.2}} \nonumber \\
\mathrm{Lip}&\rightarrow\mathrm{Lip}\:\mathrm{as\:in\:Lemma\:\ref{lemma:6.3}} \nonumber \\
(\delta_{j},\mathrm{\bf{f}}_{j})&\rightarrow(\delta_{j},\mathrm{\bf{f}}_{j})\:\mathrm{as\:in\:hypotheses\:(1)\:and\:(2)\:of\:Theorem\: \ref{main_theorem_1}} \nonumber
\end{align}
\\
We now check that the assumptions of the Proposition 5.1 are actually satisfied.
\\
\\
(a) Consistent with Definition \ref{def:6.2}. \\
(b) $\Pi[\pi_{j},\sigma_{*}]:\mathcal{F}\rightarrow(0,\infty)^{2}\times\mathbb{R}\times(0,\infty)\subset\mathbb{R}^{4}$ is continuous by Proposition \ref{prop:3.3}. \\
(c) Consistent with $\mathrm{Err}$ and $\mathrm{Lip}$ from Lemma \ref{lemma:6.2} and Lemma \ref{lemma:6.3}. \\
(d) Recall Lemma \ref{lemma:6.1}, (2). By direct calculation
\\
\begin{eqnarray*}
\sum_{l=j+1}^{\infty}\lbrace\prod_{k=j+1}^{l-1}\mathrm{Lip}(\delta_{k},\mathrm{\bf{f}}_{k})\rbrace\mathrm{Err}(\delta_{l},\mathrm{\bf{f}}_{l})&=&\sum_{l=j+1}^{\infty}\lbrace\prod_{k=j+1}^{l-1}2^{13}W(\delta_{k},\mathrm{\bf{f}}_{k})^{3}\mathrm{max}\lbrace 1,z_{k}\rbrace\rbrace 2^{42}\left(\frac{1}{\mathrm{\bf{h}}_{l}-\delta_{l}}\right)^{2}W(\delta_{l},\mathrm{\bf{f}}_{l})^{5} \\
&\times&\mathrm{exp}\left(-\frac{1}{\mathrm{\bf{h}}_{l}}2^{-9}W(\delta_{l},\mathrm{\bf{f}}_{l})^{-2}\right)\mathrm{max}\lbrace 1,z_{l}\rbrace^{2} \\
&\leq&\sum_{l=j+1}^{\infty}\lbrace\prod_{k=j+1}^{l-1}2^{16}W(\mathrm{\bf{f}}_{k})^{3}\mathrm{max}\lbrace 1,z_{k}\rbrace\rbrace 2^{49}\left(\frac{1}{\mathrm{\bf{h}}_{l}}\right)^{2}W(\mathrm{\bf{f}}_{l})^{5} \\
&\times&\mathrm{exp}\left(-\frac{1}{\mathrm{\bf{h}}_{l}}2^{-11}W(\mathrm{\bf{f}}_{l})^{-2}\right)\mathrm{max}\lbrace 1,z_{l}\rbrace^{2} \\
&=&\delta_{j}.
\end{eqnarray*}
$\square$
\\
\\
%Main Theorem 2%
\begin{theorem}
\label{main_theorem_2}
Recall Proposition \ref{prop:4.4}. Suppose the vector $\mathrm{\bf{f}}_{0}=(\mathrm{\bf{h}}_{0},w_{0},q_{0},z_{0})$ satisfies the following assumptions:
\\
\begin{align}
w_{0}&\in (0,1)\setminus\mathbb{Q} \:\:\:\:\:\:\: &\mathrm{\bf{C}}(w_{0},q_{0})& <\infty \nonumber \\
q_{0}&\in (0,\infty)\setminus\mathbb{Q} \:\:\:\:\:\:\: &0<\mathrm{\bf{h}}_{0}\leq&2^{-14}(\mathrm{\bf{C}}(w_{0},q_{0}))^{-1} \nonumber \\
z_{0}& >0 \nonumber 
\end{align}
\\
Recall the definitions of the Era Pointer, Era Counter, and of the sequences $(k_{n})_{n\in\mathbb{Z}}$, $(w_{j})_{j\in\mathbb{Z}}$, $(q_{j})_{j\geq0}$, $(\mathrm{\bf{h}}_{j})_{j\geq0}$, $(z_{j})_{j\geq0}$ from the Proposition \ref{prop:4.4}. Introduce the sequence $(\mathrm{\bf{f}}_{j})_{j\geq0}$ by
\\
\begin{equation*}
\mathrm{\bf{f}}_{j}=(\mathrm{\bf{h}}_{j},w_{j},q_{j},z_{j})\in (0,\infty)^{4}
\end{equation*}
\\
Introduce sequences $(\mathrm{\bf{H}}_{j})_{j\geq0}$, $(K_{j})_{j\geq0}$ and $(\mathcal{Z}_{j})_{j\geq0}$ by
\\
\begin{align}
\mathrm{\bf{H}}_{j}&=\mathrm{\bf{h}}_{0}\frac{1+w_{j}}{1+w_{0}}\prod_{l=0}^{N(j)-1}w_{J(l)}w_{J(l-1)} &>0 \nonumber \\
K_{j}&=\mathrm{max}\lbrace k_{N(j)-2},k_{N(j)-1},k_{N(j)},k_{N(j)+1}\rbrace &\geq 1 \nonumber \\
\mathcal{Z}_{j}&=z_{0}\prod_{l=1}^{N(j)-1}\frac{1}{w_{J(l)}w_{J(l-1)}} &>0  \nonumber  
\end{align} 
\\
Suppose:
\\
\begin{enumerate}
\item $\mathrm{\bf{H}}_{j}<2^{-21}(K_{j})^{-2}$ for all $j\geq0$. 
\item $2^{75}\left(\frac{1}{\mathrm{\bf{H}}_{j}}\right)^{2}(K_{j})^{5}\mathrm{exp}\left(-\frac{1}{\mathrm{\bf{H}}_{j}}2^{-21}(K_{j})^{-2}\right)\mathrm{max}\lbrace 1,\mathcal{Z}_{j}\rbrace^{2}<1$ for all $j\geq 0$.
\item The sequence $(\not{\delta}_{j})_{j\geq0}$ given by
\\ 
\begin{equation*}
\not{\delta}_{j}=\sum_{l=j+1}^{\infty}\lbrace\prod_{k=j+1}^{l-1}2^{28}(K_{k})^{3}\mathrm{max}\lbrace1,\mathcal{Z}_{k}\rbrace\rbrace 2^{75}\left(\frac{1}{\mathrm{\bf{H}}_{l}}\right)^{2}(K_{l})^{5}\mathrm{exp}\left(-\frac{1}{\mathrm{\bf{H}}_{l}}2^{-21}(K_{l})^{-2}\right)\mathrm{max}\lbrace1,\mathcal{Z}_{l}\rbrace^{2}
\end{equation*}
\\
satisfies $\not{\delta}_{j}\leq 2^{-3}\mathrm{min}\lbrace\mathrm{2^{-1}\bf{H}}_{j},\mathcal{Z}_{j}\rbrace$. 
\item $\pi_{0}\in S_{3}$ and $(\pi_{j})_{j\geq0}$ is the unique sequence in $S_{3}$ that satisfies $\pi_{j-1}=\Pi_{L}(\pi_{j},\mathrm{\bf{f}}_{j})$ for all $j\geq1$
\item $\sigma_{*}\in\lbrace -1,+1\rbrace^{3}$ \\
\end{enumerate}
Then $(\not{\delta_{j}},\mathrm{\bf{f}}_{j})\in B_{2}\mathcal{F}$ for all $j\geq0$ and there exists a sequence $(\mathrm{\bf{g}}_{j})_{j\geq0}$ with $\mathrm{\bf{g}}_{j}\in B[\not{\delta}_{j},\mathrm{\bf{f}}_{j}]\subset\mathcal{F}$ such that for all $j\geq1$
\\
\begin{equation*}
\mathrm{\bf{g}}_{j-1}=\Pi[\pi_{j},\sigma_{*}](\mathrm{\bf{g}}_{j})\:\:\:\:\:\:\: and \:\:\:\:\:\:\: \pi_{j-1}=\mathcal{P}_{L}(\pi_{j},\mathrm{\bf{g}}_{j})
\end{equation*}
\end{theorem}
\underline{Proof}: Recall Proposition \ref{prop:4.4} and hypotheses (1) and (2) of the Theorem \ref{main_theorem_2}. We have
\\
\begin{eqnarray*}
2^{-2}\mathrm{\bf{H}}_{j}\:\:\: \leq \:\:\: \mathrm{\bf{H}}_{j}&\leq&2^{2}\mathrm{\bf{h}}_{j} \\
2^{-1}\mathcal{Z}_{j}\:\:\: \leq \:\:\: z_{j}&\leq&2\mathcal{Z}_{j} \\
\mathrm{max}\lbrace\frac{1}{w_{j}},w_{j},\frac{1}{q_{j}},\frac{1}{\vert q_{j}-1\vert},q_{j}\rbrace&\leq& 2^{4}K_{j} \\
2^{-4}(K_{j})^{-1}&\leq&\mathrm{min}\lbrace w_{j},q_{j},\vert q_{j}-1\vert\rbrace \\
2\not{\delta}_{j}&\leq&2^{-1}\mathrm{min}\lbrace w_{j},q_{j},\vert q_{j}-1\vert,\mathrm{\bf{h}}_{j},z_{j}\rbrace
\end{eqnarray*}
\\
Hence, $B[2\not{\delta}_{j},\mathrm{\bf{f}}_{j}]\subset(0,\infty)^{4}$ for every $j\geq0$. Further, $\forall (\mathrm{\bf{h}}',w',q',z')\in B[2\not{\delta}_{j},\mathrm{\bf{f}}_{j}]\subset (0,\infty)^{4}$, we have $q'\neq 1$ and
\\
\begin{equation*}
2^{-3}\mathrm{\bf{H}}_{j}\leq  2^{-1}\mathrm{\bf{h}}_{j} \leq \mathrm{\bf{h}}_{j}-2\not{\delta}_{j} \leq \mathrm{\bf{h}}' \leq \mathrm{\bf{h}}+2\not{\delta}_{j} \leq 2\mathrm{\bf{h}}_{j} \leq 2^{3} \mathrm{\bf{H}}_{j} 
\end{equation*}
\\
and
\\
\begin{equation*}
2^{-2}\mathcal{Z}_{j}\leq 2^{-1}z_{j} \leq z_{j}-2\not{\delta}_{j} \leq z' \leq z+2\not{\delta}_{j} \leq 2z_{j} \leq 2^{3}\mathcal{Z}_{j} \nonumber
\end{equation*}
\\
and
\\
\begin{eqnarray*}
\mathrm{max}\lbrace\frac{1}{w'},w',\frac{1}{q'},\frac{1}{\vert q'-1\vert},q'\rbrace&\leq&\mathrm{max}\lbrace\frac{1}{w'-2\not{\delta}_{j}},w'+2\not{\delta}_{j},\frac{1}{q'-2\not{\delta}_{j}},\frac{1}{\vert q'-1\vert-2\not{\delta}_{j}},q'+2\not{\delta}_{j}\rbrace \\
&\leq&2\mathrm{max}\lbrace\frac{1}{w_{j}},w_{j},\frac{1}{q_{j}},\frac{1}{\vert q_{j}-1\vert,q_{j}}\rbrace \\
&\leq& 2^{5}K_{j}
\end{eqnarray*}
\\
Recall the definition of $\tau_{*}$ from Section 3. Then the estimates above imply $\tau_{*}(\mathrm{\bf{h}}',w',q',z')\geq 2^{-11}(K_{j})^{-2}$, and by assumption (2) of the Theorem \ref{main_theorem_2} we have
\\
\begin{equation*}
\mathrm{\bf{K}}(\mathrm{\bf{h}}',w',q',z')\leq2^{75}\left(\frac{1}{\mathrm{\bf{H}}_{j}}\right)^{2}(K_{j})^{5}\mathrm{exp}\left(-\frac{1}{\mathrm{\bf{H}}_{j}}2^{-21}(K_{j})^{-2}\right)\mathrm{max}\lbrace 1,\mathcal{Z}_{j}\rbrace^{2}<1
\end{equation*}
\\
Further, using the hypothesis (1) of the Theorem \ref{main_theorem_2}, we estimate
\\
\begin{equation*}
\mathrm{\bf{h}}'\leq 2^{3}\mathrm{\bf{H}}_{j}\leq 2^{-18}(K_{j})^{-2}\leq2^{-7}\tau_{*}(\mathrm{\bf{h}}',w',q',z')
\end{equation*}
\\
which is true for all $(\mathrm{\bf{h}}',w',q',z')\in B[2\not{\delta}_{j},\mathrm{\bf{f}}_{j}]$. Therefore, $B[2\not{\delta}_{j},\mathrm{\bf{f}}_{j}]\subset\mathcal{F}$, $\forall j\geq 0$, in particular $\mathrm{\bf{f}}_{j}\in\mathcal{F}$ (recall NewDef.3.19). We have $(\not{\delta}_{j},\mathrm{\bf{f}}_{j})\in B_{2}\mathcal{F}$.
\\
By the Proposition \ref{prop:4.4}, we have $\mathrm{\bf{f}}_{j-1}=\mathcal{Q}_{L}(\mathrm{\bf{f}}_{j})$ $\forall j\geq1$. Since $\delta_{j}\leq\not{\delta}_{j}$, Theorem \ref{main_theorem_2} follows from Theorem \ref{main_theorem_1}. $\square$
\\
\\
%Main Theorem 3%
\begin{theorem}
\label{main_theorem_3}
Fix constants $\mathrm{\bf{D}}\geq 1$, $\gamma\geq 0$. Suppose the vector $\mathrm{\bf{f}}_{0}=(\mathrm{\bf{h}}_{0},w_{0},q_{0},z_{0})\in (0,\infty)^{4}$ satisfies
\\
\begin{enumerate}
\item $w_{0}\in (0,1)\setminus\mathbb{Q}$ and $q_{0}\in (0,\infty)\setminus\mathbb{Q}$. 
\item $k_{n}\leq\mathrm{\bf{D}}\mathrm{max}\lbrace 1,n\rbrace^{\gamma}$ with $(k_{n})_{n\in\mathbb{Z}}$ as in Proposition \ref{prop:4.4}, i.e.
\begin{equation*}
(1+q_{0})^{-1}=\langle k_{0},k_{-1},k_{-2},...\rangle \:\:\:\:\:\:\: w_{0}=\langle k_{1},k_{2},k_{3},...\rangle
\end{equation*}
\item $0<\mathrm{\bf{h}}_{0}<\mathrm{\bf{A}}^{\sharp}$ where $\mathrm{\bf{A}}^{\sharp}=\mathrm{\bf{A}}^{\sharp}(\mathrm{\bf{D}},\gamma)=2^{-56}\mathrm{\bf{D}}^{-4}(4(\gamma+1))^{-4(\gamma+1)}$. 
\item $0\leq z_{0}<\mathrm{\bf{A}}^{\flat}$ where $\mathrm{\bf{A}}^{\flat}=2^{1-2\gamma} \frac{\mathrm{\bf{A}^{\sharp}}^{2}}{\mathrm{\bf{h}_{0}}}$. \\
\end{enumerate}
Then
\\
\begin{itemize}
\item the assumptions of the Theorem \ref{main_theorem_2} hold.
\item Set $\rho_{+}=\frac{1}{2}(1+\sqrt{5})$. The sequence $(\not{\delta}_{j})_{j\geq0}$ in Theorem \ref{main_theorem_2} satisfies for all $j\geq 0$:
\\
\begin{equation}
\label{assmp_on_ndelta}
\not{\delta}_{j}\leq\mathrm{exp}\left(-\frac{1}{\mathrm{\bf{h}}_{0}}\mathrm{\bf{A}}^{\sharp}\rho_{+}^{N(j)}\right) \:\:\:\:\:\:\: and \:\:\:\:\:\:\: N(j)\geq (\mathrm{\bf{D}}^{-1}j)^{1/(\gamma+1)}
\end{equation}
\\
where $N:\mathbb{Z}\rightarrow\mathbb{Z}$ (Era Counter) is the map in Proposition \ref{prop:4.4} \\
\end{itemize}
If $\gamma>1$ and $\mathrm{\bf{D}}>\frac{1}{\mathrm{log}2}\frac{\gamma}{\gamma-1}$, then the set of all vectors $\mathrm{\bf{f}}_{0}\in(0,\infty)^{4}$ that satisfy (1),(2),(3),(4) has positive Lebesgue measure.
\end{theorem}
\underline{Proof}.
\\
\emph{Preliminaries}. Fix $\mathrm{\bf{D}}\geq1$ and $\gamma\geq0$ as in Theorem \ref{main_theorem_3}. For all $s=(s_{1},s_{2},s_{3},s_{4},s_{5})\geq(0,0,0,1,0)$ with $s_{i}\in\mathbb{R}$, set
\\
\begin{equation}
\label{def_of_A_bold}
\mathrm{\bf{A}}(s)=2^{-s_{1}-s_{2}\gamma}\mathrm{\bf{D}}^{-s_{3}}(s_{4}(\gamma+1))^{-s_{5}(\gamma+1)}
\end{equation}
\\
Observe that $\mathrm{\bf{A}}(s)$ has properties $0<\mathrm{\bf{A}}(s)\leq 2^{-s_{1}}\leq1$ and $\mathrm{\bf{A}}(s)\leq\mathrm{\bf{A}}(s')$ if $s\geq s'$.
\\
\\
\emph{Basic smallness assumptions}. $k_{n}\leq\mathrm{\bf{D}}\mathrm{max}\lbrace 1,n\rbrace^{\gamma}$ for all $n\geq-2$ and $\mathrm{\bf{h}}_{0}<\mathrm{\bf{A}}(\kappa)$. The vector $\kappa=(\kappa_{1},\kappa_{2},\kappa_{3},\kappa_{4},\kappa_{5})\geq(0,0,0,1,0)$ will be fixed during the proof.
\\
\\
\emph{Estimates 1}. Recall Proposition \ref{prop:4.4}. All the old estimates still hold (assuming $k_{n}\leq\mathrm{\bf{D}}\mathrm{max}\lbrace 1,n\rbrace^{\gamma}$ holds), namely

\begin{eqnarray*}
\mathrm{\bf{C}}(w_{0},q_{0})&\leq&2\mathrm{\bf{D}}^{2}(\gamma+1)^{2(\gamma+1)}=\mathrm{\bf{A}}(1,0,2,1,2)^{-1} \\
J(n)&=&\sum_{l=1}^{n}k_{l}\leq\mathrm{\bf{D}}\sum_{l=1}^{n}l^{\gamma}\leq\mathrm{\bf{D}}n^{\gamma+1} \\
j&\leq&J(N(j))\leq\mathrm{\bf{D}}N(j)^{\gamma+1} \\
N(j)&\geq&(\mathrm{\bf{D}}^{-1}j)^{1/(\gamma+1)} \\
\mathrm{\bf{h}}_{j}&\leq& 2^{4}\mathrm{\bf{h}}_{0}\rho_{+}^{-2N(j)} \\
\mathrm{\bf{h}}_{j}&\geq&2^{-1}\mathrm{\bf{h}}_{0}\mathrm{max}\lbrace 1,2\mathrm{\bf{D}}N(j)^{\gamma}\rbrace^{-2N(j)} \\
K_{j}&\leq&\mathrm{\bf{D}}2^{\gamma}\mathrm{max}\lbrace 1,N(j)\rbrace^{\gamma} \\
\mathrm{\bf{h}}_{j}&\leq&2^{4+2\gamma}\mathrm{\bf{D}}^{2}\mathrm{\bf{h}}_{0}(\gamma+1)^{2(\gamma+1)}=\mathrm{\bf{h}}_{0}\mathrm{\bf{A}}(4,2,2,1,2)^{-1}
\end{eqnarray*}

and additionally, we estimate

\begin{eqnarray*}
\mathcal{Z}_{j}&=&z_{0}\prod_{l=1}^{N(j)-1}\frac{1}{w_{J(l)}w_{J(l-1)}} \\
&\leq&z_{0}\prod_{l=1}^{N(j)-1}(k_{l}+1)(k_{l+1}+1) \\
&\leq&z_{0}\prod_{l=1}^{N(j)-1}(2\mathrm{\bf{D}}(l+1)^{\gamma})^{2} \\
&\leq& z_{0}\frac{1}{\mathrm{max}\lbrace 1,2\mathrm{\bf{D}}N(j)^{\gamma}\rbrace^{-2N(j)}}
\end{eqnarray*}

Require $\kappa\geq(25,2,2,1,2)$, then $\mathrm{\bf{h}}_{j}<2^{-21}(K_{j})^{-2}$ and $\mathrm{\bf{h}}_{0}\leq 2^{-14}(\mathrm{\bf{C}}_{0}(w_{0},q_{0}))^{-1}$ as required by Theorem \ref{main_theorem_2}.
\\
\\
\emph{Estimates 2}. Let $(\not{\delta}_{j})_{j\geq0}$ be as in Theorem \ref{main_theorem_2}. Then, with proper choice of $\kappa$, we have:
\\
\\
(A) $\not{\delta}_{J(n)}\leq2^{-5}\frac{1}{\mathrm{\bf{h}}_{0}}(2\mathrm{\bf{D}}(n+1)^{\gamma})^{-2(n+1)}\mathrm{exp}\left(-\frac{1}{\mathrm{\bf{h}_{0}}}\mathrm{\bf{A}(\kappa)}\rho_{+}^{n+1}\right)$
\\
(B) $\not{\delta}_{j}\leq 2^{-3}\mathrm{max}\lbrace 2^{-1}\mathrm{\bf{H}}_{j},\mathcal{Z}_{j}\rbrace$, and $\not{\delta}_{j}\leq\mathrm{exp}\left(-\frac{1}{\mathrm{\bf{h}_{0}}}\mathrm{\bf{A}(\kappa)}\rho_{+}^{N(j)+1}\right)$
\\
\\
The fact that (A)$\Rightarrow$(B) follows from the vacuum case is the direct consequence of the assumption (4) in Theorem \ref{main_theorem_3} and definitions of $\mathrm{\bf{H}_{j}}$ and $\mathcal{Z}_{j}$ from Theorem \ref{main_theorem_2}. Recall the following properties from Proposition \ref{prop:4.4}: $N\circ J$ is the identity, and $J(m+1)=J(m)+k_{m+1}$. 
\\
\\
Estimate for $n\geq0$:
\\
\begin{eqnarray*}
\not{\delta}_{J(n)}&=&\sum_{m=n}^{\infty}\sum_{l=J(m)+1}^{J(m+1)}\lbrace\prod_{k=J(n)+1}^{l-1}2^{28}(K_{k})^{3}\mathrm{max}\lbrace 1,\mathcal{Z}_{k}\rbrace\rbrace 2^{75}\left(\frac{1}{\mathrm{\bf{h}}_{l}}\right)^{2}(K_{l})^{5}\mathrm{exp}\left(-(2^{21}\mathrm{\bf{h}}_{l}K_{l}^{2})^{-1}\right)\mathrm{max}\lbrace1,\mathcal{Z}_{l}\rbrace^{2} \\
&\leq&\sum_{m=n}^{\infty}\sum_{l=J(m)+1}^{J(m+1)}(2^{15}\mathrm{max}_{1\leq k\leq l} K_{k})^{3l+2}\left(\frac{1}{2\mathrm{\bf{h}}_{l}}\right)^{2}\mathrm{exp}\left(-(2^{21}\mathrm{\bf{h}}_{l}K_{l}^{2})^{-1}\right)\mathrm{max}_{1\leq k\leq l}\lbrace1,\mathcal{Z}_{k}\rbrace^{l+2} \\
&\leq&\sum_{m=n}^{\infty}k_{m+1}(2^{15+\gamma}\mathrm{\bf{D}}(m+1)^{\gamma})^{3J(m+1)+2}\frac{1}{2^{2}}\left(\frac{2}{\mathrm{\bf{h}}_{0}\mathrm{max}\lbrace1,2\mathrm{\bf{D}}(m+1)^{\gamma}\rbrace^{-2(m+1)}}\right) \\
&\times&\mathrm{exp}\left(-2^{-25-2\gamma}\mathrm{\bf{D}}^{-2}\frac{1}{\mathrm{\bf{h}}_{0}}\rho_{+}^{2(m+1)}(m+1)^{-2\gamma}\right)\mathrm{max}\lbrace1,z_{0}(2\mathrm{\bf{D}}(m+1)^{\gamma})^{2(m+1)}\rbrace^{J(m+1)+2} \\
&\leq&\left(\frac{1}{\mathrm{\bf{h}}_{0}}\right)^{2}\sum_{m=n+1}^{\infty}(2^{15+\gamma}\mathrm{\bf{D}}m^{\gamma})^{10\mathrm{\bf{D}}m^{\gamma+1}}\mathrm{exp}\left(-2^{-25-2\gamma}\mathrm{\bf{D}}^{-2}\frac{1}{\mathrm{\bf{h}}_{0}}\rho_{+}^{2m}m^{-2\gamma}\right)\mathrm{max}\lbrace1,z_{0}(2\mathrm{\bf{D}}m^{\gamma})^{2m}\rbrace^{\mathrm{\bf{D}}m^{\gamma+1}+2} \\
&\leq&\left(\frac{1}{\mathrm{\bf{h}}_{0}}\right)^{2}\sum_{m=n+1}^{\infty}(2^{15+\gamma}\mathrm{\bf{D}}m^{\gamma})^{14\mathrm{\bf{D}}m^{\gamma+2}}\mathrm{exp}\left(-2^{-25-2\gamma}\mathrm{\bf{D}}^{-2}\frac{1}{\mathrm{\bf{h}}_{0}}\rho_{+}^{2m}m^{-2\gamma}\right)\mathrm{max}\lbrace1,z_{0}\rbrace^{\mathrm{\bf{D}}m^{\gamma+1}+2} \\
&\leq&\left(\frac{1}{\mathrm{\bf{h}}_{0}}\right)^{2}\sum_{m=n+1}^{\infty}(2^{15+\gamma}\mathrm{\bf{D}}m^{\gamma})^{14\mathrm{\bf{D}}m^{\gamma+2}}\mathrm{exp}\left(-2^{-25-2\gamma}\mathrm{\bf{D}}^{-2}\frac{1}{\mathrm{\bf{h}}_{0}}\rho_{+}^{2m}m^{-2\gamma}\right)\mathrm{max}\lbrace1,z_{0}\rbrace^{3\mathrm{\bf{D}}m^{\gamma+1}}
\end{eqnarray*}
\\
Observe that $2^{5}\frac{1}{\mathrm{\bf{h}}_{0}}(2\mathrm{\bf{D}}(n+1)^{\gamma})^{2(n+1)}\leq\frac{1}{\mathrm{\bf{h}}_{0}}(2^{6}\mathrm{\bf{D}}m^{\gamma})^{2m}$ for all $m\geq n+1$. Therefore,
\\
\begin{eqnarray*}
\mathrm{\bf{S}}(n)&\overset{\mathrm{def}}=&\not{\delta}_{J(n)}2^{5}\frac{1}{\mathrm{\bf{h}}_{0}}(2\mathrm{\bf{D}}(n+1)^{\gamma})^{2(n+1)} \\
&\leq&\left(\frac{1}{\mathrm{\bf{h}}_{0}}\right)^{3}\sum_{m=n+1}^{\infty}(2^{15+\gamma}\mathrm{\bf{D}}m^{\gamma})^{16\mathrm{\bf{D}}m^{\gamma+2}}\mathrm{exp}\left(-2^{-25-2\gamma}\mathrm{\bf{D}}^{-2}\frac{1}{\mathrm{\bf{h}}_{0}}\rho_{+}^{2m}m^{-2\gamma}\right)\mathrm{max}\lbrace1,z_{0}\rbrace^{3\mathrm{\bf{D}}m^{\gamma+1}} \\
&\leq&\left(\frac{1}{\mathrm{\bf{h}}_{0}}\right)^{3}\sum_{m=n+1}^{\infty}\mathrm{exp}\left(16\mathrm{\bf{D}}m^{\gamma+2}\mathrm{log}(2^{15+\gamma}\mathrm{\bf{D}}m^{\gamma})-2^{-25-2\gamma}\mathrm{\bf{D}}^{-2}\frac{1}{\mathrm{\bf{h}}_{0}}\rho_{+}^{2m}m^{-2\gamma}\right)\mathrm{max}\lbrace1,z_{0}\rbrace^{3\mathrm{\bf{D}}m^{\gamma+1}} \\
&\leq&\left(\frac{1}{\mathrm{\bf{h}}_{0}}\right)^{3}\sum_{m=n+1}^{\infty}\mathrm{exp}\left(2^{9}\mathrm{\bf{D}}^{2}(\gamma+1)m^{\gamma+3}-2^{-25-2\gamma}\mathrm{\bf{D}}^{-2}\frac{1}{\mathrm{\bf{h}}_{0}}\rho_{+}^{2m}m^{-2\gamma}\right)\mathrm{max}\lbrace1,z_{0}\rbrace^{3\mathrm{\bf{D}}m^{\gamma+1}} \\
&\leq&\left(\frac{1}{\mathrm{\bf{h}}_{0}}\right)^{3}\sum_{m=n+1}^{\infty}\mathrm{exp}\left(2^{9}\mathrm{\bf{D}}^{2}(\gamma+1)m^{\gamma+3}-2^{-25-2\gamma}\mathrm{\bf{D}}^{-2}\frac{1}{\mathrm{\bf{h}}_{0}}\rho_{+}^{2m}m^{-2\gamma}+3\mathrm{\bf{D}}m^{\gamma+1}\mathrm{log}(\mathrm{max}\lbrace 1,z_{0}\rbrace)\right)
\end{eqnarray*}
\\
Observe that if we require $\kappa\geq(35,2,4,\frac{3}{2},3)$, the absolute value of the second term is at least twice the absolute value of the first term, that is,
\\
\begin{eqnarray*}
2\cdot2^{9}\mathrm{\bf{D}}^{2}(\gamma+1)m^{\gamma+3}\cdot 2^{25+2\gamma}\mathrm{\bf{D}}^{2}\rho_{+}^{-2m}m^{2\gamma}&=&2^{35+2\gamma}\mathrm{\bf{D}}^{4}(\gamma+1)\rho_{+}^{-2m}m^{3\gamma+3} \\
&\leq& 2^{35+2\gamma}\mathrm{\bf{D}}^{4}\left(\frac{3}{2}(\gamma+1)\right)^{3(\gamma+1)} \\
&=&\mathrm{\bf{A}}(35,2,4,\frac{3}{2},3)^{-1} \\
&\leq&\mathrm{\bf{A}}(\kappa)^{-1} \\
&\leq&\frac{1}{\mathrm{\bf{h}}_{0}}
\end{eqnarray*}
\\
Therefore,
\\
\begin{equation*}
\mathrm{\bf{S}}(n)\leq\left(\frac{1}{\mathrm{\bf{h}}_{0}}\right)^{3}\sum_{m=n+1}^{\infty}\mathrm{exp}\left(-2^{-26-2\gamma}\mathrm{\bf{D}}^{-2}\frac{1}{\mathrm{\bf{h}}_{0}}\rho_{+}^{2m}m^{-2\gamma}+3\mathrm{\bf{D}}m^{\gamma+1}z_{0}\right)
\end{equation*}
\\
Now, using the estimate $2^{26+2\gamma}\mathrm{\bf{D}}^{2}\mathrm{sup}_{m\geq1}\rho_{+}^{-m}m^{2\gamma}\leq 2^{26+2\gamma}\mathrm{\bf{D}}^{2}(2(\gamma+1))^{2(\gamma+1)}=2^{-2}\mathrm{\bf{A}}_{*}^{-1}$ with $\mathrm{\bf{A}}_{*}=\mathrm{\bf{A}}(28,2,2,2,2)$, and requiring $\kappa\geq(28,2,2,2,2)$, we get $\mathrm{\bf{h}}_{0}\leq\mathrm{\bf{A}}_{*}$. 
\\
\\
Further, 
\\
\begin{equation*}
3\mathrm{\bf{D}}m^{\gamma+1}z_{0}=3\mathrm{\bf{D}}\rho_{+}^{m}\rho_{+}^{-m}m^{\gamma+1}z_{0}\leq3\mathrm{\bf{D}}\rho_{+}^{m}(\gamma+1)^{\gamma+1}z_{0}
\end{equation*}
\\
and
\\
\begin{equation*}
3\mathrm{\bf{D}}(\gamma+1)^{\gamma+1}<2^{2}\mathrm{\bf{D}}(\gamma+1)^{\gamma+1}=\mathrm{\bf{A}}(2,0,1,1,1)^{-1}\leq\mathrm{\bf{A}_{*}^{-1}}
\end{equation*}
\\
Therefore,
\\
\begin{eqnarray*}
\mathrm{\bf{S}}(n)&\leq&\left(\frac{1}{\mathrm{\bf{h}}_{0}}\right)^{3}\sum_{m=n+1}^{\infty}\mathrm{exp}\left(-4\frac{1}{\mathrm{\bf{h}_{0}}}\mathrm{\bf{A}_{*}}\rho_{+}^{m}+z_{0}\frac{1}{\mathrm{\bf{A}_{*}}}\rho_{+}^{m}\right) \\
&\leq&\left(\frac{1}{\mathrm{\bf{h}}_{0}}\right)^{3}\sum_{m=n+1}^{\infty}\mathrm{exp}\left(-4\rho_{+}^{m}\left(\frac{1}{\mathrm{\bf{h}}_{0}}\mathrm{\bf{A}_{*}}-z_{0}\frac{1}{4\mathrm{\bf{A}_{*}}}\right)\right)
\end{eqnarray*}
\\
By assumption (4) in Theorem \ref{main_theorem_3}, we have $z_{0}<2\mathrm{\bf{A}_{*}^{2}}/\mathrm{\bf{h}_{0}}$. Therefore,
\\
\begin{equation*}
\mathrm{\bf{S}}(n)\leq\left(\frac{1}{\mathrm{\bf{h}}_{0}}\right)^{3}\mathrm{exp}\left(-\frac{1}{\mathrm{\bf{h}_{0}}}\mathrm{\bf{A}_{*}}\rho_{+}^{n+1}\right)\mathrm{exp}\left(-\frac{1}{\mathrm{\bf{h}_{0}}}\mathrm{\bf{A}_{*}}\right)\sum_{m=1}^{\infty}\mathrm{exp}\left(-2\rho_{+}^{m}\right)
\end{equation*}
\\
Using $\sum_{m=1}^{\infty}\mathrm{exp}\left(-2\rho_{+}^{m}\right)\leq 1$ and requiring $\kappa\geq(56,4,4,2,4)$, we get $\mathrm{\bf{h}_{0}}\leq\mathrm{\bf{A}_{*}^{2}}$ and  \\ $\left(\frac{1}{\mathrm{\bf{h}}_{0}}\right)^{3}\mathrm{exp}\left(-\frac{1}{\mathrm{\bf{h}_{0}}}\mathrm{\bf{A}_{*}}\right)\leq 1$. Therefore,
\\
\begin{equation*}
\mathrm{\bf{S}}(n)\leq\mathrm{exp}\left(-\frac{1}{\mathrm{\bf{h}_{0}}}\mathrm{\bf{A}_{*}}\rho_{+}^{n+1}\right)
\end{equation*}
\\
The rest of the argument is identical with the vacuum case.
\\
\\
\emph{Lebesgue measure of the set of admissible $\mathrm{\bf{f}}_{0}$}. The set of all $(\mathrm{\bf{h}}_{0},w_{0},q_{0})\in(0,\infty)^{3}$ that satisfy the assumptions (1)-(4) of the Theorem \ref{main_theorem_3}, is a product $(0,\mathrm{\bf{A}}^{\sharp})\times F_{w}\times F_{q}$ with $F_{w}\subset(0,1)\setminus\mathbb{Q}$, $F_{q}\subset(0,\infty)\setminus\mathbb{Q}$. Further, for $z_{0}$, $F_{z}$ is the area under $\mathrm{\bf{A}}^{\flat}$. Note that $(0,\mathrm{\bf{A}}^{\sharp})$, $F_{q}$ and $F_{z}$ have positive measure.
\\
\\
The rest of the proof stays the same as for the vacuum case in \cite{RT}. Suppose $\gamma>1$ and $\mathrm{\bf{D}}>\frac{1}{\mathrm{log}2}\frac{\gamma}{\gamma-1}$. Let $G(x)=\frac{1}{x}-\lfloor\frac{1}{x}\rfloor$ be the Gauss map from $(0,1)\setminus\mathbb{Q}$ to itself. Let $\mu_{G}$ be the probability measure on $(0,1)\setminus\mathbb{Q}$ with $d\mu_{G}(x)=\frac{1}{(1+x)\mathrm{log}2}dx$, and with a well-known property $\mu_{G}(X)=\mu_{G}(G^{-1}(X))$ for all measurable $X\subset(0,1\setminus\mathbb{Q})$. Observe that $k_{n+1}=\lfloor\frac{1}{G^{n}(w_{0})}\rfloor$ for all $n\geq0$.
\\
\\
For all $n\geq0$ define
\\
\begin{equation*}
\mu_{G}(X_{n})=\mu_{G}\left((0,\mathrm{\bf{D}}^{-1}(n+1)^{-\gamma})\setminus\mathbb{Q}\right)=\frac{1}{\mathrm{log}2}\mathrm{log}\left(1+\frac{1}{\mathrm{\bf{D}}(n+1)^{\gamma}}\right)\leq\frac{1}{\mathrm{log}2}\frac{1}{\mathrm{\bf{D}}(n+1)^{\gamma}}
\end{equation*}
\\
and let $X^{c}_{n}$ be the complement of $X_{n}$ in $(0,1)\setminus\mathbb{Q}$. We then have $\bigcap_{n\geq0}X_{n}^{c}\subset F_{w}$ and
\\
\begin{equation*}
\mu_{G}(F_{w})\geq\mu_{G}\left(\bigcap_{n\geq0}X_{n}^{c}\right)=1-\mu_{G}\left(\bigcup_{n\geq0}X_{n}\right)\geq1-\sum_{n\geq0}\mu_{G}(X_{n})\geq1-\frac{1}{\mathrm{\bf{D}}\mathrm{log}2}\frac{\gamma}{\gamma-1}>0
\end{equation*}
\\
Positive Gauss measure implies positive Lebesgue measure.
\\
$\square$.

\end{document}